\documentclass[fleqn,usenatbib]{mnras}

\usepackage{newtxtext,newtxmath}

\usepackage[T1]{fontenc}
\usepackage{ae,aecompl}

\usepackage{lineno,xcolor}
\usepackage{graphicx}	
\usepackage{amsmath}	
\usepackage{amssymb}	
\usepackage{float}
\usepackage{subfigure}
\usepackage{nccmath}

\newcommand{\citepbox}[1]{\mbox{\citep{#1}}}
\newcommand{\citealtbox}[1]{\mbox{\citealt{#1}}}

\newcommand{\flask}{\texttt{FLASK }}
\newcommand{\healpix}{\texttt{HEALPix }}
\newcommand{\pliny}{\texttt{PLINY }}
\newcommand{\uclcl}{\texttt{UCLCL }}
\newcommand{\uclci}{\texttt{UCLPI }}
\newcommand{\class}{\texttt{CLASS }}
\newcommand{\camb}{\texttt{CAMB }}
\newcommand{\mangle}{\texttt{MANGLE }}
\newcommand{\EQ}[2]{\begin{ceqn}\begin{equation} \label{Eq:#1} #2 \end{equation}\end{ceqn}}
\renewcommand{\Re}{\operatorname{Re}}

\title[$C_{\ell}$ Analysis of BOSS DR12 Tomography]{Cosmological Measurements from Angular Power Spectra Analysis of BOSS DR12 Tomography }

\author[Loureiro, Moraes, Abdalla, Cuceu et al.]{Arthur Loureiro$^{1}$\thanks{E-mail: arthur.loureiro.14@ucl.ac.uk}, Bruno Moraes$^{1}$\thanks{E-mail: b.moraes@ucl.ac.uk }, Filipe B. Abdalla$^{1,2}$\thanks{E-mail: fba@star.ucl.ac.uk}, Andrei Cuceu$^{1}$\thanks{E-mail: andrei.cuceu.14@ucl.ac.uk}, \newauthor Michael McLeod$^{1}$, Lorne Whiteway$^{1}$, Sreekumar T. Balan$^{1}$, Aur\'elien Benoit-L\'evy$^{3}$, \newauthor Ofer Lahav$^{1}$, Marc Manera$^{4,5}$, Richard P. Rollins$^{6}$, Henrique S. Xavier$^{7}$ \\
$^{1}$Department of Physics \& Astronomy, University College London, Gower Street, London WC1E 6BT, UK\\
$^{2}$Department of Physics and Electronics, Rhodes University, PO Box 94, Grahamstown, 6140, South Africa\\
$^{3}$CNRS, UMR 7095, Institut d'Astrophysique de Paris, F-75014, Paris, France\\
$^{4}$Institut de F\'isica d'Altes Energies, The Barcelona Institute of Science and Technology, Campus UAB, 08193 Bellaterra (Barcelona), Spain\\
$^{5}$Kavli Institute for Cosmology, University of Cambridge, Madingley Road, Cambridge CB3 0HA, UK \\
$^{6}$Jodrell Bank Centre for Astrophysics, University of Manchester, Oxford Road, Manchester M13 9PL, UK \\
$^{7}$Instituto de Astronomia, Geof\'isica e Ci\^encias Atmosf\'ericas, Universidade de S\~ao Paulo, Rua do Mat\~ao, S\~ao Paulo 05508-090, Brazil 
}

\date{Accepted XXX. Received YYY; in original form ZZZ}

\pubyear{2018}
\usepackage{breakcites}

\begin{document}
\label{firstpage}
\pagerange{\pageref{firstpage}--\pageref{lastpage}}
\maketitle

\begin{abstract}
We constrain cosmological parameters by analysing the angular power spectra of the Baryon Oscillation Spectroscopic Survey DR12 galaxies, a spectroscopic follow-up of around 1.3 million SDSS galaxies over 9,376 deg$^2$ with an effective volume of $\sim 6.5$ (Gpc $h^{-1}$)$^3$ in the redshift range $0.15 \leq  z  < 0.80$. We split this sample into 13 tomographic bins ($\Delta z = 0.05$); angular power spectra were calculated using a Pseudo-$C_{\ell}$ estimator, and covariance matrices were estimated using log-normal simulated maps. Cosmological constraints obtained from these data were combined with constraints from Planck CMB experiment as well as the JLA supernovae compilation. Considering a $w$CDM cosmological model measured on scales up to $k_{max} = 0.07h$ Mpc$^{-1}$, we constrain a constant dark energy equation-of-state with a $\sim 4\%$ error at the 1-$\sigma$ level: $w_0 = -0.993^{+0.046}_{-0.043}$, together with $\Omega_m = 0.330\pm 0.012$, $\Omega_b = 0.0505 \pm 0.002$, $S_8 \equiv \sigma_8 \sqrt{\Omega_m/0.3} = 0.863 \pm 0.016$, and $h = 0.661 \pm 0.012$. For the same combination of datasets, but now considering a $\Lambda$CDM model with massive neutrinos and the same scale cut, we find: $\Omega_m = 0.328 \pm 0.009$, $\Omega_b = 0.05017^{+0.0009}_{-0.0008}$, $S_8 = 0.862 \pm 0.017$, and $h = 0.663^{+0.006}_{-0.007}$, and a 95\% credible interval (CI) upper limit of $\sum m_{\nu} < 0.14$ eV for a normal hierarchy. These results are competitive if not better than standard analyses with the same dataset, and demonstrate this should be a method of choice for future surveys, opening the door for their full exploitation in cross-correlations probes.

\end{abstract}




\section{Introduction}
Since the discovery that the expansion of the Universe is accelerating, made by two independent Type Ia Supernovae analyses in the late 1990s \citepbox{1998Riess,1999Perlmutter}, the main questions in cosmology have concerned the nature of dark energy and dark matter. Uncertainties in some of the properties of these components have been reduced to the few-percent level, and recent results from an array of different probes support a standard cosmological model with no spatial curvature and a cosmological constant driving acceleration \citep{2017MNRAS.465.1454H,2017arXiv170801530D,PlanckCosmology2018}. Despite some suggested external and internal tensions in cosmological data sets \citep{Riess2018, PlanckCosmology2018, Efstathiou2018}, this so-called flat $\Lambda$CDM model is supported by observations from galaxy photometric surveys such as KiDS \citepbox{2017MNRAS.465.1454H} and DES \citepbox{2017arXiv170801530D}, cosmic microwave background (CMB) fluctuations \citepbox{PlanckCosmology2018}, type Ia supernovae \citepbox{Scolnic2018}, and galaxy cluster counts \citepbox{deHaan2016}, among others. At present, there is no overwhelming statistical evidence from cosmological data requiring any model extensions.

Cosmology has proven a fertile area for progress in measuring known matter components of our Universe. Measuring neutrino masses and the neutrino hierarchy is one of the great modern challenges in physics \citepbox{Kajita2016, McDonald2016}. The best current constraints from particle physics experiments place a lower limit of 0.06 $\mathrm{eV}$ to the sum of neutrino masses in the normal hierarchy \citep[e.g.][]{Esteban2017}. Due to their effect of smoothing matter perturbations on the primordial Universe, the neutrino masses can potentially be measured using probes of the large-scale structure of the Universe \citepbox{Lesgourgues2013}. Upper bounds on the sum of their masses are currently reaching a level where strong constraints on their hierarchy are possible \citepbox{PlanckCosmology2018}, and a measurement of their masses is tantalisingly close. Innovative strategies for the analysis of cosmological data could help to overcome the remaining hurdles.

Recent years have seen increased interest in measuring cross-correlations of distinct cosmological probes. Simultaneously modelling and fitting auto- and cross-correlations of observable cosmological fields can improve the dark energy figure-of-merit of surveys \citepbox{2008PhRvD..77l3525W}, provide better control of systematic errors, and potentially unveil new physics \citep[e.g.][]{Kirk2015}. Examples of this approach include combinations of CMB primary and secondary anisotropies with galaxy clustering and cosmic shear signals that help to constrain galaxy bias and intrinsic alignments \citep{Giannantonio2016, Hand2015}, `3x2pt' correlations between galaxy clustering and lensing signals which provide the strongest low-redshift constraints on cosmological models \citep{2017MNRAS.465.1454H, 2017arXiv170801530D}, and also between galaxy clustering and CMB \citepbox{2016NicolaA, 2017Nicola, Doux2017}.

A consistent treatment of all probes requires a common theoretical framework for the analysis of the data and covariance matrices across the different correlations. A natural candidate for this is the angular power spectrum. It has been in widespread use by the CMB community for decades \citep{COBE,Healpix,Polspice0,PolSpiceSzapudi2001,PolSpice2001}, providing several advantages over other statistical estimators. Spherical harmonic decompositions are particularly suited to the analysis of data on the sphere, as they are easily connected to the underlying linear cosmological perturbations in a statistically isotropic and homogeneous Universe, and possess a simple covariance structure for most practical cases despite mode mixing from partial sky observations. Construction of the estimator from galaxy survey data does not require any de-projection using cosmological information, and covariance estimation from log-normal simulations can be estimated in a cosmology-independent way. This allows for a consistent end-to-end analysis. Last, but not least, self-calibration of photometric redshift distributions using cross-correlations with spectroscopic surveys is more readily implemented, and more robust to potential systematic errors \citep{McQuinnWhite2013, 2016McLeod} when compared to other methods such as $P(k)$, $\xi(r)$ and $w(\theta)$ \citepbox{2017RossBOSS,2017SalazarBOSSwTheta}.  We argue this should be the case because methods which live in angular space such as angular power spectra and $w(\theta)$ can be naturally binned finely and hence more information about the redshift evolution can be extracted without further modelling and further assumptions. We further argue that non-linearities are better separated in our angular power spectra method than they would be if using the data in configuration space.

Spectroscopic surveys give precise information about the radial distances to galaxies, since the redshifts can be precisely measured from the spectra. In light of the precision in redshift for such galaxy surveys, the usual cosmological approach is the use the 3D power spectrum, $P(k)$, or the 3D correlation function in real space, $\xi(r)$ \citep{2001Percival,2017RossBOSS,2017BeutlerBOSS,2017WangBOSS}. Although these approaches have some advantages related to exploring the full radial information from spectroscopic surveys, a fiducial cosmology always needs to be assumed in order to translate from redshift space to real space. This choice of fiducial cosmology may potentially bias cosmological measurements, justifying once more the choice of a tomographic angular power spectra analysis.

However, there are difficulties involved in using angular power spectrum estimators on a spectroscopic galaxy survey. Firstly, it is not simple to ensure that all of this radial information is contained in the angular power spectra of projected redshift bins - even if a fine redshift binning strategy is employed. A second and more relevant issue is that spectroscopic surveys have a much lower galaxy density due to necessarily long integration times and targeting of specific galaxies with fibre spectrographs. This leads to a low signal-to-noise ratio of galaxies to Poisson noise once the data are projected in several tomographic redshift bins. A judicious choice of redshift bin width and Fourier scales can ensure that all relevant linear cosmological information is retrieved \citep{Asorey2012, Gaztanaga2012, Eriksen2015, Kirk2015}, but no consistent application of 2D angular power spectra tomography with multiple narrow bins has been attempted on real spectroscopic survey data.\footnote{\citet{2017SalazarBOSSwTheta} perform a similar analysis in real space with the BOSS DR12 galaxies.}

In this work, we apply the angular power spectrum formalism to the Baryon Oscillation Spectroscopic Survey (BOSS) 12th and final public data release. The Baryon Oscillation Spectroscopic Survey (BOSS) is one of the components of the third phase of the Sloan Digital Sky Survey (SDSS-III). Its main aim is to measure the preferred scale of baryonic acoustic oscillations in the primordial baryon-photon plasma, as imprinted in the late-time galaxy distribution. The DR12 data release contains the largest spectroscopic catalogue to date \citepbox{BOSS2015}. It is based on observations of around 2.5 million objects of which around 1.5 million were classified as galaxies, which are further selected to form a large-scale structure galaxy sample ready for cosmological analysis \citepbox{BOSSCatalogue2016}. We choose to work with this data set because of its constraining power, its public availability, and because of the possibility of comparing our results to those previously obtained by the BOSS collaboration with this same data set (see \citealtbox{2016BOSSCosmology} and the BOSS publications website\footnote{http://www.sdss3.org/science/boss\underline{~~}publications.php} for a list of cosmological publications from the collaboration).

Using the BOSS large-scale structure sample, we show that it is possible not only to measure the full shape of the angular power spectra in very thin tomographic redshift bins, but also to obtain reliable cosmological constraints for $\Lambda$CDM, $w$CDM and $\Lambda$CDM with $\sum m_{\nu}$ cosmologies using such a survey alone. The method presented here uses the full shape of the angular power spectra -- not just the BAO scale. This is achieved by separating the galaxy samples into tomographic redshift bins with $\Delta z = 0.05$, and using both the auto power spectra and the cross power spectra of adjacent bins to extract information from the radial correlation of galaxies. 
Further combination with external CMB \citepbox{PlanckCosmology2016}\footnote{The Planck 2018 likelihood was not available by the time we submitted this work.} and SNIa {ox{\citepbox{JLAdata}}} data sets achieves competitive constraints on the models mentioned above.

This paper is organised as follows: Section \ref{Sec:Data} describes the BOSS LSS sample selection criteria, the mask creation, and the construction of the galaxy overdensity maps. Section \ref{Sec:Measurements} describes the Pseudo-$C_{\ell}$ estimator used for the angular power spectrum analysis. Section \ref{Sec:Theory} describes the theoretical modelling of the angular power spectrum and the use of log-normal mocks for covariance matrix estimation. Section \ref{Sec:Systm} describes our analysis of potential systematic errors using the cross-power spectra between the data and different sources of systematic effects. Section \ref{Sec:CosmoBananas} explains the Bayesian modelling for cosmological parameter estimation, describes a series of consistency checks performed on the data, the covariance matrix, and the pipelines, and finally presents cosmological parameter constraints for flat $\Lambda$CDM, $w$CDM, and $\Lambda$CDM + $\sum m_{\nu}$ models using the BOSS $C_{\ell}$s alone and in combination with external CMB results from the Planck collaboration \citepbox{PlanckLikelihood2015} and type Ia supernovae results from JLA \citepbox{JLAdata}. {This is a publication by the \textsf{ZXCorr Collaboration}. All data products, data vectors, covariances and cosmological chains will be made available in the \textsf{ZXCorr} website: \url{http://zxcorr.org}.}

\section{BOSS DR12 Data}\label{Sec:Data}

The BOSS Data Release 12 (BOSS DR12) is the result of one of the experiments in the third phase of the Sloan Digital Sky Survey (SDSS-III); it is the largest spectroscopic redshift galaxy survey to date. See \cite{BOSS2015} for a full description of BOSS DR12 (and in particular for descriptions of the target selection criteria and of the object weighting scheme for offsetting various systematic effects).

The BOSS DR12 is subdivided into two main samples: LOWZ and CMASS. The BOSS collaboration created these samples by applying colour-magnitude and colour-colour cuts to the SDSS photometric catalogue in order to generate lists of targets for spectroscopic observation. The LOWZ sub-sample is designed as a simple extension of the original SDSS Luminous Red Galaxy (LRG) sample \citepbox{2001Eisenstein} at low redshifts, while the CMASS  sample is defined to select a stellar mass-limited sample of galaxies of all colours - hence its name, for ``constant stellar mass" - complemented by a colour cut whose goal is to select higher-redshift objects. The targets were then observed spectroscopically and objects that revealed themselves not to be galaxies (e.g. stars or quasars) were discarded. For a comprehensive discussion of the photometric cuts, selection criteria, and the terminology used, see \cite{BOSS}.

\subsection{Galaxy Catalogues}

\begin{figure}
	\includegraphics[width=\columnwidth]{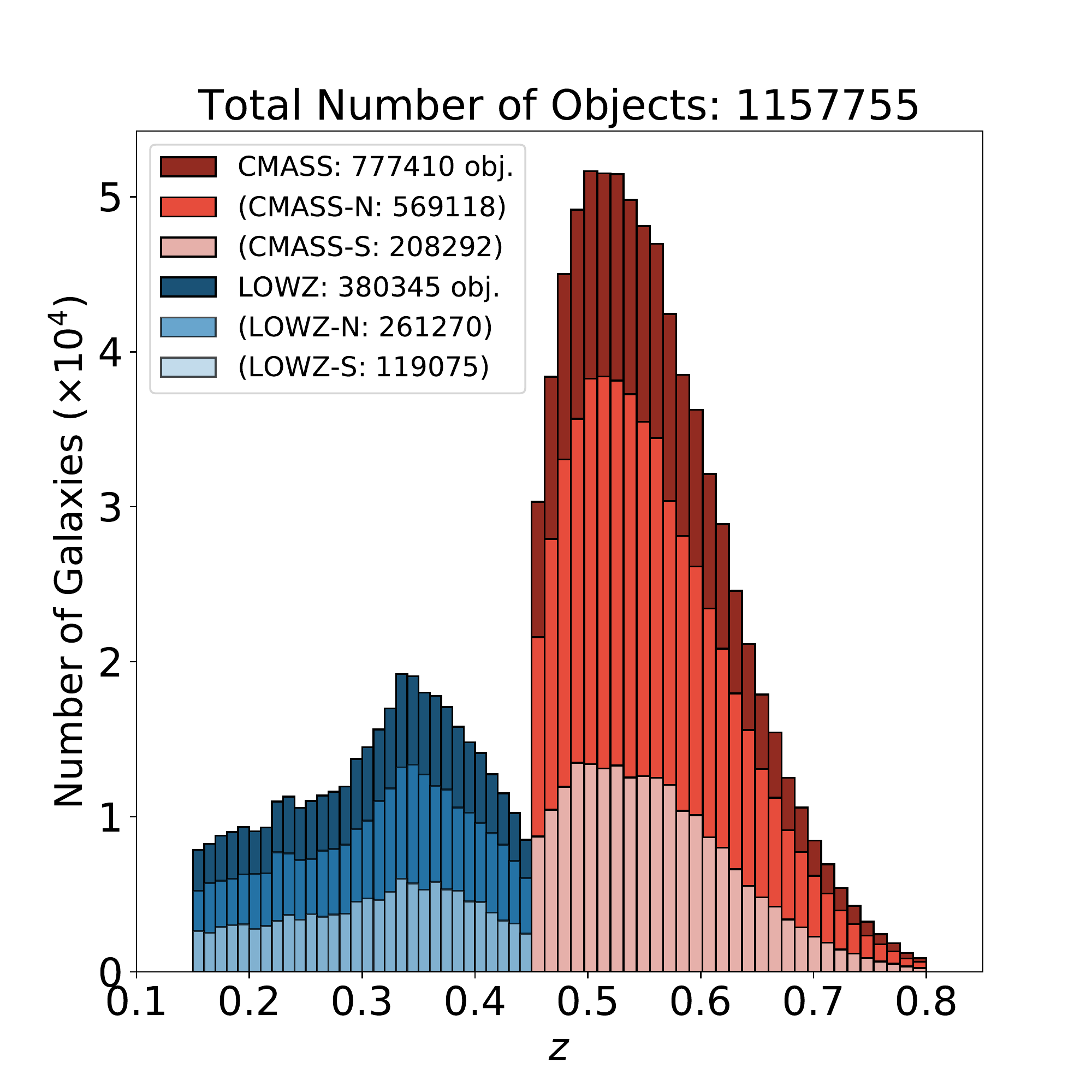}
    \caption{The redshift distribution of the BOSS samples. The darker histograms correspond to the total samples for LOWZ ($0.15\leq z <0.45$) and CMASS ($0.45\leq z < 0.80$). The overlap between samples was excluded using the redshift information, leaving a total number of 1,157,755 galaxies.}
    \label{fig:NZ_BOSS}
\end{figure}

To facilitate comparison of our results with the official BOSS collaboration results \citepbox{2016BOSSCosmology}, the construction of the catalogues used in this work followed a procedure similar to that outlined in \cite{BOSSCatalogue2016}. The data set was downloaded from the BOSS DR12 website.\footnote{http://data.sdss3.org/sas/dr12/boss/} We have further restricted these samples by applying redshift cuts of $0.15 \leq z < 0.45$ for LOWZ and $0.45 \leq z < 0.80$ for CMASS. These cuts ensure that our LOWZ and CMASS samples do not overlap in redshift, which simplifies our tomographic analysis. We use $z = 0.45$ (and not a lower $z$) as the dividing point between the two samples because the LOWZ sample has around $12\%$ more galaxies in $0.4 < z < 0.45$ than does CMASS. See figure \ref{fig:NZ_BOSS} for the resulting redshift distributions. Note also that our upper limit of $z < 0.8$ for CMASS is greater than the $z < 0.75$ limit used in \cite{BOSSCatalogue2016}. As a result of these factors our redshift ranges differ from those quoted in \cite{BOSSCatalogue2016} and \cite{2016BOSSCosmology}. The subsections that follow outline the main characteristics of the samples.

\subsubsection{LOWZ sample}
The LOWZ sample contains luminous red galaxies (LRGs) with redshifts up to around $0.45$ as a extension of the SDSS-I/II LRG Cut I sample \citepbox{2001Eisenstein}. The targets are selected at low redshifts by a cut around the predicted colour locus (Equation \ref{Eq:ColourCutLZ}), and a selection of bright red objects is done at each redshift by a variable colour-magnitude cut in the \textit{r}-band (Equation \ref{Eq:RedSelecLZ}). This is the main cut in the LOWZ sample as it produces a constant number density on the redshift range of this sample. According to \cite{BOSSCatalogue2016}, the number of galaxies in the sample is extremely sensitive to this cut (see also \cite{2013ROSS}). Star-galaxy separation is done, for LRGs, with a cut on the \textit{r}-band magnitudes as shown in Equation \eqref{Eq:StarGalLZ}. Finally, to guarantee a high spectroscopic redshift success rate, a cut is performed on the \textit{r}-band to impose a brightness limit, as shown in Equation \eqref{Eq:FinalCutLZ}.

In summary, the photometric target selection criteria for the LOWZ sample are:

\begin{ceqn}
\begin{align}
 & |c_{\perp}| < 0.2 \label{Eq:ColourCutLZ} \\
 & r_{cmod} < 13.5 + c_{\parallel}/0.3 \label{Eq:RedSelecLZ} \\
 & r_{psf} - r_{cmod} > 0.3 \label{Eq:StarGalLZ} \\
 & 16 < r_{cmod} < 19.6 \label{Eq:FinalCutLZ}
\end{align}
\end{ceqn}

\begin{figure*}
\begin{center}
	\subfigure[]{\includegraphics[scale=0.35]{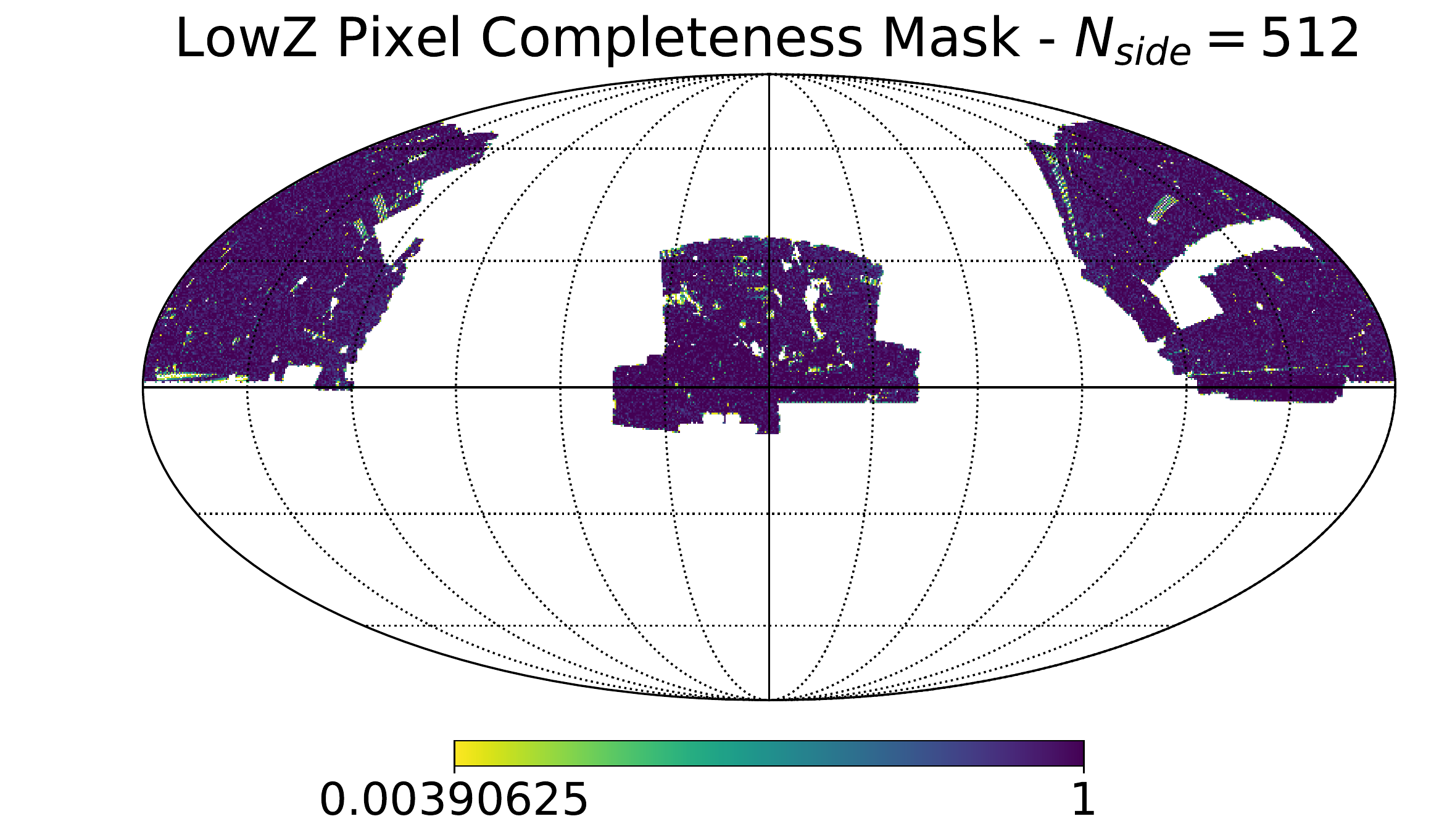}\label{fig:LOWZ_Mask}
    }
	\subfigure[]{\includegraphics[scale=0.35]{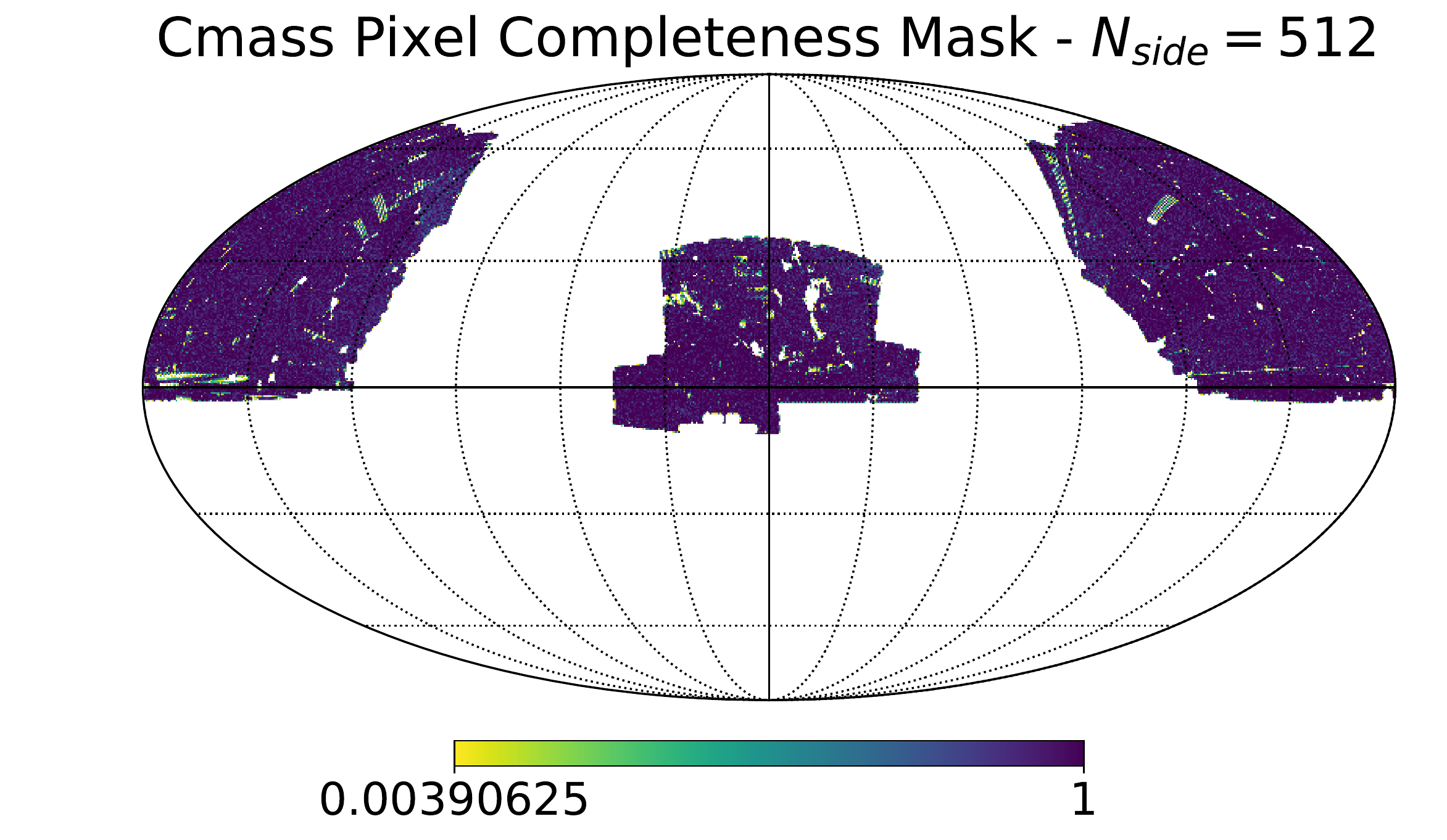}
    \label{fig:CMASS_Mask}
    }
    \caption{\textit{(a)} LOWZ final pixel completeness angular mask with $N_{side}= 512$. We excluded the LOWZE2 and LOWZE3 regions (the holes in the NGC) due to the non-standard $N(z)$ in these regions (the result of an initially different observing strategy that affected these regions). After performing a pixel completeness cut of $0.8$, the total used area of the mask is around $8529.58\deg^2$ ($f_{sky} = 0.2067$). \textit{(b)} CMASS final pixel completeness angular mask with $N_{side}=512$. After performing a pixel completeness cut of $0.8$, the total used area of the mask is around $9444.63\deg^2$ ($f_{sky} = 0.2286$).}
    \label{fig:masks}
\end{center}
\end{figure*}

In the first months of observation, the BOSS collaboration used a different star-galaxy separation criterion compared to that used later (see Appendix A from \citealt{BOSSCatalogue2016}). As a result, some sky regions from the LOWZ sample (specifically LOWZE2 and LOWZE3) have a redshift distribution that differs to that in other regions. Our method relies on having a consistent redshift distribution across the sky, and therefore we excluded these regions from our LOWZ sample (see Figure \ref{fig:LOWZ_Mask}).

\subsubsection{CMASS sample}
The CMASS sample was designed to be closer to a mass limited sample, extending the Cut-II LRGs from SDSS-I/II \citepbox{2001Eisenstein} to bluer and fainter objects using a sliding colour-magnitude cut as shown in Equation \eqref{Eq:CMCut2}. The cut in the quantity $d_{\perp}$ (Equation \ref{Eq:RedCutCM}) results in an increase in the number density of objects for the redshift range of $0.45 < z < 0.80$ (see Figure \ref{fig:NZ_BOSS}). Model and magnitude limit cuts (Equations \ref{Eq:CMCut3} and \ref{Eq:CMCut4}) ensure high redshift success rates while preventing low redshift objects from being erroneously targeted. Outliers and problematic blended objects are excluded using cuts in \textit{i}- and \textit{r}-band magnitudes (Equations \ref{Eq:CMCut5} and \ref{Eq:CMCutPix}). Finally, star-galaxy separation was done by performing a varying cut in $i_{psf} - i_{mod}$ and $z_{psf} - z_{mod}$ based on \cite{2006Cannon2Slaq} (Equations \ref{Eq:CMStarGal1} and \ref{Eq:CMStarGal2}).

In summary, the CMASS sample photometric target selection criteria for most of the survey are:

\begin{ceqn}
\begin{align}
& i_{mod} < \min (19.86 + 1.6(d_{\perp}-0.8),19.9) \label{Eq:CMCut2} \\
& d_{\perp} > 0.55 \label{Eq:RedCutCM} \\
& 17.5 < i_{cmod} < 19.9 \label{Eq:CMCut3} \\
& i_{fib2} < 21.5\label{Eq:CMCut4} \\
& r_{mod} - i_{mod} < 2 \label{Eq:CMCut5} \\
& r_{dev,i} < 20.0 \text{ pix} \label{Eq:CMCutPix} \\
& i_{psf} - i_{mod} > 0.2(21-i_{mod}) \label{Eq:CMStarGal1} \\
& z_{psf} - z_{mod} > 0.46(19.8 - z_{mod}) \label{Eq:CMStarGal2}
\end{align}
\end{ceqn}

Although around 25,000 galaxies were targeted with slightly different selection criteria (see Section 3.3.1 from \cite{BOSSCatalogue2016} for further details), this does not affect significantly the sample's redshift distribution (in the way that it did for LOWZE2 and LOWZE3 samples), and therefore we retain these galaxies.

\subsubsection{Total galaxy weights}\label{Sec:GalWeights}
Various observational effects, such as fibre collisions, will introduce bias into clustering statistics calculated from raw DR12 data. To offset this, the BOSS collaboration provides a weighting scheme for each object; clustering statistics calculated using object counts weighted by this scheme are then expected to be unbiased by such effects. The scheme is described in \cite{BOSSCatalogue2016}, which in turn was based on \cite{2014Anderson}. We use the same weighting scheme.

For each targeted galaxy the BOSS collaboration provides three components to the weighting scheme, corresponding to different observational effects \citep{2014Anderson,BOSSCatalogue2016,2017RossBOSS}:
\begin{itemize}
\item{$w_{\text{systot}}$, a combination of stellar density with airmass, sky flux, reddening, and other seeing conditions;}
\item{$w_{\text{cp}}$, which is due to close-pair objects, i. e., pairs of objects that can not have both their spectra measured due to fibre collisions;}
\item{$w_{\text{noz}}$, which takes into account nearest neighbours following a redshift failure by up-weighting such galaxies.}
\end{itemize}

We follow Equation 50 in \cite{BOSSCatalogue2016} to combine these into a single weight for each galaxy \citepbox{2017RossBOSS}:

\begin{ceqn}\begin{equation}
w_{\text{tot}} = w_{\text{systot}}(w_{\text{cp}}+w_{\text{noz}}-1) \ .
\label{Eq:Weights}
\end{equation}\end{ceqn}

The default values of $w_{\text{cp}}$ and $w_{\text{noz}}$ are unity. By construction the term inside the parentheses in Equation \eqref{Eq:Weights} conserves the total number of targeted galaxies. A more detailed study of the impact of observational systematics is presented in Section \ref{Sec:Systm}.

\subsection{Masks and Map Making}\label{Sec:MaskMaps}
We now describe the construction of the maps and masks that are our final data products. In this construction we rely on the \texttt{HEALPix}\footnote{\url{http://healpix.sourceforge.net}} software for pixelising the celestial sphere \citepbox{Healpix}. The procedures described here were used for both CMASS and LOWZ.

\subsubsection{Masks and angular selection function}\label{Sec:Masks}

The BOSS collaboration provides\footnote{\url{http://data.sdss3.org/sas/dr12/boss/lss/}} an acceptance mask and several veto masks; these are in \mangle format \citepbox{2008Mangle}.

The acceptance mask is continuous (i.e. takes values between 0 and 1), the value for a given region reflecting the completeness of observations there i.e. the extent to which spectra were obtained for all targets. The precise value is:

\begin{ceqn}\begin{equation}
C_{BOSS} = \frac{N_{obs}+N_{cp}}{N_{obs}+N_{cp}+N_{\text{missed}}} \ ,
\end{equation}\end{ceqn}

\noindent where:

\begin{itemize}
\item $N_{obs}$ is the number of spectroscopically observed objects including galaxies, stars, and unclassified objects; 
\item $N_{cp}$ is the number of close-pair objects;
\item $N_{\text{missed}}$ is the number of targeted objects with no spectra.
\end{itemize}

The veto masks are binary maps (i.e., regions are marked as either good or bad); these maps mask out regions affected by observational factors such as centerpost collisions, collision priorities, bright stars, bright objects, seeing cuts, extinction cuts, and others (see Section 5.1 in \cite{BOSSCatalogue2016}).

We transform the BOSS acceptance and veto masks into a high resolution \healpix pixelisation with $N_{side} = 16384$. Using this pixelisation scheme, we combine the BOSS masks to yield a high resolution binary mask. This is done by accepting pixels in which the acceptance mask value $C_{BOSS}$ exceeds 0.7 and which are not marked as bad in any of the veto masks; other pixels are rejected. This choice of completeness cut is based on the BOSS LSS catalogue algorithm from \cite{BOSSCatalogue2016}. This high resolution binary mask is then degraded to a lower resolution ($N_{side} = 512$) continuous mask with values $C_{pix}$ (the \textit{pixel completeness factor}), defined for a given pixel to be the fraction of high resolution sub-pixels that are marked as good in the high resolution binary mask. This is our final mask product and can be seen in Figures \ref{fig:LOWZ_Mask} for LOWZ and \ref{fig:CMASS_Mask} for CMASS. The masks used for the pseudo angular power spectrum estimator (PCL) measurements in Section \ref{Sec:Measurements} contains a hard cut in $C_{pix} \geq 0.8$: values $< 0.8$ are set to 0 and values $\geq 0.8$ are set to 1.

\begin{table}
\centering
\caption{Details of each tomographic redshift bin: redshift limits, number of objects, and shot noise. Note that shot noise is calculated after applying the galaxy weights (Section \ref{Sec:GalWeights}, Equation \eqref{Eq:Weights}).}
\label{Tb:Shells}
\begin{tabular}{lllcl}
\hline
Sample Bin & $z_{min}$ & $z_{max}$ & Num galaxies & Shot noise \\ & & & & (gal/strd)$^{-1}$ \\
\hline 
LOWZ--0  & 0.15      & 0.20      &   43,265   & $6.143 \times 10^{-5} $\\
LOWZ--1  & 0.20      & 0.25      &   51,271   & $5.156 \times 10^{-5} $\\
LOWZ--2  & 0.25      & 0.30      &   59,713   & $4.416 \times 10^{-5} $\\
LOWZ--3  & 0.30      & 0.35      &   85,394   & $3.064 \times 10^{-5} $\\
LOWZ--4  & 0.35      & 0.40      &   83,537   &  $3.136\times 10^{-5} $\\
LOWZ--5  & 0.40      & 0.45      &   57,165   &  $4.605 \times 10^{-5} $\\
CMASS--6 & 0.45      & 0.50      &   177,383  &  $1.577\times 10^{-5} $\\
CMASS--7 & 0.50      & 0.55      &   217,636  &  $1.275\times 10^{-5}$ \\
CMASS--8 & 0.55      & 0.60      &   179,571  &  $1.545\times 10^{-5}$ \\
CMASS--9 & 0.60      & 0.65      &   114,398  &  $2.435 \times 10^{-5}$ \\
CMASS--10 & 0.65      & 0.70      &   57,537   &  $4.850 \times 10^{-5}$ \\
CMASS--11 & 0.70      & 0.75      &   23,631   &  $1.182 \times 10^{-4}$ \\
CMASS--12 & 0.75      & 0.80      &    7,253   &  $3.839 \times 10^{-4}$\\
\hline
\end{tabular}
\end{table}
\subsubsection{\healpix galaxy overdensity maps}\label{Sec:Maps}

From the galaxy catalogues, we generate the final data products to be used in our analysis: the galaxy overdensity \healpix maps. First, we bin both data catalogues into tomographic redshift bins of $\Delta z = 0.05$. This gives six tomographic bins for LOWZ ($0.15 \leq z < 0.45$) and seven for CMASS ($0.45 \leq z < 0.80$). Details about each redshift bin can be found in Table \ref{Tb:Shells}. {According to \cite{Asorey2012}, $\Delta z = 0.05$ is the thickest possible redshift bin size a spectroscopic redshift survey with $z < 1$ can have in order to keep sufficient radial information without suppressing the radial BAO information due to averaging originating from mode projection. Smaller bin sizes could improve the quality of radial information; however, the trade-off between bin size and shot-noise per bin for the case considered in this work is such that the shot-noise would then be too high for the considered scales. The use of the cross-power spectra between adjacent bins allows for RSD information to be properly probed as explained in Section \ref{Sec:RSD}.} 
 
Next, we create a \textit{weighted number counts} map which contains the number of objects in each \healpix pixel, $n_p$, weighted by the \textit{total galaxy weight} ($w_{tot}$) given by Equation \eqref{Eq:Weights}. To create the final galaxy overdensity maps, we up-weight the maps by the inverse of the pixel completeness factor from the masks, $C_{pix}$. Here, objects in pixels with $C_{pix} <  0.8$ are now considered outside the footprint, i.e. the pixel value is set to zero. Thus, the expression for the overdensity maps is:

\begin{ceqn}
\begin{equation}
\delta_{i,p}^g = 
\begin{cases}
\left(\frac{1}{C_{pix,p}}\frac{n^g_{i,p}}{\bar{n}_i}\right) - 1 & \text{if } C_{pix,p} > 0.8 \\
0 & \text{otherwise,}
\end{cases}
\label{Eq:OverDMaps}
\end{equation}
\end{ceqn}
where $\bar{n}_i$ is the mean number of weighted galaxies per observed pixel, in each tomographic redshift bin. Note that the weights to which we are referring here are the ones mentioned in Equation \ref{Eq:Weights}; the $\bar{n}_i$ are not weighted by the pixel completeness weight.  After these procedures are applied to all 13 tomographic redshift bins, we are ready measure the power spectra of these maps using the Pseudo-$C_{\ell}$ estimator described in the next sections.

\section{Angular Power Spectra Estimators and Measurements}\label{Sec:Measurements}
The first proposed method for estimating the angular power spectrum $C_{\ell}$  \citepbox{Peebles1973} consists of projecting the density field onto the celestial sphere, decomposing this projected field into spherical harmonics, and then analysing statistically the coefficients of this decomposition. We refer to this method of estimating the power spectrum as a \textit{pseudo power spectrum estimator} (PCL).

In Appendix \ref{Apx:PCL1}, we describe the PCL estimator for overdensity in detail, following recent approaches as presented in e.g. \cite{ScharfLahav1992}, \cite{FisherLahav1994}, \cite{Wright1994}, \cite{2001Huterer}, \cite{PolSpice2001}, \cite{BlakeFerreira2004}, \cite{Blake2007}, \cite{Thomas2010Neutr}, \cite{Thomas2011b}, \cite{Thomas2011}, and \cite{Ho2012} and \cite{2018-FreeCitation}. In this section, we will briefly summarise how the estimator works and its main properties.

\begin{figure*}
\begin{center}
\includegraphics[scale=0.32]{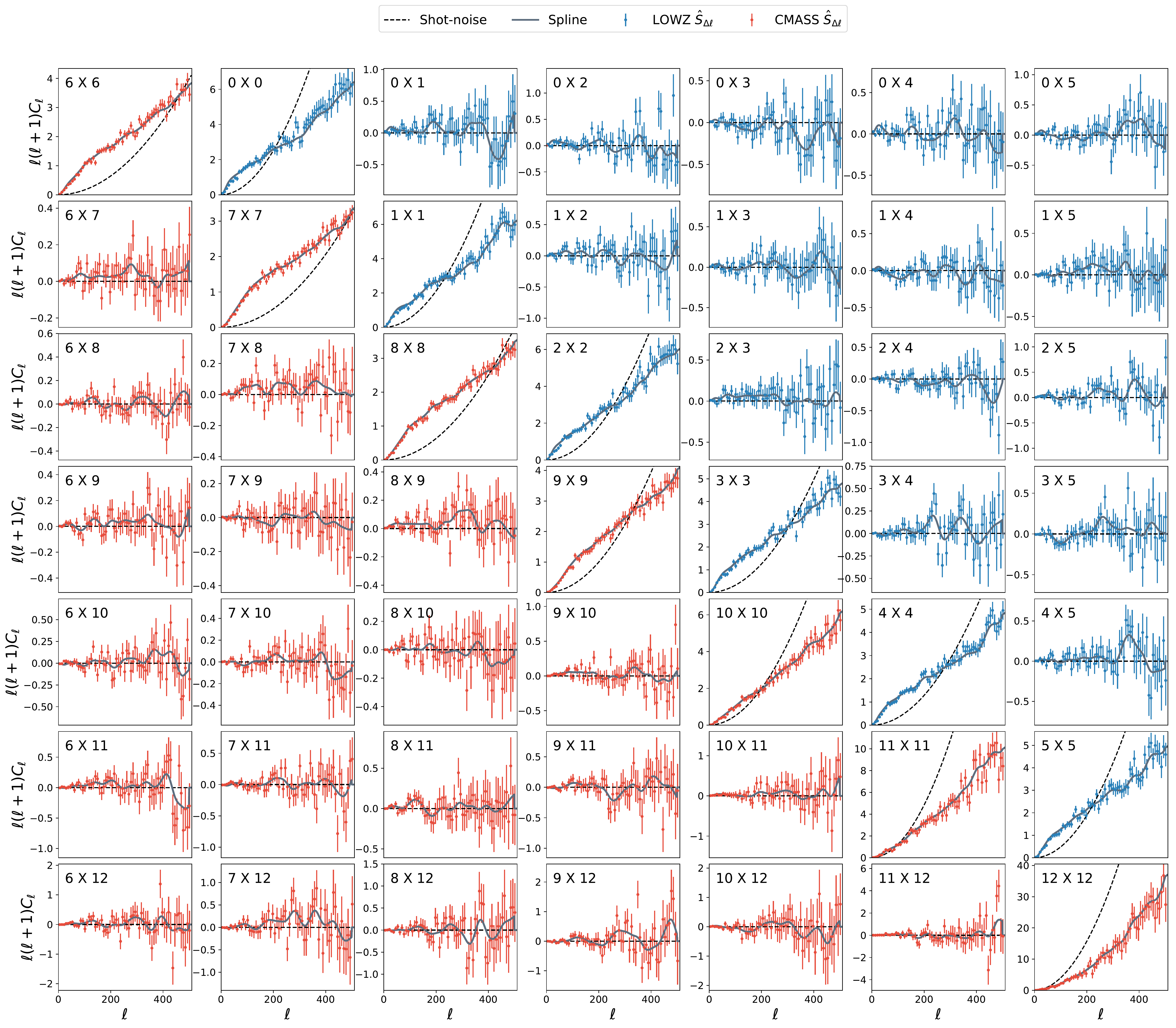}
\caption{Measured signal auto- and cross-power spectra for the LOWZ (blue) and CMASS (red) samples (Equation \ref{Eq:S_delta_ell}). The black dashed lines show the estimated Poissonian shot noise (Equation \ref{Eq:NoiseNl}). The solid grey line shows the deconvolved spline used in Section \ref{Sec:Cov} to generate the log-normal mocks for covariance matrix estimation from which the error bars in this figure were estimated. Even though the measured $\hat{S}_{\Delta\ell}$ had the shot noise removed, note that the last two CMASS bins have a significant part of their signals below the level of Poissonian shot noise.}
\label{fig:PCLs}
\end{center}
\end{figure*}

\subsection{Pseudo-$C_{\ell}$ Estimator}\label{Ref:PCL}


Our observations contain both signal and (Poisson) noise; spatial variations in the latter contribute to the measured auto power spectrum and this effect must be removed when estimating the power spectrum of the underlying signal. 

As the signal and noise are uncorrelated, the angular power spectra of the signal ($S_{\ell}$), data ($D_{\ell}$) and noise ($N_{\ell}$) are related by:

\EQ{SignalNoise}{
S_{\ell} = D_{\ell} - N_{\ell}.}

\noindent {Here $D_{\ell}$ is measured from performing a decomposition in spherical harmonics of the pixelised galaxy overdensity field (Equations \ref{Eq:AlmPix} and \ref{Eq:D_lm_ij})}. For most tomographic bins we can (using the variance of a Poisson distribution) approximate the angular power spectrum of the noise as:

\EQ{NoiseNl}{
N_{\ell} \approx \frac{\Delta\Omega_{tot}}{n^g_{tot}} = \frac{1}{\bar{n}},}

\noindent where $\bar{n}$ is the mean number of galaxies per steradians.

Amending Equation \eqref{Eq:D_hat} to account for pixelisation and shot noise yields an estimator $\hat{S}^{ij}_{\ell}$ for the partial sky signal power spectrum between two redshift bins \textit{i} and \textit{j}:

\EQ{Sl_wl}{
\hat{S}^{ij}_{\ell} = \frac{1}{w_{\ell}^2} \left[ \left(\frac{1}{(2\ell+1)}\sum_{m=-\ell}^{\ell} D^{ij}_{\ell m}\right) - N_{\ell}\delta_{ij}\right],}

\noindent {where $w_{\ell}$ is the pixel window function given by Equation \eqref{Eq:pixwin}}. Note that the estimator in Equation \eqref{Eq:Sl_wl} is symmetric in $i$ and $j$. Note also that there is no shot noise contribution for the cross-power spectra ($i \neq j$). The PCL estimator described here uses galaxy overdensity maps instead of the galaxy counts maps used in \cite{Blake2007,Thomas2011} and others; Appendix \ref{Apx:PCL2} describes the correspondence between the two approaches. This estimator is unbiased \citepbox{Peebles1973} but does not have minimum variance: maximum likelihood estimators such as QML \citep[e.g.][]{Efstat2004} have smaller variance. However, these maximum likelihood estimators are computationally expensive to use; this is why we use PCL.

\subsection{Bandwidth Binning and Measurements}

We bin the $\ell$ values into bins $\Delta\ell$ of width $8$ (so e.g. the first bin is $2 \leq \ell \leq 9$). For each bin we calculate a weighted average of the $\hat{S}_{\ell}^{ij}$ (weighted by the number of spherical harmonic coefficients):

\EQ{S_delta_ell}{
\hat{S}_{\Delta\ell}^{ij} = \frac{\sum_{\ell \in \Delta\ell}(2\ell+1)\hat{S}_{\ell}^{ij}}{\sum_{\ell \in \Delta\ell}(2\ell+1)} \ .}

\noindent This binning acts on the measurement in a way that decorrelates mixed modes (that arise from the convolution of the true measurement and the survey's angular window function). 

We measure the PCL estimator up to $\ell_{max} = 510$; Figure \ref{fig:PCLs} shows the results for the auto- and cross-power spectra for LOWZ and CMASS. We do not consider in this work any cross-correlations between the two samples. The figure also shows error bars given by the diagonal of the covariance (estimated in Section \ref{Sec:Cov}), as well as the splines used to generate the log-normal mocks (Section \ref{Sec:Cov}). The figure shows that the last two CMASS bins are dominated by shot noise (due to their small density of galaxies). Uncertainty in the characterisation of this noise will be included into the theoretical forward modelling presented in Section \ref{Sec:Theory} and marginalised over during the cosmological parameter estimation (Section \ref{Sec:CosmoBananas}).

\section{Modelling of Theory and Estimation of Covariance Matrices}\label{Sec:Theory}

\subsection{Theoretical Angular Power Spectra}
\begin{figure}
\begin{center}
\includegraphics[width=\columnwidth]{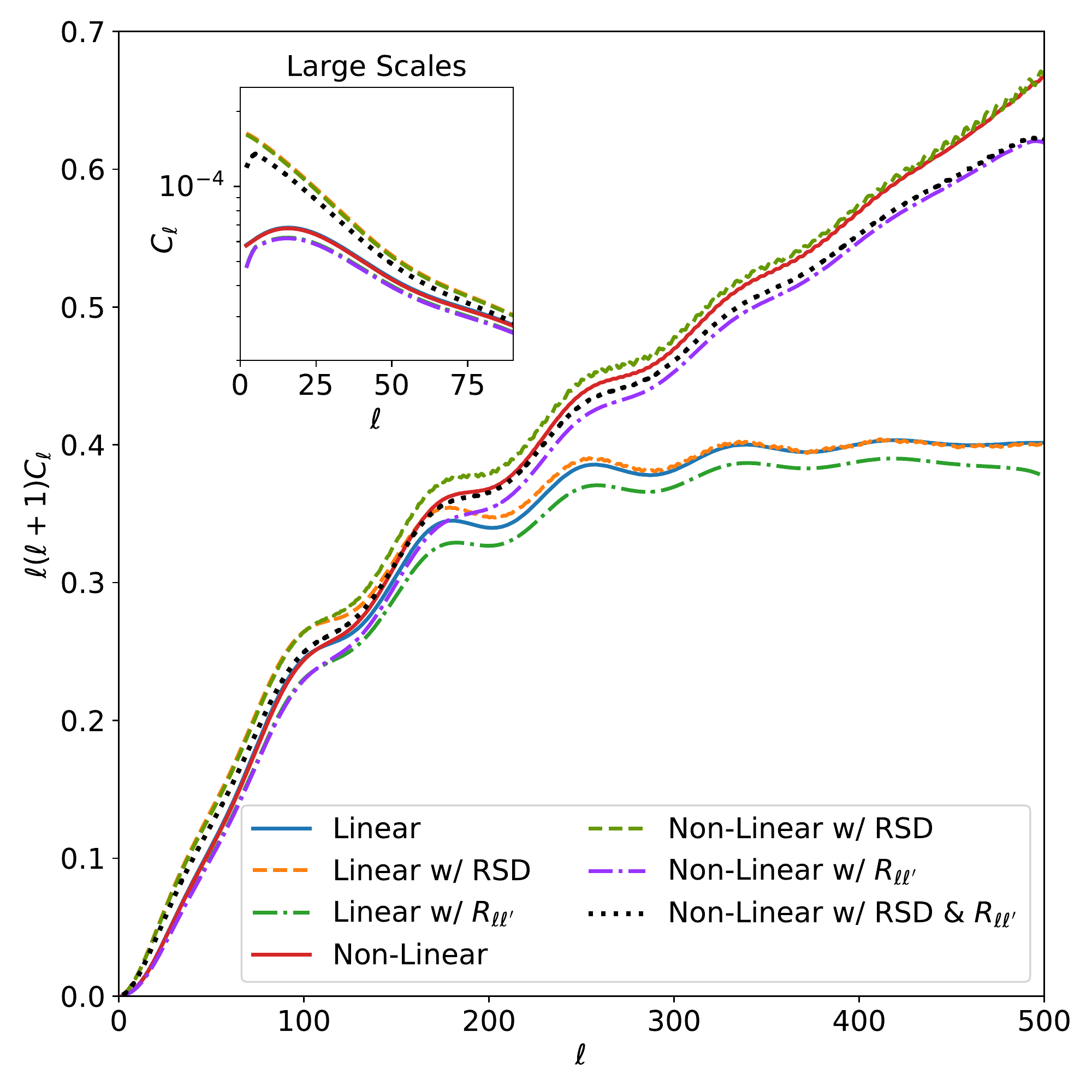}
\caption{A series of effects that impact the angular power spectrum ($0.45 < z \leq 0.50$) in a variety of ways. The two \textit{solid lines} are the linear and non-linear power spectra (Section \ref{Sec:NonLin}); they diverge as a function of scale, $\ell$. \textit{Dashed lines} include the redshift space distortions effect (Section \ref{Sec:RSD}), which increases the power for larger scales (see sub-panel). \textit{Dot-dashed lines} show the effect of the mixing matrix convolution (Section \ref{Sec:MixingMat}), which tends to suppress power on all scales. Finally, the black dotted line is a combination of all effects: RSDs, non-linearities, and mixing matrix convolution. Theoretical power spectra were calculated using the \uclcl pipeline (Cuceu et al., \textit{in prep.}), using a flat cosmology: $b=1$, $h = 0.6725$, $\Omega_b = 0.0492$, $\Omega_{cdm} = 0.265$, $w_0=-1.0$, $\tau_r = 0.079$, $\log A_s = 3.093 \times 10^{-10}$, $n_s = 0.965$.}
\label{fig:Cl_Theory}
\end{center}
\end{figure}

Our goal is to use observations to constrain cosmological parameters; as part of this we describe the theory that connects the statistics of the underlying matter field with the measured angular power spectra. Our treatment is similar to that found in the literature \citep{ScharfLahav1992,2001Huterer,Padm2007,Thomas2011,Asorey2012}.

Let $\delta_{g}(\textbf{x},z)$ denote the galaxy density function. Let $\delta_{g}(\textbf{k},z)$ be its Fourier transform; we can write this in terms of the growth function $D(z)$, the bias $b(z)$ (assumed here to be scale-independent), and the Fourier components $\delta(\textbf{k},0)$ of the underlying matter distribution at the current time:

\begin{ceqn}\begin{equation}
\delta_g(\textbf{k},z) = D(z)\delta_g(\textbf{k}) = D(z)b(z)\delta(\textbf{k},0).
\end{equation}\end{ceqn}

\noindent The correlation structure of the Fourier transform is

\begin{ceqn}\begin{equation}
\langle \delta_{g}(\textbf{k},z) \delta_{g}^*(\textbf{k}',z) \rangle = (2\pi)^3\delta^{(D)}(\textbf{k}-\textbf{k}')P_g(k,z) \
\end{equation}\end{ceqn}

\noindent where $P_g(k,z) = b(z)^2P(k,z)$ is the power spectrum of the galaxy density field and $P(k,z)$ is the power spectrum of the underlying matter density field. 

Integrating the galaxy density along the line of sight, $\hat{\textbf{n}}$, yields:

\begin{ceqn}\begin{equation}
\delta_g(\hat{\textbf{n}}) = \int_0^\infty \delta_g(\chi(z)\hat{\textbf{n}}, z) n(z) dz
\end{equation}\end{ceqn}
where $n(z)$ is the normalised redshift-dependent selection function and $\chi(z)$ is the comoving distance. The spherical harmonic components $a_{\ell m}$ of this projected galaxy distribution are:

\begin{ceqn}
\begin{align}
a_{\ell m} &= \int Y_{\ell m}(\hat{\textbf{n}}) \delta_g(\hat{\textbf{n}}) d\Omega \\
		&= \int Y_{\ell m}(\hat{\textbf{n}}) \int \delta_g(\chi(z)\hat{\textbf{n}},z) n(z) dz d\Omega \\
        &= \frac{4\pi}{(2\pi)^3} \int b(z)n(z)D(z) \int \delta(\textbf{k},0) i^{\ell} j_{\ell}(k\chi(z)) Y_{{\ell},m}(\hat{\textbf{k}}) d^3k dz\label{Eq:ThisOne}
\end{align}
\end{ceqn}
\noindent where $j_{\ell}(k\chi(z))$ are the spherical Bessel functions \citepbox{Thomas2010Neutr, Thomas2011}. The final step uses the plane wave expansion and the spherical harmonic addition theorem. 

We may collect the $z$ dependencies into a window function:

\begin{ceqn}\begin{equation}
W_{g,\ell}(k) = \int b(z) n(z)D(z)j_{\ell}(k\chi(z)) dz.\
\end{equation}\end{ceqn}

Using the window function in Equation \eqref{Eq:ThisOne} yields a simple expression for the angular power spectrum:

\begin{ceqn}\begin{align}
C_{\ell}^{ij} & \equiv \left\langle a_{\ell m}^i a_{\ell m}^{j*} \right\rangle \\
& = \frac{2}{\pi} \int W^i_{g,\ell}(k)W^j_{g,\ell}(k)k^2 P(k,0) dk.
\label{Eq:ClTheoretical}
\end{align}\end{ceqn}
Here we have introduced superscripts $i$ and $j$ to denote different redshift shells and the equation above therefore defines both auto-$C_{\ell}$ (for $i = j$) and cross-$C_{\ell}$ (for $i \neq j$). The same formalism can be used to obtain the $C_{\ell}$ between two different tracers, between photometric and spectroscopic redshift shells, etc.

This work uses the {\textit{Unified Cosmological Library for $C_{ell}$s}}, also referred to as the \uclcl code (Cuceu et al., \textit{in prep.}). This code obtains the primordial power spectra and transfer functions from the \texttt{CLASS} Boltzmann code \citepbox{Class}, and then applies Equation \eqref{Eq:ClTheoretical} to obtain the angular power spectrum. \uclcl deals with the redshift distribution in more flexible ways than does \class and \camb \citepbox{CAMB}: it allows for the input $n(z)$ distribution to be a spline and also allows it to be convolved with a Gaussian error function to take into account redshift systematic effects (Equation \ref{Eq:GaussianErrNz} in Section \ref{Sec:SpecNz}). A comparison between these codes is presented in Appendix \ref{Apx:Code_Comparison}.

\subsubsection{Spectroscopic redshift distribution and shot-noise modelling}\label{Sec:SpecNz}
The spectroscopic selection provides a full (un-normalised) $n(z)$ function for both the LOWZ and CMASS samples (see Figure \ref{fig:NZ_BOSS}). Binning is achieved by hard cuts on each of these samples in intervals of $\Delta z = 0.05$, with no overlap or gaps between bins. Despite the impressive precision of spectroscopy, to suggest that these bins have no overlap (i.e. that there is no error in the spectroscopic measurement) is unrealistic; there is overlap, and it has a significant impact on the cross correlations between bins. Spectroscopic errors are modelled within the distribution functions by a convolution with a narrow Gaussian function representing the uncertainty on a given measurement. Such a convolution is given by

\begin{ceqn}\begin{equation}
n^i(z) = \int n_*^i(z-z^\prime) \exp\left({-\frac{z^{\prime 2}}{2\sigma_s^2}}\right)\mathop{dz^\prime},
\label{Eq:GaussianErrNz}
\end{equation}\end{ceqn}
where $n_*^i(z)$ is the raw redshift distribution, $\sigma_s$ (the standard deviation of the Gaussian) is a proxy for the spectroscopic measurement error, and $n^i(z)$ is the final redshift distribution to be used in calculations. In practice, the convolution is achieved by means of a fast Fourier transform (FFT) algorithm, multiplication of the functions in Fourier space, and reverse transform. 

This convolution is also used to account for the Fingers-of-God (FoG) effect \citep{Kang2002,Percival-FoG2011}, which could dominate the measurements on $\sigma_s(z)$. This effect, which is similar to that of redshift-space distortion (RSD), arises from the peculiar motions of galaxies within virialised structures. These motions elongate structures in redshift space, smearing out the redshift distribution by the addition of Doppler shift to cosmological redshift \citepbox{1987Kaiser}. The convolution width $\sigma_s$ models the combined impact of spectroscopic redshift errors and of the FoG effect; $\sigma_s$ is then varied and marginalised over during the cosmological analysis. Due to the sensitivity of the cross-angular power spectra to these effects, a separate $\sigma_s$ is used for each redshift bin (for more details see Section \ref{Sec:LikelihoodsPriors}).

\subsubsection{Redshift space distortions}\label{Sec:RSD}

\begin{figure}
\includegraphics[width=\columnwidth]{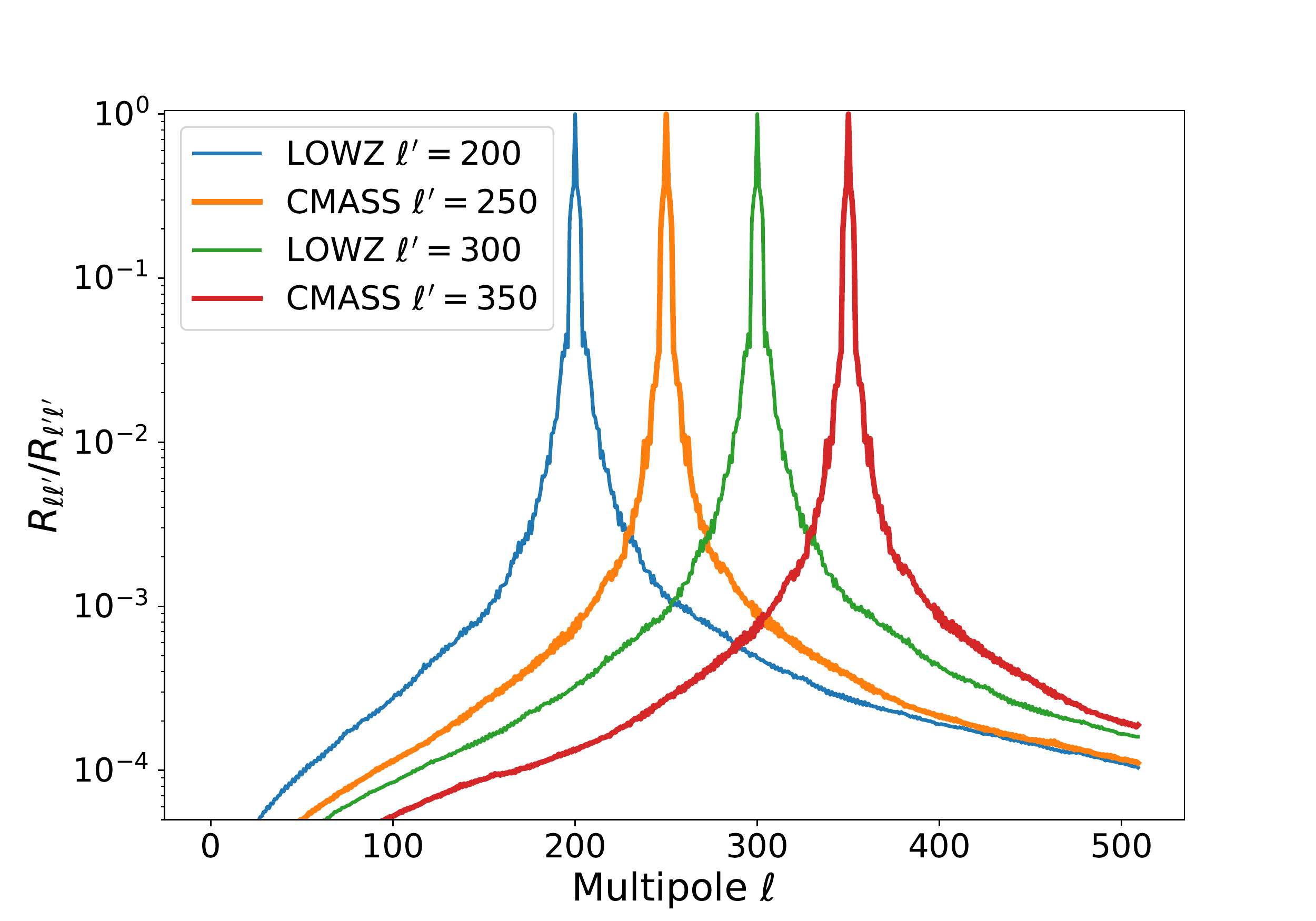}
\caption{Slices through the mixing matrices for LOWZ (resp. CMASS) using $\ell'=200$ ($250$) and $\ell'=300$($350$). Amplitudes were normalized by $R_{\ell\ell}$. As expected, the maximum amplitude peaks in the fixed $\ell'$ and approaches zero in a given $\Delta\ell$. The shape of both matrices remains the same as a function of $\ell$. This shape indicates the correlation introduced due to $\ell$-mixing by the convolution between the mask and the true signal, present in the PCL measurements (Figure \ref{fig:PCLs}).}
\label{fig:Rll_slice}
\end{figure}

The full large scale structure window function needs to take into account Redshift Space Distortions (RSD) \citep{Padm2007,Thomas2011}. This effect tends to increase the power for large scales, $\Delta \ell < 60$, due to the mix of redshift and peculiar velocities of galaxies. This creates the illusion that local peculiar motion of galaxies moving towards us appear closer (i. e. appear to be at lower redshifts); while galaxies with peculiar motion moving away from us appear to be ever further away (i. e., they appear to be at higher redshifts). This effect can be easily taken into account by adding the RSD window function  \citep{ScharfLahav1992, FisherLahav1994, Padm2007,Kirk2015, 2016McLeod} to equation \eqref{Eq:ClTheoretical}:

\begin{ceqn}\begin{equation}\label{Eq:Window_counts}
W^{Tot,i}_{\ell}(k) = W^i_{g, \ell}(k) + W^i_{RSD,\ell}(k) \ .
\end{equation}\end{ceqn}
Here the RSD window function is given by:
\begin{ceqn}\begin{equation}
\begin{split}
W^i_{RSD,\ell}(k) & = \frac{\beta^i}{k} \int d\chi \frac{dn^i}{d\chi} j'_{\ell}(k\chi (x)) \\
& = \beta^i \int \, n^i(\chi(z))\left[ \frac{(2\ell^2 + 2\ell -1)}{(2\ell + 3)(2\ell -1)}j_{\ell}(k\chi(z)) \right. \\
& + \frac{\ell(\ell-1)}{(2\ell-1)(2\ell+1)}j_{\ell-2}(k\chi(z)) \\ 
& - \left. \dfrac{(\ell+1)(\ell+2)}{(2\ell+1)(2\ell+3)}j_{\ell+2}(k\chi(z)) \right] d\chi \ ,
\end{split}
\end{equation}\end{ceqn}
where we defined the redshift distortion parameter, $\beta^i(z) = (d \ln D(z)/d\ln a)/b^{i}(z) \approx \Omega_m^{\gamma}(z)/b^i(z)$, to be dependent on the bias of the given redshift shell or tracer. The RSD window function does not account for the FoG effect, which affects small scales due to the virial motion of galaxies inside clusters \citepbox{Kang2002}; instead, as discussed in \ref{Sec:SpecNz}, the FoG effect is subsumed into the spread of the spectroscopic redshift distribution.  

Figure \ref{fig:Cl_Theory} shows the impact on the angular power spectrum of different effects considered in this section: redshift space distortions, non-linearities (Section \ref{Sec:NonLin}), and partial sky convolution with the mixing matrix (Section \ref{Sec:MixingMat}). Note that for some of these effects, the scale at which they have impact varies with redshift.

\subsubsection{Non-linear angular power spectra: Halofit}\label{Sec:NonLin}
In the \uclcl pipeline, the $C_{\ell}$ estimation may be extended some way into the non-linear regime by introducing the scale-dependent non-linear overdensity $\delta_{NL}(k,\chi)$, and therefore the corresponding non-linear growth function 

\begin{ceqn}\begin{equation}
D_{NL}(k,\chi) = \frac{\delta_{NL}(k,\chi)}{\delta_{NL}(k,0)}.
\end{equation}\end{ceqn}
The calculation of this non-linear density is extracted from the \textsc{class} code (see \cite{Class,CLASSgal}), which expresses a ratio

\begin{ceqn}\begin{equation}
R_{NL}(k,\chi) = \frac{\delta_{NL}(k,\chi)}{\delta_L(k,\chi)} = \left( \frac{P_{NL}(k,\chi)}{P_L(k,\chi)} \right)^{\frac{1}{2}}
\end{equation}\end{ceqn}
of the non-linear perturbations to the linear ($\delta_L(k,\chi)$); the second equality follows from $P = \langle \delta \delta^* \rangle$. This ratio is calculated using the modified \textsc{halofit} of \cite{Takahashi2012} (also employed by \textsc{camb sources} \citepbox{CambSources}), with additional corrections from \cite{Bird2012} for neutrino effects. 

The window function in Equation \eqref{Eq:Window_counts} contains both the selection function \emph{and} the growth function, which tracks the ratio of the power spectrum at different redshifts. The non-linear power spectrum is related to the linear, present day power spectrum by:

\begin{ceqn}\begin{equation}
\begin{split}
P_{NL}(k,\chi) & = R^2_{NL}(k,\chi) P_L(k,\chi) \\
& = R^2_{NL}(k,\chi) D_L^2(\chi) P_L(k,0).
\end{split}\end{equation}\end{ceqn}
This means that the window functions in Equation \eqref{Eq:Window_counts} should have an additional factor of $R_{NL}(k,\chi)$ inside the integral over $\chi$. In the case of these very narrow spectroscopic redshift bins, we take

\begin{ceqn}\begin{equation}
R_{NL}(k,z) = R_{NL}(k,\bar{z})
\end{equation}\end{ceqn}
where $\bar{z}$ is the mean of the redshift bin i.e. we assume that the non-linear ratios vary negligibly over the width of a single bin (but may vary between different bins). This simplifies the calculation of the window function considerably, and is a good approximation when the width of the bin is small. In this case the window functions for the redshift bins are related in a straightforward way to their linear counterparts:

\begin{ceqn}\begin{equation}
W^i_{NL,\ell}(k) = R_{NL}(k,\bar{z}^i)W^i_{g,\ell}(k).
\end{equation}\end{ceqn}
The rest of the calculation may proceed as usual. 

\subsubsection{Partial sky: mixing matrix convolution}\label{Sec:MixingMat}
When dealing the PCL estimator measurements, partial sky effects mean that we must calculate the convolution of the theory and the survey's angular selection function. It is computationally expensive and unstable to deconvolve this effect from the measurements. This leads to forward modelling, where the experimental systematics are modelled and introduced into the theoretical predictions \citep{ScharfLahav1992,FisherLahav1994,Thomas2011}. This effect is taken into account through a convolution with the mixing matrix, $R_{\ell \ell'}$ \citep{Peebles1973_2,PolSpice2001,PolSpice2005,Blake2007}:

\begin{ceqn}\begin{equation}
S_{\ell} = \sum_{\ell'}R_{\ell \ell'} C_{\ell'} \ .
\label{Eq:Cl_Conv}
\end{equation}\end{ceqn}

The mixing matrix itself depends only on the survey's geometry through the mask's angular power spectrum

\begin{ceqn}\begin{equation}
W_{\ell} = \sum_{m=-\ell}^{\ell} \frac{|I_{\ell m}|^2}{(2\ell +1)}
\end{equation}\end{ceqn}
where (see Appendix \ref{Apx:PCL2})

\EQ{Ilm}{ I_{\ell m} = \sum_p^{N_{pix}} Y^*_{\ell m} (\theta_p,\phi_p)\Delta\Omega_p;}
the mixing matrix is then

\begin{ceqn}\begin{equation}
R_{\ell \ell'} = \dfrac{2\ell' + 1}{4\pi}\sum_{\ell ''}(2\ell'' + 1)W_{\ell ''}\begin{pmatrix} \ell & \ell' & \ell'' \\ 0 & 0 & 0 \end{pmatrix}^2.
\label{Eq:MixMat}
\end{equation}\end{ceqn}

\noindent The $2 \times 3$ matrix above is the Wigner \textit{3j} function; these coefficients were calculated using the \texttt{WIGXJPF} library \citepbox{Wig3j}. The mixing matrices are shown in detailed slices in Figure \ref{fig:Rll_slice} which gives an intuition about the size of $\Delta\ell$-bands used to bin the \textit{measured} $\hat{S}_{\ell}$s as it shows the range of multipoles that are mixed due to the survey's mask. These small correlations between the multipoles can be ``washed away" by binning our measurements. In addition, as can be seen in Figure \ref{fig:Cl_Theory}, the mixing matrix convolution tends to suppress power in all scales.

Finally, after being convolved with the mixing matrix (Equation \ref{Eq:Cl_Conv}), the theoretical $S_{\ell}$ is binned in the same way as the data in Equation \eqref{Eq:S_delta_ell}:

\begin{ceqn}\begin{equation}S_{\Delta\ell}^{ij} = \frac{1}{\sum_{\ell'}^{\ell'+\Delta\ell}(2\ell+1)}\sum_{\ell'}^{\ell'+\Delta\ell}(2\ell+1)S_{\ell}^{ij}  \ .
\label{Eq:S_delta_ell2}
\end{equation}\end{ceqn}

\subsection{Data Theoretical Covariance}\label{Sec:TheoCov}

\begin{figure}
\includegraphics[width=\columnwidth]{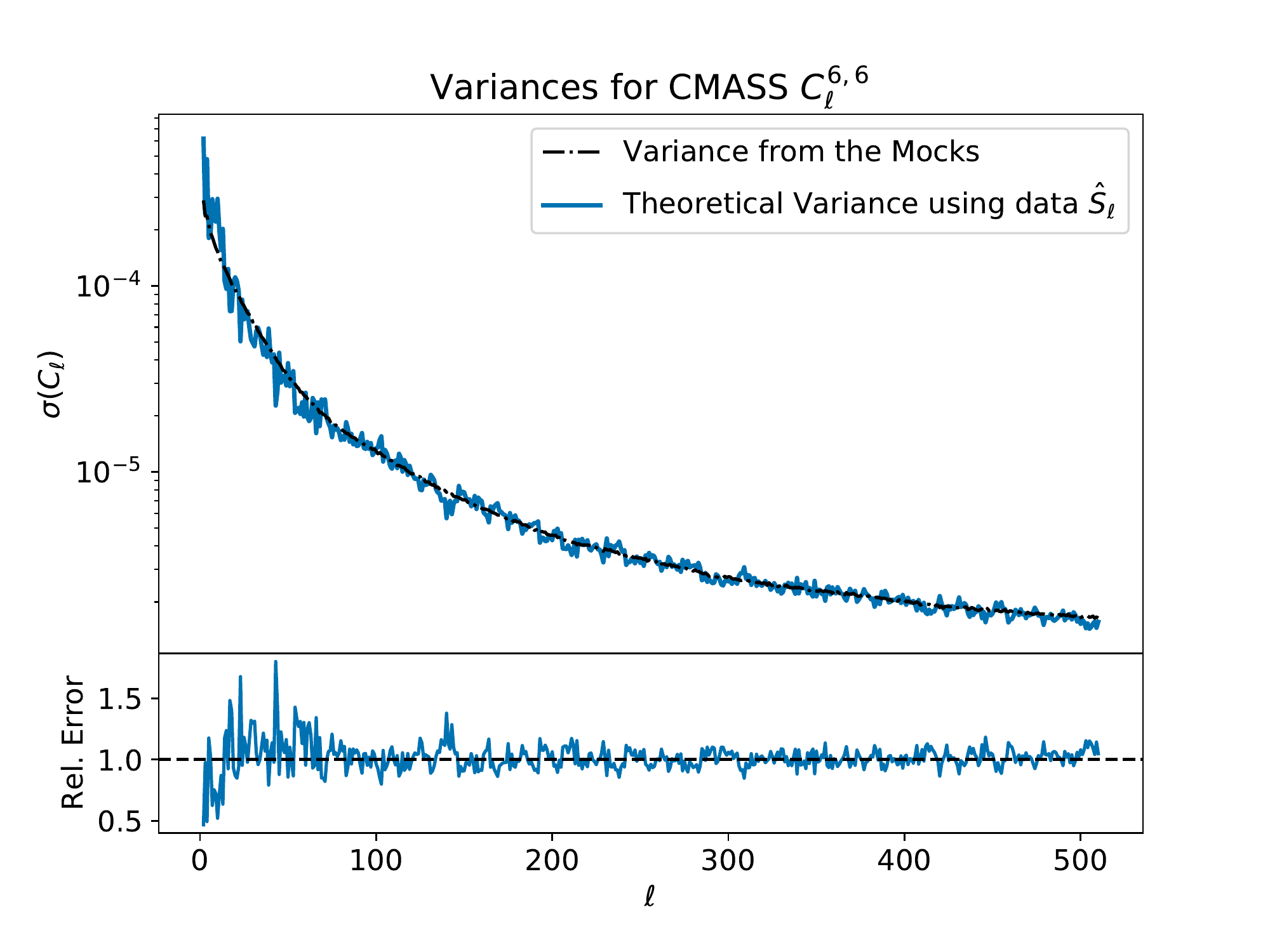}
\caption{A typical example of the validation of the covariance matrix obtained from the \flask log-normal simulations; here we show the result for the first auto-power spectrum for CMASS. We show the analytical expression for the angular power spectrum variance (Equation \ref{Eq:TheoVariance}) and the variance from the simulations. The bottom panel shows the relative error for this example. All 49 measured $C_{\ell}$s from Figure \ref{fig:PCLs} were validated in this way and no trends were apparent.}
\label{fig:Mocks_Variance}
\end{figure}
We follow here the formalism developed in \cite{2008DahlenSimons} for the covariance of spectral estimation on a sphere. For clarity we first re-derive some of the results from Section \ref{Sec:Measurements} from a different perspective, that of projectors in pixel space. 

Consider a data vector $\textbf{d}$ that is a sum of signal and noise ($\textbf{d}(\textbf{r}) = \textbf{s}(\textbf{r}) + \textbf{n}(\textbf{r})$) and that has a covariance, $\mathcal{D}$, that is a combination of signal covariance $\mathcal{S}$ and a noise covariance $\mathcal{N}$. In pixel space, the data covariance can be expressed as:

 \EQ{DataCov}{\mathcal{D} = \langle \textbf{s} \textbf{s}^T \rangle + \langle \textbf{n} \textbf{n}^T\rangle = \sum_{\ell} (S_{\ell} + N_{\ell})\mathcal{P}_{\ell}}
 where $\mathcal{P}_{\ell}$ is the projector in pixel space, defined as:
 
\EQ{Projector}{\mathcal{P}_{\ell} = \sum_{m}Y_{\ell m}(\textbf{r})Y^*_{\ell m}(\textbf{r}).}
The projector satisfies the following identity in the full sky case:

\EQ{ProjIdentFullSky}{tr(\mathcal{P}_{\ell}\mathcal{P}_{\ell'}) = (\Delta\Omega_p)^{-2}(2\ell + 1)\delta_{\ell\ell'}.}
Using this identity, the Pseudo-$C_{\ell}$ estimator from Equation \eqref{Eq:Sl_wl} can be written in terms of the projector and data covariance from \eqref{Eq:DataCov}:

\EQ{}{\hat{S}_{\ell} = \frac{\Delta\Omega_p^2}{(2\ell + 1)}\left[ \textbf{d}^T \mathcal{P}_{\ell}\textbf{d} -tr(\mathcal{N}\mathcal{P}_{\ell})\right].}

\noindent where $\Delta\Omega_p$ is the area of the pixels.

Assuming a Gaussian signal \citepbox{Blake2007}, the covariance matrix for the angular power spectra estimator between different multipoles $\ell$ and $\ell'$ can be expressed as:

\begin{ceqn}
\begin{align}
\Sigma_{\ell \ell'} & = \mathcal{C}ov(\hat{S}_{\ell}, \hat{S}_{\ell'}).
\label{Eq:ThCovSimple}
\end{align}
\end{ceqn}
The symmetry of the $\mathcal{P}_{\ell}$ and $\mathcal{D}$ matrices, together with the definition $\mathcal{C}ov(\textbf{X},\textbf{X}') = \langle \textbf{XX}'\rangle - \langle \textbf{X}\rangle\langle \textbf{X}'\rangle$, allows us to rewrite the covariance as:

\begin{ceqn}
\begin{align}
\Sigma_{\ell \ell'}& = \frac{2(\Delta\Omega_p)^4}{(2\ell+1)(2\ell'+1)}tr(\mathcal{D}\mathcal{P}_{\ell}\mathcal{D}\mathcal{P}_{\ell'})
\label{Eq:CovTrace}
\end{align}
\end{ceqn}

This expression works for both full and partial sky cases. The difference between the two cases appears on the projector identity from Equation \eqref{Eq:ProjIdentFullSky}. Using the definition of the pixel space projector (Equation \ref{Eq:Projector}), the $I_{\ell m}$ expression from Equation \eqref{Eq:Ilm}, and the fact that the spectra we consider are \textit{moderately coloured}, which means that the spectra do not vary drastically within the range considered \citepbox{2008DahlenSimons}, one can rewrite Equation \eqref{Eq:ThCovSimple} for a partial sky observation with area $\Delta\Omega_{tot}$ as:

\begin{ceqn}
\begin{align}
\Sigma_{\ell \ell'}   = & \frac{1}{2\pi}\left(\frac{4\pi}{\Delta\Omega_{tot}}\right)^2 (S_{\ell} + N_{\ell})(S_{\ell '}+ N_{\ell '}) \\ \nonumber
& \times \sum_{\ell ''}(2\ell ''+1)W_{\ell''}\begin{pmatrix} \ell & \ell' & \ell'' \\ 0 & 0 & 0 \end{pmatrix}^2 \\ \nonumber
& = \frac{2}{f_{sky}(2\ell' + 1)}(S_{\ell} + N_{\ell})(S_{\ell '}+ N_{\ell '})R_{\ell\ell'}
\label{Eq:CovRll}
\end{align}
\end{ceqn}
where the last equality uses the definition of the mixing matrix from Equation \eqref{Eq:MixMat} and $f_{sky}$ is the observed fraction of the sky. This expression is similar to ones used in \cite{Blake2007,Padm2007,Thomas2011}, but has been extended to account for the mixing of modes due to the mask. 

However, this expression (derived by \cite{2008DahlenSimons}) accounts neither for the pixel window function effect nor for cross-correlations between tomographic redshift bins. To include these effects, we generalise the data angular power spectra by changing $S_{\ell} + N_{\ell}$ to $w_{\ell}^2S_{\ell}^{ij} + N_{\ell}\delta_{ij} = D^{ij}_{\ell}$ and we include the effect of cross-correlation in the covariance by changing $(S_{\ell} + N_{\ell})(S_{\ell'} + N_{\ell'})$ to $\frac{1}{2}[D^{ij}_{\ell}D^{ij}_{\ell'} + D^{ii}_{\ell}D^{jj}_{\ell'}]$ in Equation \eqref{Eq:CovRll} \citepbox{Rassat2007}. 

The final expression for the angular power spectra theoretical covariance matrix is therefore:

\begin{ceqn}
\begin{align}
\Sigma_{\ell \ell'}^{ij}  = & \frac{1}{f_{sky}(2\ell'+1)}[D^{ij}_{\ell}D^{ij}_{\ell'} + D^{ii}_{\ell}D^{jj}_{\ell'}] R_{\ell \ell'} \\ \label{Eq:TheoVariance}
= & \frac{R_{\ell\ell'}}{f_{sky}(2\ell'+1)} \left[(w_{\ell}^2S_{\ell}^{ij} + N_{\ell}\delta_{ij})(w_{\ell'}^2S_{\ell'}^{ij} + N_{\ell'}\delta_{ij}) \right. \\ \nonumber
& \left. + \, (w_{\ell}^2S_{\ell}^{ii} + N_{\ell}\delta_{ii})(w_{\ell'}^2S_{\ell'}^{jj} + N_{\ell'}\delta_{jj}) \right]
\end{align}
\end{ceqn}

By performing these modifications, Equation \eqref{Eq:TheoVariance} recovers the variance expression for cross-power spectra from \cite{Rassat2007} when considering just the diagonal; and recovers the original expression by \cite{2008DahlenSimons} when considering just the auto-power spectrum. Figure \ref{fig:Mocks_Variance} shows a comparison between the variance (the diagonal) of equation \eqref{Eq:TheoVariance} with the variance from the estimated covariance matrix from Section \ref{Sec:Cov}. {Note that the covariance between different angular power spectra is considered to be zero, i. e. $\Sigma^{ij,i'j'}_{\ell\ell'} = \Sigma^{ij}_{\ell\ell'} \delta_{ii'}\delta_{jj'}$.}


\subsection{Covariance Matrices using Log-Normal Mocks}\label{Sec:Cov}
We seek to constrain cosmological parameters using observations; one of the requirements of this process is accurate covariance matrices. Covariances can be estimated using galaxy clustering simulations that reflect not only the cosmology but also systematic effects and observational artefacts. Previous works have used either Gaussian realisations \citep{Blake2007,Thomas2011,2016Nicola} or the mocks provided by the BOSS Collaboration \citep{2016BOSSMocks,Manera2013}. However this work instead uses log-normal simulations. The decision not to use the official BOSS \texttt{PATCHY} mocks from \cite{2016BOSSMocks} was made due to the different choice of redshift ranges for our samples: the CMASS \texttt{PATCHY} mocks do not contain galaxies beyond redshift $z = 0.75$ whereas the samples we selected extend to $z = 0.80$ (as described in Section \ref{Sec:Data}). 

We generated our mocks using \texttt{FLASK}\footnote{\url{http://www.astro.iag.usp.br/~flask/}} \citepbox{Flask2016}, a publicly available code that produces log-normal simulations of correlated fields on the sphere. We used the data $\hat{S}_{\ell}$ measurements (Section \ref{Sec:Measurements}) as inputs for the simulations; this allows us to reproduce systematic effects, RSD, non-linear power spectra, and other known and unknown effects that may be present in the data (with no need to model the effects nor to assume any fiducial cosmology). This is a main benefit of this approach to covariance estimation: any effects present in the measured angular power spectra will be reproduced in \texttt{FLASK}'s simulations via the $S_{\ell}$s measured from the data. 

For each sample we produced 6,000 log-normal mocks to estimate the data covariance matrix. These mocks were also Poisson sampled to reproduce noise properties and radial and angular selection effects. 

The data covariance matrix was produced as follows:

\begin{enumerate}
\item[\textbf{1.}] Produce a spline, $\tilde{S}(\ell)$, using the $\hat{S}_{\Delta\ell}$ measurements (Figure \ref{fig:PCLs}) and a Gaussian filter to smooth the measurements.
\item[\textbf{2.}] Deconvolve the mixing matrix $R_{\ell\ell'}$ from the splines to obtain

\begin{ceqn}\begin{align}
\tilde{C}^{ij}(\ell) = \sum_{\ell'}R_{\ell\ell'}^{-1}\tilde{S}^{ij}(\ell).
\end{align}\end{ceqn}
\item[\textbf{3.}] Linearly extrapolate the splines to $\ell_{max} = 8192$ (necessary to allow \texttt{FLASK} to create high resolution \healpix maps).
\item[\textbf{4.}] For each tomographic redshift bin, produce \texttt{FLASK} partial sky galaxy number count mocks with $N_{side} = \ell_{max} = 2048$.\footnote{The signal realisation maps were sampled using a log-normal transformation. Due to the transformation's non-linearity, we had to generate mocks with a higher $N_{side} \quad \& \quad\ell_{max}$ than the data as the log-normal realisations introduce a damping after a certain $\ell$ (see figure 18 from \cite{Flask2016}). The simulated data maps also used a $N_{side}=2048$ version of the masks presented in \ref{Sec:Masks}.}
\item[\textbf{5.}] Degrade the mocks to $N_{side}=512$ to match the $N_{side}$ used when analysing the data.
\item[\textbf{6.}] Produce up-weighted galaxy overdensity maps using the pixel completeness factor (as described in Section \ref{Sec:Maps}).
\item[\textbf{7.}] Run the partial sky PCL estimator; include here the pixel window function correction $w_{\ell}^2$ (as described in Equations \eqref{Eq:Sl_wl} and \eqref{Eq:S_delta_ell}) that arises from the degrading of the maps at step \textbf{5}.
\item[\textbf{8.}] Measure the covariance of the ensemble of angular power spectra obtained from the simulated data:

\begin{ceqn}\begin{equation}
\mathcal{C}^{ij}_{\Delta\ell\Delta\ell'} \equiv \frac{1}{N_S-1}\sum^{N_S}_{s=1}\left(S_{\Delta\ell}^{ij,s} - \langle S_{\Delta\ell}^{ij} \rangle \right)\left(S_{\Delta\ell'}^{ij,s} - \langle S_{\Delta\ell'}^{ij} \rangle \right)^T.
\label{Eq:Covariance}
\end{equation}\end{ceqn}
\end{enumerate}
\noindent Here $N_S$ is the number of simulations. 
To validate the estimated covariance matrix, we compared the diagonal of the covariance matrix in Equation \eqref{Eq:Covariance} with the expression for the theoretical variance for the measured angular power spectra in Equation \eqref{Eq:TheoVariance}; Figure \ref{fig:Mocks_Variance} shows a typical result.

\section{Systematics Tests}\label{Sec:Systm}
Large-scale survey observations, spread over thousands of observation hours, are taken under a variety of conditions. Turbulence in the atmosphere, sky background brightness and telescope inclination angle are amongst the factors that can influence image quality and object detection. Other than those atmospheric effects, galactic properties are also at play: extinction from dust within the Milky Way and variations of stellar density, as well as the presence of bright stars, are position-dependant and also have an impact on our ability to detect galaxies. Jointly, those observational factors can create small density fluctuations in the galaxy distribution that can imprint a statistical signal easily confused with the cosmological large-scale structure fluctuations that we are attempting to measure. This effect has been detected and corrected for in several previous analyses with a range of datasets \citep{Blake2007, 2011MNRAS.417.1350R, Thomas2011, Boris2013,2014BorisSyst2,Ho2012, 2016ElsnerSyst1,2017ElsnerPCLModeProj,Doux2017}.

In this section, we present the analysis performed on the data to ensure that the measured power spectra are not significantly dominated by any known observational systematic effects. We consider a systematic to have a significant effect on the observed power spectra if the cross-power spectra between them deviates from zero, with a deviation that is bigger than both the data variance and the cross-power spectra variance. We describe the systematic effects considered in our analyses, describe our methods for map creation and cross-spectrum measurement, and give some representative results. 

\subsection{Systematic Maps}\label{Sec:SystMaps}
The Sloan Digital Sky Survey monitors and records observational conditions for every tile of the survey. This information is available as a combined set of two files, one that defines a pixelisation of the observed sky in \mangle format and another that records the observational information for each \mangle polygon.\footnote{The files, \texttt{window\textunderscore unified.fits} and \texttt{window\textunderscore flist.fits}, can be found in \url{http://www.sdss.org/dr12/algorithms/resolve/}, together with a detailed description of the construction of the survey geometry and of the \texttt{score} quantity described further in the text.} The first step is to reconstruct the \mangle maps for each observational systematic from these files. We use the \mangle python wrapper\footnote{\url{https://github.com/mollyswanson/manglepy}} to perform this transformation. Since there is potentially more than one observation in a given region of the sky, there can be multiple values for a given polygon. The SDSS files indicate which amongst multiple options is to be taken as the primary value for the field. We select the IDs from those primary fields, match them to their observational properties in the fields list, and create a new \mangle mask for each of those properties, which are recorded in the weight of the masks.

\begin{figure*}
\begin{center}
\includegraphics[scale=0.31]{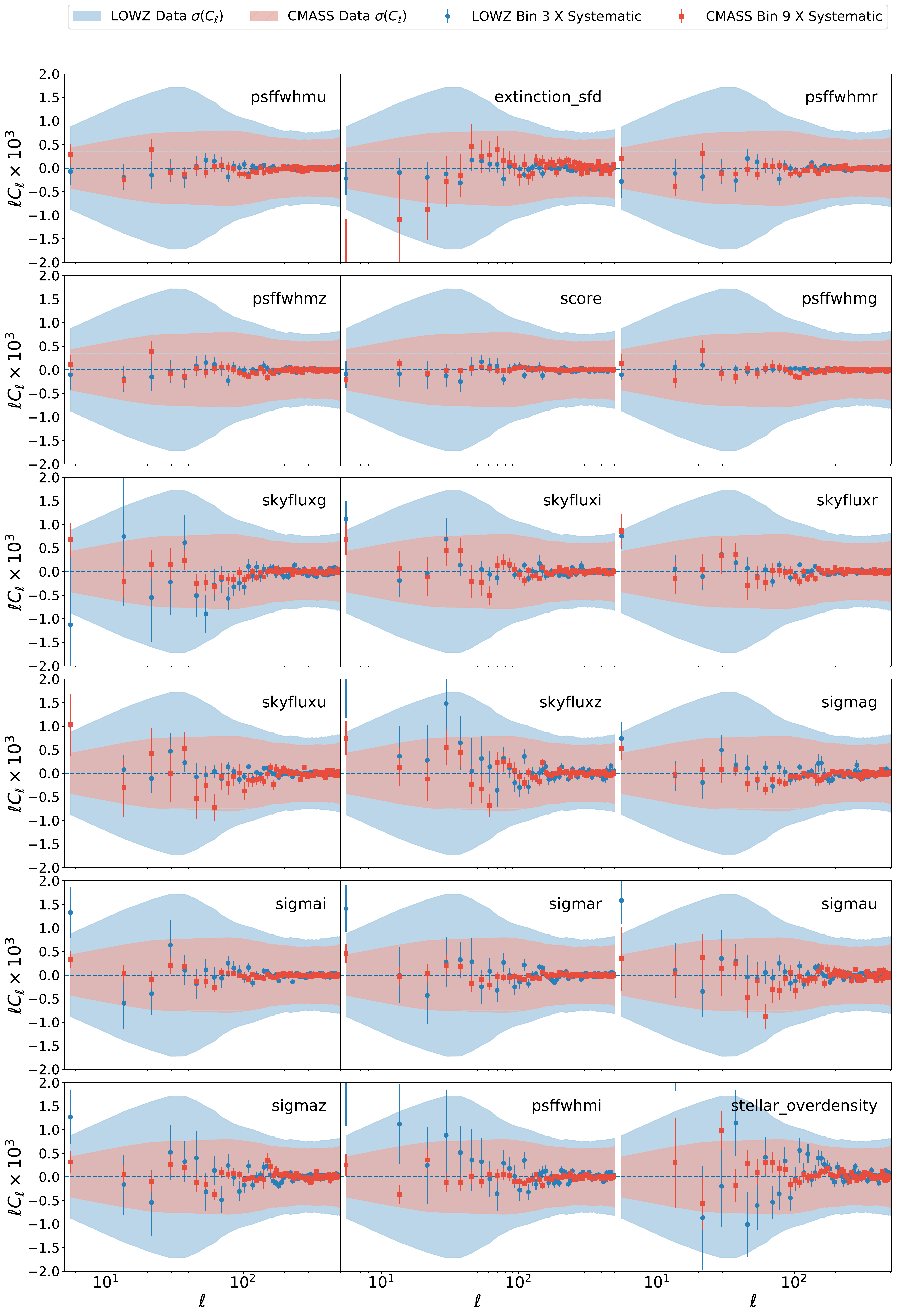}
\caption{An example of the systematics analysis described in Section \ref{Sec:SystCls}. Here, we show the cross-power spectra between the 18 systematics overdensity maps produced in \ref{Sec:SystMaps}, and LOWZ--3(CMASS--9) tomographic bins in blue dots (red squares). The error-bars were obtained by cross-correlating the $\delta^{Sys}$ maps with the \texttt{FLASK} mocks produced in Section \ref{Sec:Cov}; the shaded region shows the variance of the data, which was also obtained from the same mocks. This figure indicates that the shape of the measured power spectra in Figure \ref{fig:PCLs} is not dominated by any of the systematics considered, as the variance of the cross-power spectra between data and systematics is consistent with the variance of the data's auto-power spectra. The results for the other bins are similar to the results shown in this figure. Note also that the first $\ell-$band in the stellar overdensity cross-$C_{\ell}$ is completely out of the acceptable range, which leads us to exclude this data point on all bins for both samples.}
\label{fig:SystBin3}
\end{center}
\end{figure*}

We create \mangle masks of sky background flux, sky variance and average PSF FWHM in all five photometric bands. We also create a mask of the \textit{score} of each field, defined by the SDSS collaboration to express ``observational quality" as an empirical combination of observational values with processing status flags. Additional observational properties can be found in the Field Table, available from the SDSS SkyServer Schema Browser.\footnote{\url{http://skyserver.sdss.org/dr12/en/help/browser/browser.aspx}} The choice of which systematics to take into account is somewhat arbitrary, as there are correlations between observational properties that make information redundant \citepbox{Boris2013}. We choose to add stellar density and galactic extinction to the systematics listed above, as those have been shown to correlate with galaxy density in several previous analyses \citep[e.g.][]{Thomas2011, ElvinPoole2017}. We construct a bright star catalogue from the SDSS object catalogue with the following cuts:

\begin{ceqn}
\begin{eqnarray}
18 < r_{\rm psf} < 19.5,\nonumber\\
{\rm type} = 6,\\
r_{\rm psf} - r_{\rm model} < 0.25,\nonumber
\end{eqnarray}
\end{ceqn}
where the extinction-corrected magnitude cut ensures robust star selection \citepbox{Padm2007}, the type selection is the standard SDSS star-galaxy classifier,\footnote{\url{http://www.sdss.org/dr12/algorithms/classify/}} and the magnitude-difference cut is an additional point-source selection performed by the \texttt{GAMA} survey \citep[e.g.][]{Christodoulou2012}. For galactic extinction, we create a map directly in \healpix format. For simplicity, we take advantage of a Python implementation of extinction $E(B-V)$ value retrieval and map creation.\footnote{\url{https://github.com/kbarbary/sfdmap}} We use the original SFD scaling \citepbox{Schlegel1998}.

The \mangle masks created from the SDSS FITS files are not appropriately snapped, pixelised and balkanised, which breaks the local character of the \mangle procedure \citepbox{2008Mangle}. As a consequence, further operations suffer from impractically large processing times. We therefore run all the steps of the \mangle pixelisation anew, which corrects whatever imperfections remained in the first pass. From these masks, we create full-sky \healpix maps at resolution $N_{\rm side}=16384$, which defines an angular scale much smaller than the average resolution of the mask features. For each observational systematic, we populate the sub-resolution \healpix pixels with values from the associated \mangle mask. The resulting \healpix maps encapsulate all the information contained in the original footprint description.

Once the \healpix systematics maps are created, the next step is to transform them into overdensity maps using the same procedure outlined in Section \ref{Sec:Maps} for the data \citepbox{Boris2013}. The idea is to treat the systematic maps in the same way as the data in order to apply the statistical estimators consistently. Therefore, we degrade the high-resolution maps to the data resolution ($N_{side} = 512$) and up-weight the maps according to the pixel completeness mask that takes into account the holes in the footprint  (see Section \ref{Sec:Masks}); we then perform a cut in pixel completeness $C_{pix} = 0.8$ in the maps. This up-weighting is due to the pixelisation of the degraded mask, as explained in Section \ref{Sec:Masks}.  From these post-processed maps, we create the systematics overdensity maps as: 

\begin{ceqn}
\begin{equation}
\delta_{p}^{Sys} = 
\begin{cases}
\left(\frac{1}{C_{pix,p}}\frac{n^{Sys}_{p}}{\bar{n}^{Sys}}\right) - 1 & \text{, if } C_{pix,p} \geq 0.8 \\
0 & \text{, otherwise}
\end{cases}
\label{Eq:OverDMapsSyst}
\end{equation}
\end{ceqn}
where $n^{Sys}_{p}$ is the pixel value for a given systematic and $\bar{n}^{Sys}$ is the mean value of the map in the observed fraction of the sky. The systematics overdensity maps were created using both the CMASS and LOWZ masks presented in Section \ref{Sec:Masks}. The resulting systematics overdensity maps will be available at the \textsf{ZXCorr Collaboration} website.

\subsection{Cross-power spectra between data and systematic maps}\label{Sec:SystCls}

For the systematics analysis using cross-power spectra, we will follow a data analysis similar to that performed for the galaxy overdensity maps in Section \ref{Sec:Measurements}.  Using Equation \eqref{Eq:AlmPix} we decompose the systematics overdensity maps, $\delta^{Sys}$, into spherical harmonics. 

The estimator for the cross-power spectra between the data overdensity maps, $\delta^{g}$, and the systematics can be written as a modified version of Equation \eqref{Eq:Sl_wl}:

\begin{ceqn}\begin{equation}
\hat{S}^{gs}_{\ell} = \frac{1}{(2\ell+1)w_{\ell}^2}\sum_{m=-l}^l  \frac{\frac{1}{2}\left|d_{}^{g}  d_{\ell m}^{s*} + d_{\ell m}^{g*}  d_{\ell m}^{s}\right|}{J_{\ell m}}
\label{Eq:Sl_wlSyst}
\end{equation}\end{ceqn}
where the index \textit{g} stands for a data map, and \textit{s} for a systematics map. We then obtained the estimates for the variance of the systematics cross-power spectra by measuring the $\hat{S}^{gs}_{\ell}$ (Equation \ref{Eq:Sl_wlSyst}) between the systematics maps and the data mocks described in Section \ref{Sec:Cov}.

We cross-correlated all 13 tomographic redshift bins with all 18 systematic maps, resulting in a total of 234 cross-power spectra. Figure \ref{fig:SystBin3} shows an example for LOWZ--3 and CMASS--9 bins and cross-power spectra for all systematics.
From all of these measurements, the majority are consistent with the variance of the data measured from the log-normal simulation (Section \ref{Sec:Cov}), which lead us to be confident in using the full shape of our measured $C_{\ell}$s. Note, however, that a few of the large scale measurements (low-$\ell$) in CMASS sample have a small excess in cross-power spectra with stellar overdensity. The first point on the cross-power spectra between some of the systematic maps and most BOSS bins is clearly more than one sigma away from the data's variance. Due to this excess in correlation with stellar overdensity and the level of cosmic variance on the first $\ell$-band, we decided to exclude this first point from our cosmological analysis (see Section \ref{Sec:CosmoBananas} for details on the $\ell$ range used). As for the second $\ell$-band ($\ell = 13.5$) presenting an excess of correlation between a few bins and stellar overdensity: we found it to be sub-dominant, with no significant impact from this measurement in our cosmological analysis; therefore, we decided to keep it.

\section{Cosmological Analysis}\label{Sec:CosmoBananas}

In this section, we present the cosmological implications from the measured angular power spectra of BOSS galaxies for flat $\Lambda$CDM, $w$CDM, and $\Lambda$CDM with $\sum m_{\nu}$ models. Using the theoretical framework and having estimated covariance matrices as described in Section \ref{Sec:Theory}, we performed a Bayesian analysis using the \texttt{PLINY} (\citealt{PlinyRichardThesis} and Rollins et al., \textit{in prep}) nested sampler and the {\textit{Unified Cosmological Library for Parameter Inference}} code, or \texttt{UCLPI} (Cuceu et al., \textit{in prep.}). {All analyses considered in this section use the auto-power spectra and the cross-power spectra from adjacent tomographic bins using the measurements presented in Section \ref{Sec:Measurements}. Cross-power spectra between distant bins are not a part of our final BOSS-$C_{\ell}$ data vector.}

\subsection{Likelihoods, Priors \& Bayes Factor}\label{Sec:LikelihoodsPriors}
\begin{table}
\centering
\caption{Maximum multipole considered in the cosmological analysis for each tomographic redshift bin. All the samples start at $\ell_{min} = 13.5$ and have a bandwidth of $\Delta\ell = 8$. When considering the cross-power spectra between bins, the lower $\ell_{max}$ is used. The $\ell_{max}^{5\%}$ column corresponds to a $k_{max}  \lesssim 0.07 h$ Mpc$^{-1}$, and the $\ell_{max}^{10\%}$ column corresponds to a  $k_{max}  \lesssim 0.10 h$ Mpc$^{-1}$.}
\label{Tb:EllCuts}
\begin{tabular}{lllll}
\hline
Sample Bin & $z_{min}$ & $z_{max}$ & $\ell_{max}^{5\%}$ & $\ell_{max}^{10\%}$\\
\hline 
LOWZ--0  & 0.15      & 0.20      & 53	&	69 \\
LOWZ--1  & 0.20      & 0.25      & 77	&	93 \\
LOWZ--2  & 0.25      & 0.30      & 93	&	109\\
LOWZ--3  & 0.30      & 0.35      & 109	&	133\\
LOWZ--4  & 0.35      & 0.40      & 125	&	157\\
LOWZ--5  & 0.40      & 0.45      & 141	&	173\\
CMASS--6 & 0.45      & 0.50      & 157	&	221\\
CMASS--7 & 0.50      & 0.55      & 165	&	237\\
CMASS--8 & 0.55      & 0.60      & 189	&	261\\
CMASS--9 & 0.60      & 0.65      & 197	&	277\\
CMASS--10 & 0.65      & 0.70      & 213	&	317\\
CMASS--11 & 0.70      & 0.75      & 245	&	333\\
CMASS--12 & 0.75      & 0.80      & 261	&	381\\
\hline
\end{tabular}
\end{table}
The cosmological analysis performed in this work follows a standard Bayesian analysis framework as commonly performed in the literature, e.g. \cite{Blake2007,Thomas2011,2017MNRAS.465.1454H,2017arXiv170801530D}. 

The posterior distribution of the cosmological parameters, $\pmb{\Theta}$, given the measured angular power spectra, $\hat{S}_{\Delta\ell}$, and a model $\mathcal{M}$ can be written as a marginalisation of the full posterior over the nuisance parameters, $\pmb{\nu}$:

\EQ{MargPost}{\Pr (\pmb{\Theta}|\hat{S}_{\Delta\ell}, \mathcal{M}) = \int \Pr(\pmb{\Theta}, \pmb{\nu} | \hat{S}_{\Delta\ell}, \mathcal{M})d\pmb{\nu}  }
The full posterior distribution can be written as:

\EQ{FullPost}{
\Pr (\pmb{\Theta}, \pmb{\nu} | \hat{S}_{\Delta\ell}, \mathcal{M}) = \frac{\mathcal{L}(\hat{S}_{\Delta\ell}|\pmb{\Theta}, \pmb{\nu}, \mathcal{M}) \pi(\pmb{\Theta}, \pmb{\nu})}{Z({\hat{S}_{\Delta\ell}}| \mathcal{M})}
}
where $\mathcal{L}(\hat{S}_{\Delta\ell}|\pmb{\Theta}, \pmb{\nu}, \mathcal{M})$ is the likelihood, $\pi(\pmb{\Theta}, \pmb{\nu})$ is the prior on the sampled parameters, and $Z({\hat{S}_{\Delta\ell}}| \mathcal{M})$ is the evidence, which is calculated using \pliny , a nested sampler used in our analysis (\citealt{PlinyRichardThesis} and Rollins et al., \textit{in prep}; \cite{2008FerozHobson}).

If we had access to the true covariance matrix $\Sigma$, the likelihood, assumed here to be Gaussian, would be:

\begin{ceqn}\begin{align}
\mathcal{L_G}(\hat{S}_{\Delta\ell}|\pmb{\Theta}, \pmb{\nu}) = & \frac{1}{\sqrt[]{\vert 2\pi \Sigma \vert}} \exp\bigg\{ - \frac{1}{2} \big[ \hat{S}_{\Delta\ell} - S^{th}_{\Delta\ell}(\pmb{\Theta}, \pmb{\nu})\big]^T \nonumber \\ 
\times \, & \Sigma^{-1} \big[ \hat{S}_{\Delta\ell} - S^{th}_{\Delta\ell}(\pmb{\Theta}, \pmb{\nu})\big]\bigg\}
\label{Eq:GaussianLike}
\end{align}\end{ceqn}
where $S^{th}_{\Delta\ell}(\pmb{\Theta}, \pmb{\nu})$ is the theoretical angular power spectra after being convolved with the mixing matrix (Equation \ref{Eq:Cl_Conv}) and binned into the same bandwidths as the data (Equation \ref{Eq:S_delta_ell2}). 

However, this is not the case when estimating the covariance matrix $\mathcal{C}$ from simulations (Equation \ref{Eq:Covariance}). Even though $\mathcal{C}$ can be an unbiased estimator of the true covariance $\Sigma$, its inverse $\mathcal{C}^{-1}$ is not necessarily an unbiased estimator of the inverse covariance $\Sigma^{-1}$, needed to estimate the likelihood in Equation \eqref{Eq:GaussianLike}. \cite{Hartlap2007} proposed to keep using the Gaussian likelihood, and to apply a simple rescaling to the inverse of the estimated covariance matrix in order to de-bias it \citepbox{AndersonBook}.

\begin{ceqn}\begin{equation}
\Sigma^{-1} \rightarrow \alpha \mathcal{C}^{-1}
\end{equation}\end{ceqn}
where 

\begin{ceqn}\begin{equation}
\alpha = \frac{N_s - p - 2}{N_s - 1}
\end{equation}\end{ceqn}
and $N_s$ is the number of simulations and $p$ is the size of the data vector. 

More recently, \cite{2016SellentinHeavens} (hereafter, SH16) showed that when replacing the true covariance $\Sigma$ with an estimated $\mathcal{C}$, one should marginalise over the true covariance conditioned on the estimated one from simulations.  The resulting likelihood is no longer Gaussian; instead, the likelihood is now given by a multivariate t-distribution (SH16):

\begin{ceqn}\begin{equation}
\mathcal{L_{SH}}(\hat{S}_{\Delta\ell}|\pmb{\Theta}, \pmb{\nu}) = \frac{\overline{c}_p}{\vert \mathcal{C} \vert^{1/2}} \Big[1 + \frac{(\hat{S}_{\Delta\ell} - S^{th}_{\Delta\ell})^T \mathcal{C}^{-1} (\hat{S}_{\Delta\ell} - S^{th}_{\Delta\ell})}{N_s + 1}\Big]^{\frac{-N_s}{2}}
\end{equation}\end{ceqn}
where

\begin{ceqn}\begin{equation}
\overline{c}_p = \frac{\Gamma\big(\frac{N_s}{2}\big)}{\big[\pi(N_s - 1)\big]^{p/2} \Gamma\big(\frac{N_s - p}{2}\big)}
\end{equation}\end{ceqn}
and $\Gamma$ is the gamma function.

\begin{figure}
\begin{center}
\includegraphics[width=\columnwidth]{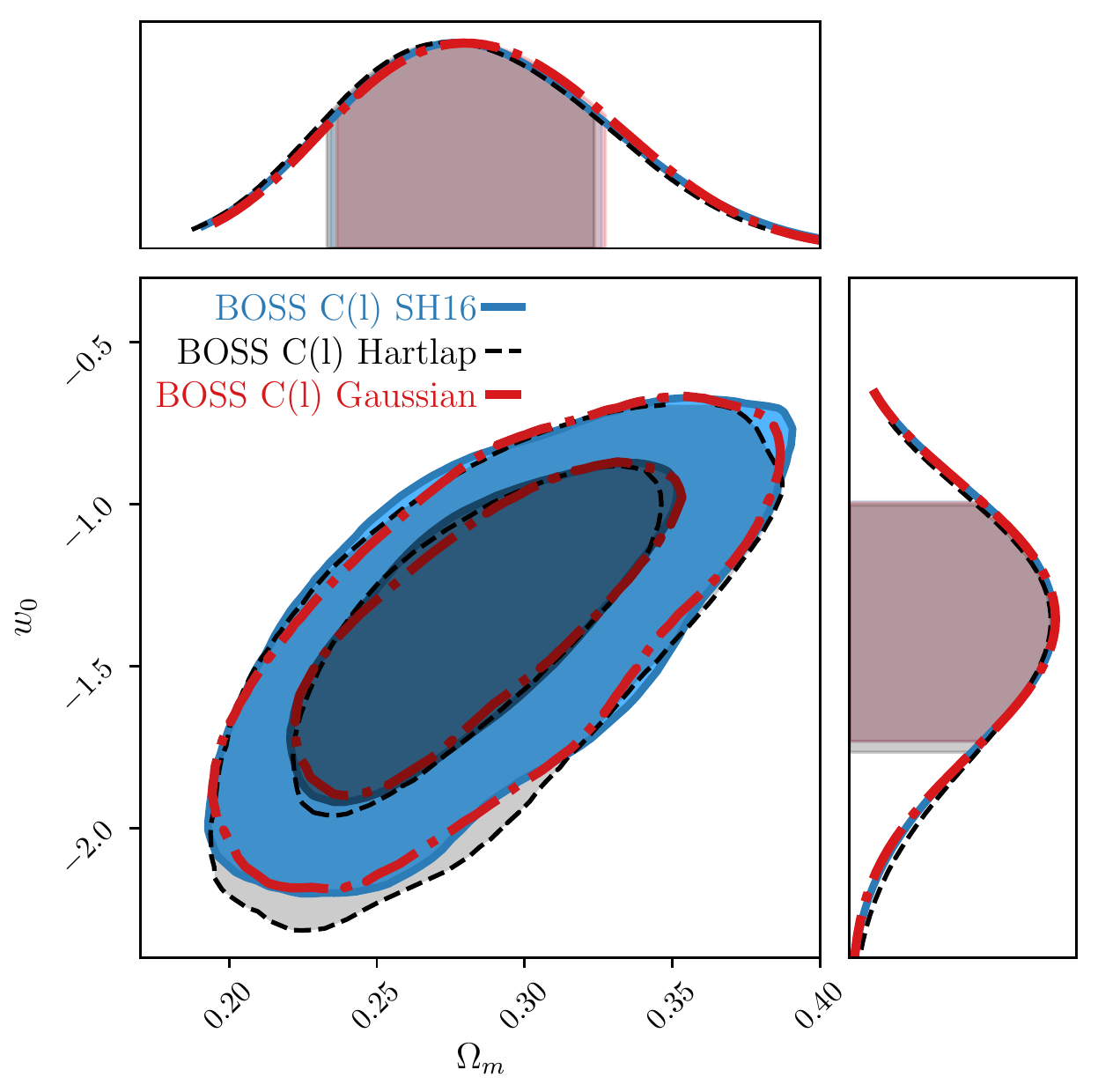}
\caption{Comparison between the three likelihood methods mentioned in Section \ref{Sec:LikelihoodsPriors} using the BOSS $C_{\ell}$ data only for $\Omega_{m}$ and $w_0$ in a $w$CDM model: Gaussian (red), Gaussian with Hartlap correction (black) and the SH16 (blue) likelihoods. Note how, given the high number of log-normal simulation used to estimate the inverse of the covariance, the Hartlap correction likelihood, SH16, and Gaussian have equivalent contours even though the sampled parameters and likelihood values are different. It is clear from this analysis that our estimated covariance matrix from Section \ref{Sec:Cov} is robust and was estimated with a sufficient number of simulations.}
\label{fig:LikelihoodCompare}
\end{center}
\end{figure}

\begin{figure}
\includegraphics[width=\columnwidth]{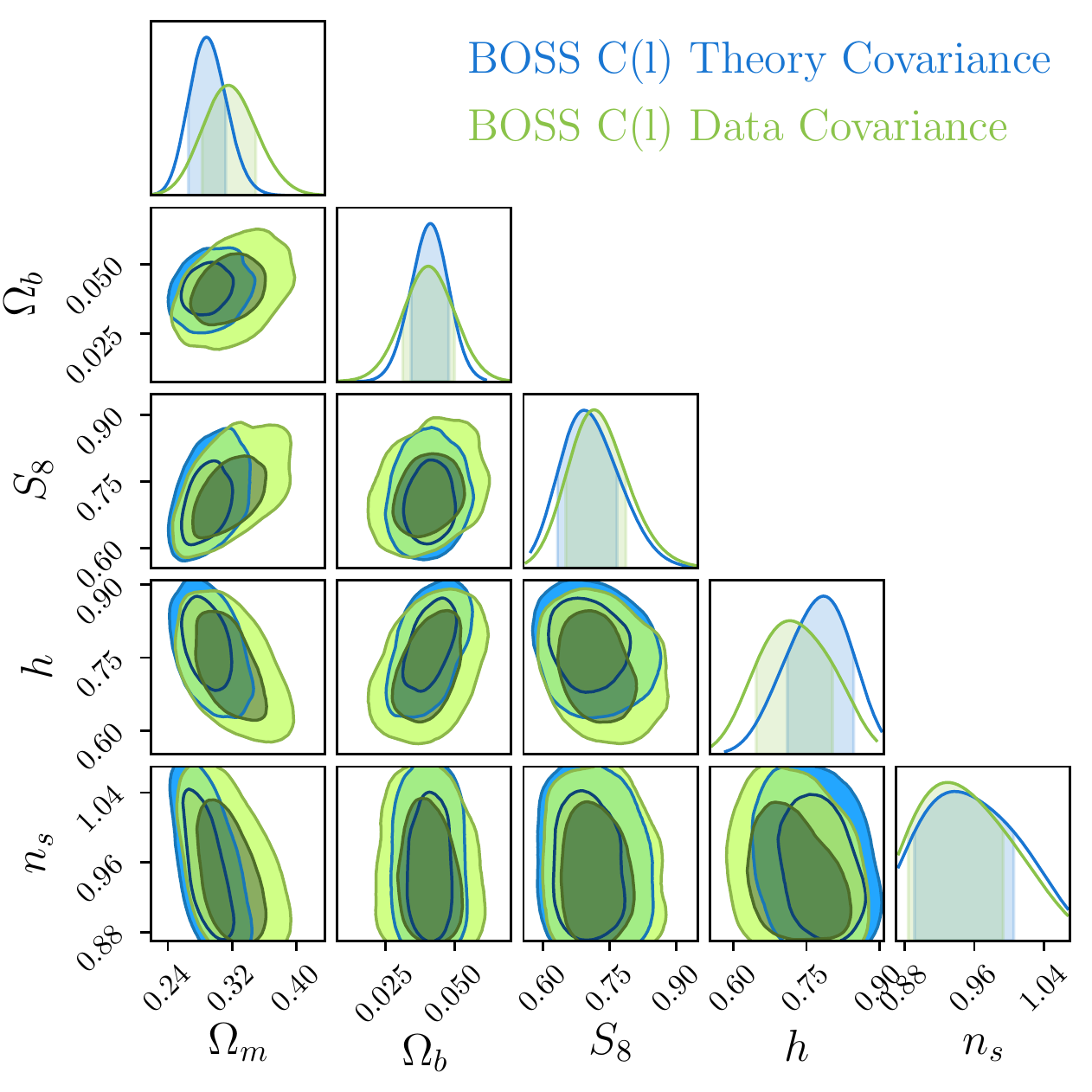}
\caption{Comparison between $\Lambda$CDM cosmologies recovered using the covariance matrix estimated in Section \ref{Sec:Cov} (\textit{green contours}, same as the ones from Section \ref{Sec:LCDM}), and the cosmology dependent theoretical covariance matrix from Equation \eqref{Eq:TheoVariance} (\textit{blue contours}). Note that the same parameters were sampled in both cases as the theoretical covariance also depends on the same nuisance parameters. These marginalised credible intervals (CI) 1- and 2-$\sigma$ plots indicate both the estimated covariance matrix and the \uclci pipeline robustness.}
\label{fig:TheoryCovTriangle}
\end{figure}
\begin{table}
  \centering
  \caption{Ranges on the flat priors used in all Bayesian analysis. Parameters are divided into two groups: cosmological and nuisance.}
  \label{Tb:Priors}
  \begin{tabular}{cc}
    \hline
    Parameter & Prior Range \\
    \hline
     $\Omega_b$ & $1 \times 10^{-3}, \, 0.3$    \\
     $\Omega_{cdm}$ & $0.0, \, 0.8$    \\[0.1cm]
     $\ln 10^{10} A_s$ & $2.0, \, 4.0$    \\
     $n_s$ & $0.87, \, 1.07$    \\
     $h$ & 0.55, 0.91 \\
     $w_0$ & -3, -0.3\\
     $\sum m_{\nu}$ & 0.0, 1.0 eV\\[0.1cm]
     $\tau^{Planck}_{reio}$  & 0.0, 0.8 \\
     \hline
     $b(z)$  & 1.1, 3.3 \\
     $\sigma_s(z)$ & $1 \times 10^{-6}, 9 \times 10^{-3}$ \\
     $\mathcal{N}_{9}, \, \mathcal{N}_{10}$ & $0.0, \, 1\times 10^{-4}$ \\
     $\mathcal{N}_{11}$ & $0.0, \, 8\times 10^{-5}$ \\
     $\mathcal{N}_{12}$ & $0.0, \, 4\times 10^{-4}$ \\
     $y_{cal}^{Planck}$ & $0.99, \, 1.01$ \\
     $M_B^{JLA}$ & -20.0,\, -18.5.\\
     \hline
  \end{tabular}
\end{table}

Even though the non-linear model described in Section \ref{Sec:NonLin} is sufficiently reliable, we performed cuts in $\ell_{max}$ for each of the tomographic redshift bins in order to exclude non-linear scales. In order to make this choice, we used a fiducial cosmology (the same used in Section \ref{Sec:ConsistControl}) to generate theory $C_{\ell}$s and performed a preliminary cut in $\ell_{max}$ where the percent deviation between the linear and non-linear models were smaller than $5\%$. We performed robustness checks on the $5\%$ deviation cut choice  by extending the cuts to $\ell_{max}$ which had a deviation up to $20\%$ and concluded that our cosmological results could be trusted up to a $15\%$ deviation between linear and non-linear theories for this fiducial test. In this paper we present results where this percentage cut is $5\%$ and $10\%$. For avoidance of doubt, for the fiducial cosmology of choice, applying a $5\%$ implies rejecting the majority of modes $k \gtrsim 0.07 h$ Mpc $^{-1}$, whereas $10\%$ implies rejecting modes where $k \gtrsim 0.1 h$ Mpc $^{-1}$. The resulting cuts can be found in table \ref{Tb:EllCuts}. As for the $\ell$ cuts for cross-power spectra, we chose the lowest $\ell_{max}$ between the two relevant bins in order to keep a consistent and conservative cut for each bin.

We used a total of 28 nuisance parameters ($\pmb{\nu}$) in the BOSS $C_{\ell}$ likelihood analysis in most of our results for a $5\%$ cut on $\ell_{max}$. These parameters are: a scale independent bias, $b(z)$, for each redshift bin; a redshift error dispersion, $\sigma_s(z)$ (Equation \ref{Eq:GaussianErrNz}), for each redshift bin that takes into account spectroscopic redshift error and Finger-of-God effects due to shell-crossing (see Section \ref{Sec:SpecNz} for details); and an extra shot noise term for bins 11 and 12 that is forward modelled into the theoretical angular power spectrum inside the likelihood as:

\EQ{Eq:ExtraNoise}{\hat{S}^{th}_{\Delta\ell} \rightarrow \hat{S}^{th}_{\Delta\ell} + \mathcal{N}}
where $\mathcal{N}$ is a constant that takes into account extra shot noise due to the very low number of galaxies in these two redshift bins. In the case of a $10\%$ cut we used two further shot noise nuisance parameters for bins 9 and 10 as we go further into the non-linear regime where the shot noise in those bins dominates over the signal.

We chose flat priors in all Bayesian analysis. These were based on priors used in \cite{JLAdata,2016BOSSCosmology,PlanckCosmology2016,2017arXiv170801530D} and were set equally for all analyses. The prior ranges can be found in table \ref{Tb:Priors} for all parameters considered in the cosmological analysis in this section: the baryonic matter density ($\Omega_b$), the cold dark matter density ($\Omega_{cdm}$), the amplitude of the primordial power spectrum ($A_s$), the spectral index ($n_s$), the Hubble constant ($h$), the equation-of-state of dark energy ($w_0$), the sum of neutrino mass species ($\sum m_{\nu}$), the optical depth at reionisation epoch ($\tau_{reio}$), the bias of BOSS galaxies as a function of redshift ($b(z)$), the redshift dispersion parameter for BOSS galaxies ($\sigma_s(z)$), the extra shot-noise for BOSS galaxies ($\mathcal{N}_i$), the Planck absolute calibration parameter ($y_{cal}^{Planck}$), and the absolute magnitude of SNe Ia at peak light in blue band ($M_B^{JLA}$).

Finally, to perform consistency checks between BOSS DR12 and the external datasets described in Section \ref{Sec:ExternalData}, we used the Bayes factor. The Bayes factor for the consistency of two datasets (A and B) is given by:

\begin{ceqn}\begin{equation}
R_{A,B} = \frac{P(\vec{A},\vec{B}|M)}{P(\vec{A}|M)P(\vec{B}|M)}
\label{Eq:BayesFactor1}
\end{equation}\end{ceqn}
\noindent or, for three datasets (A, B, and C):

\begin{ceqn}\begin{equation}
R_{A,B,C} = \frac{P(\vec{A},\vec{B},\vec{C}|M)}{P(\vec{A}|M)P(\vec{B}|M)P(\vec{C}|M)}
\label{Eq:BayesFactor}
\end{equation}\end{ceqn}

\noindent where $M$ is the model, $P(\vec{A},\vec{B}|M)$ is the evidence when the model is fitted to both datasets simultaneously and $P(\vec{A}|M)P(\vec{B}|M)$ is the product of the evidences when the model is fitted to each dataset individually. Since \pliny is a nested sampler, all our cosmological estimations lead to values for the evidences of each model, dataset, and combination of datasets. We are aware that this method has received some criticism as the value of $R_{A,B}$ depends crucially on the prior volume chosen for the sampling. Although this is true, choosing priors which are physically motivated should lead to values of $R_{A,B}$ which can be reflective of the real consistency between datasets $A$ and $B$. However, a host of new methods have been proposed in the literature which we have not fully investigated in this paper, but which we intend to implement in further analysis with this dataset. These methods involve a range of overlap integrals between the separate posteriors and can be better suited to determine the consistency between two or more datasets. We refer the reader to these methods in \cite{2017CharnockTension,2018HuTension,2018FeeneyTension}

In order to perform a robust analysis, we implemented the three likelihoods mentioned above: Gaussian, Gaussian plus the Hartlap correction, and the SH16 t-distribution likelihood. We observed that the Gaussian + Hartlap correction, the SH16, and the Gaussian likelihoods led to very similar cosmological contours. This is most likely as we have sufficient mocks to reliably estimate our covariance matrix. Figure \ref{fig:LikelihoodCompare} shows a comparison between the three likelihoods for a $w$CDM model, using the BOSS $C_{\ell}$s only, for $w_0$ and $\Omega_m$. In all of the following results in this section, we use the SH16 likelihood.

\subsection{Consistency Checks}
\begin{figure}
\begin{center}
\includegraphics[width=\columnwidth]{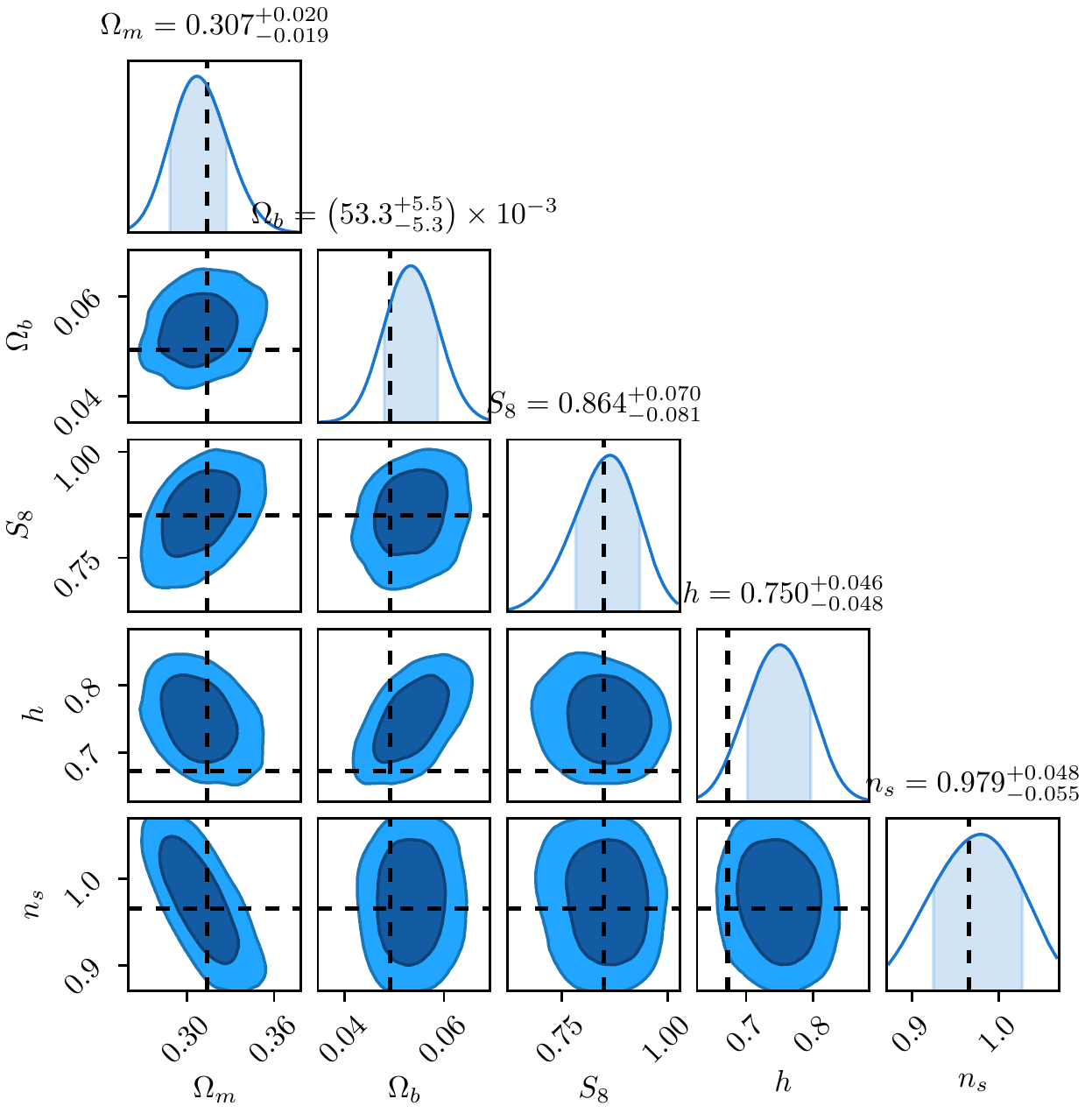}
\caption{Cosmological constraints recovered from a controlled cosmology pipeline test. The dashed lines denote the Planck-like cosmology used as input in the simulations analysed in the blue contours. All parameters agree within the estimated errorbars.}
\label{fig:Controlledtest}
\end{center}
\end{figure}

\begin{figure*}
\begin{center}
\subfigure[]{\includegraphics[width=\columnwidth]{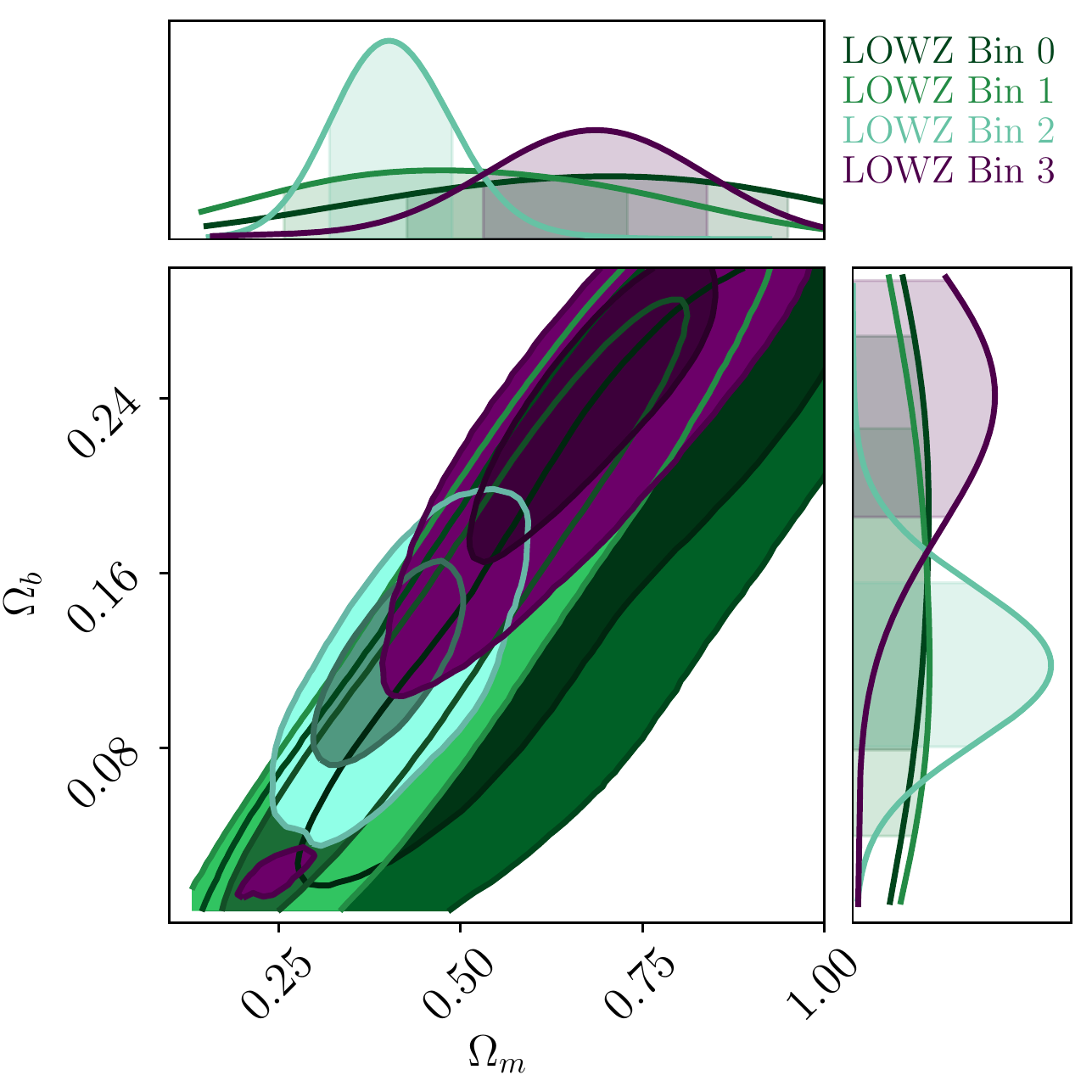}
\label{fig:SingleBin_LOWZ0}}
\subfigure[]{\includegraphics[width=\columnwidth]{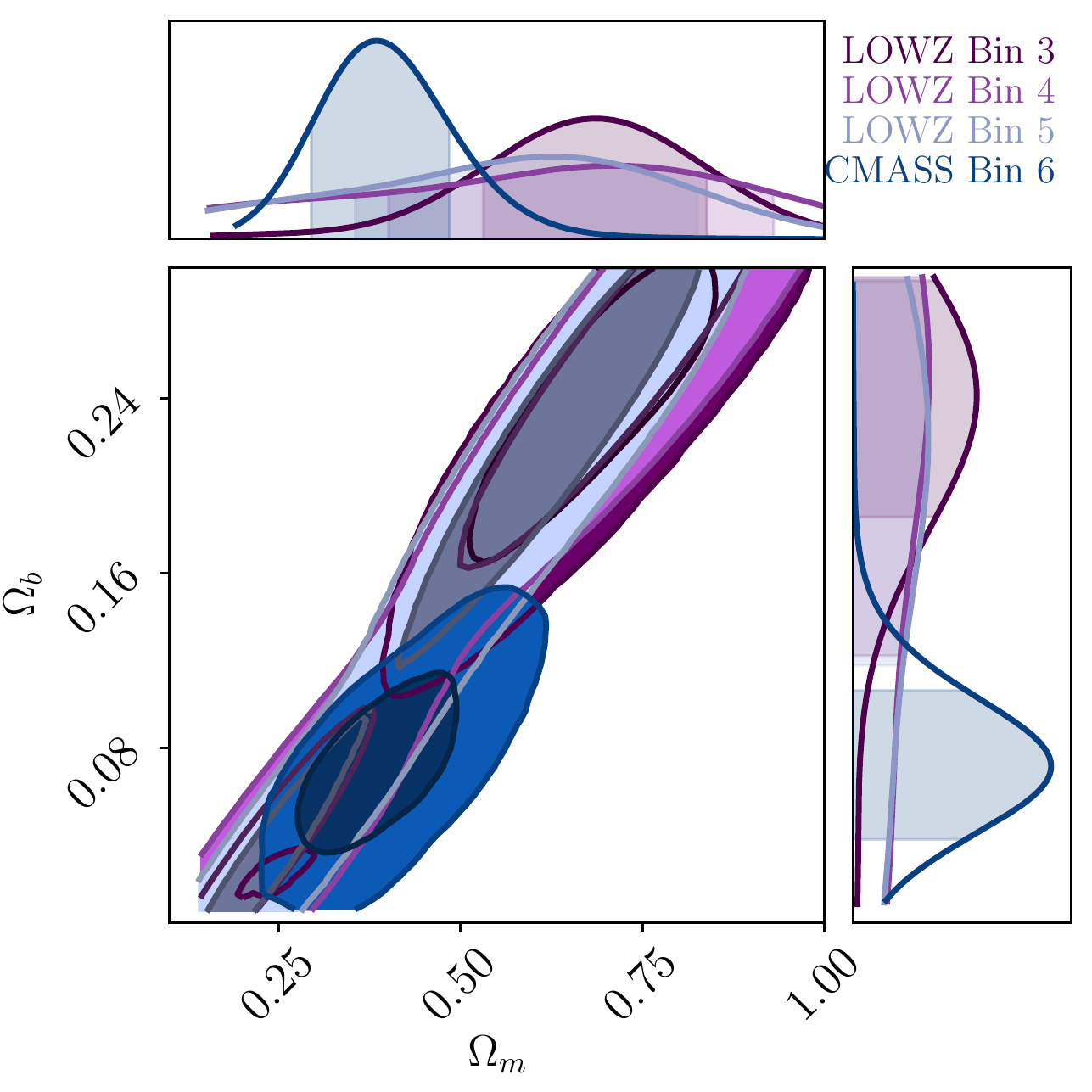}
\label{fig:SingleBin_LOWZ1}}
\subfigure[]{\includegraphics[width=\columnwidth]{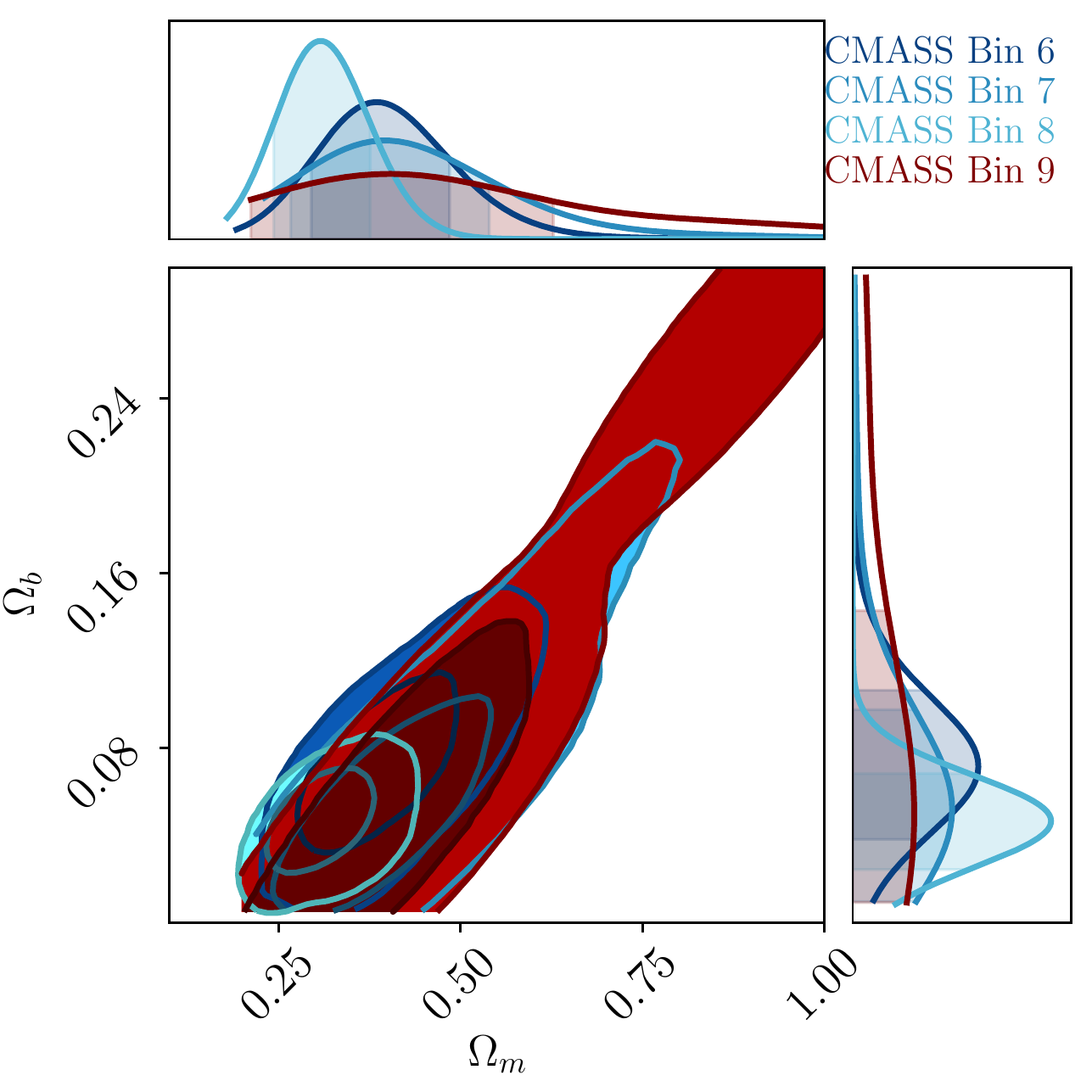}
\label{fig:SingleBin_CMASS0}}
\subfigure[]{\includegraphics[width=\columnwidth]{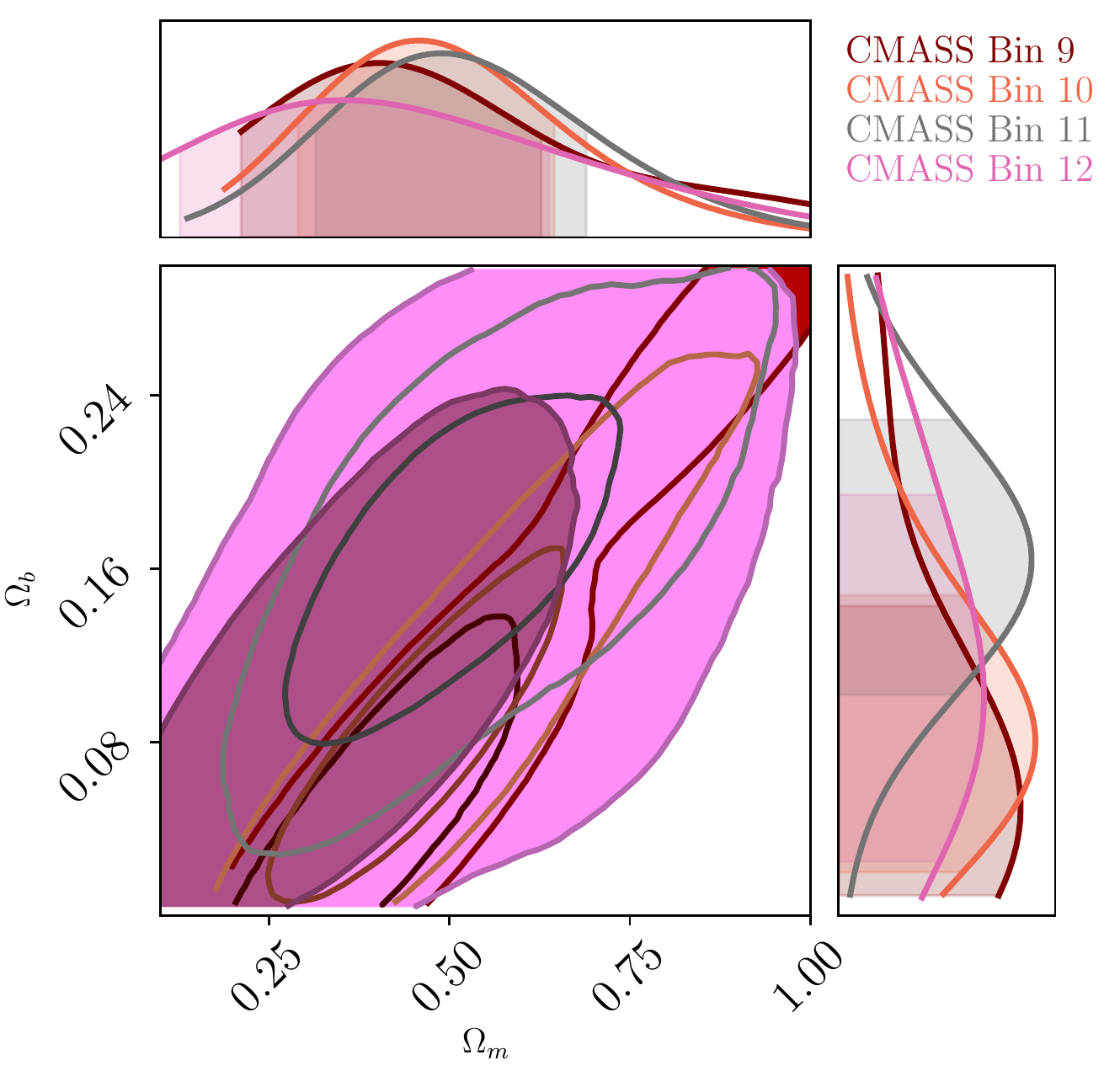}
\label{fig:SingleBin_CMASS1}}
\caption{Consistency checks for single tomographic redshift bins for all LOWZ and CMASS bins. Here, we show the $\Omega_b\,  \times \, \Omega_{m}$ contours taken from a $\Lambda$CDM cosmological inference, i. e. varying the same cosmological parameters as the ones from Section \ref{Sec:LCDM}. \textit{(a)} Shows LOWZ-0, LOWZ-1, LOWZ-2, and LOWZ-3 tomographic bins; \textit{(b)} shows LOWZ-3, LOWZ-4, LOWZ-5, and CMASS-6 tomographic bins; \textit{(c)} shows CMASS-6, CMASS-7, CMASS-8, and CMASS-9 tomographic bins; and \textit{(d)} shows CMASS-9, CMASS-10, CMASS-11, and CMASS-12 tomographic bins.}
\label{fig:SiingleBinAnaly}
\end{center}
\end{figure*}

\begin{figure}
\begin{center}
\includegraphics[scale=0.45]{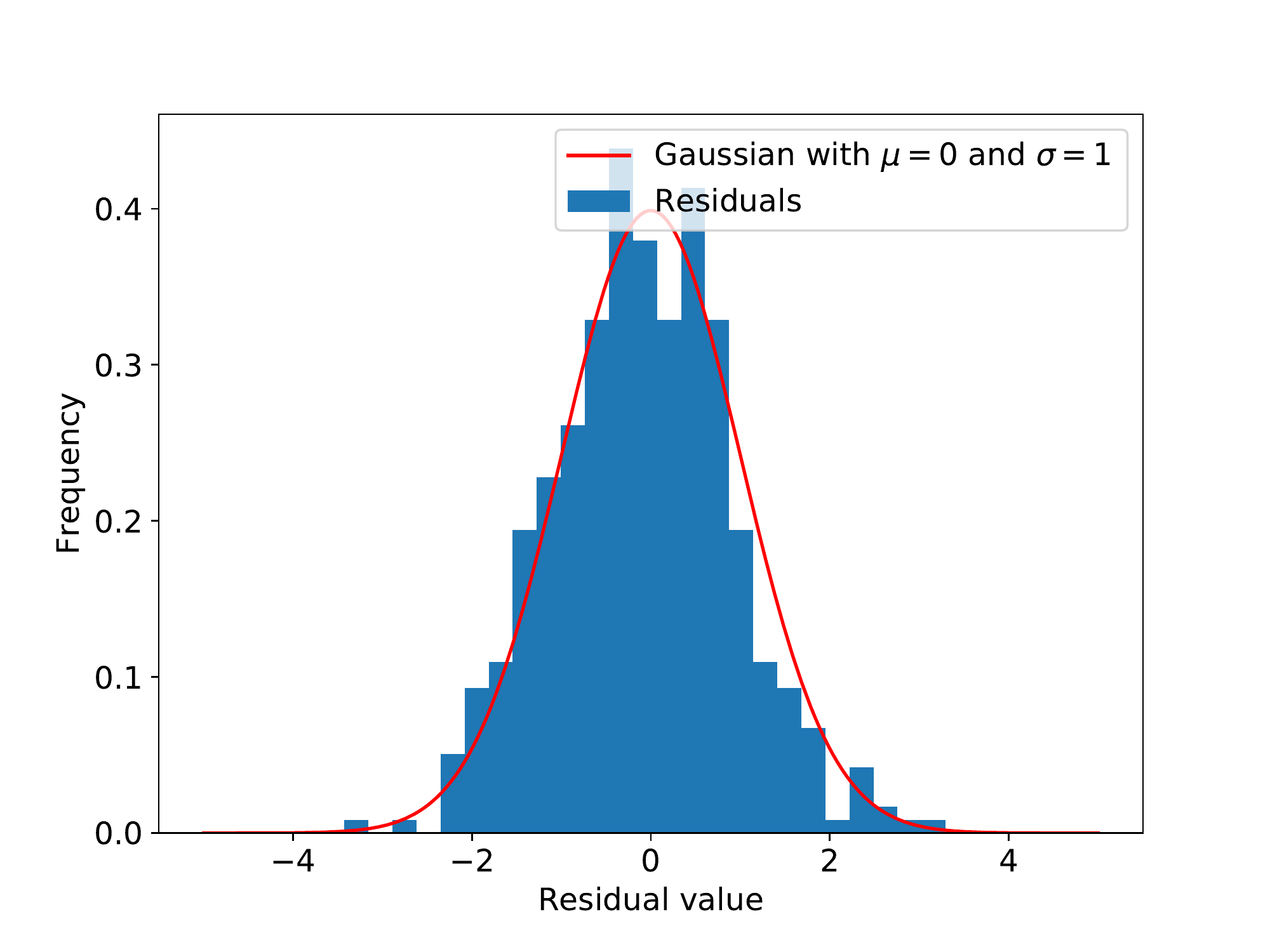}
\caption{Comparison between the histogram of the distribution of residuals (given by Equation \ref{Eq:Residuals_2}) calculated at the best-fit point in a flat $\Lambda$CDM cosmology (see Section \ref{Sec:LCDM}) and a Gaussian with $\mu=0$ and $\sigma=1$. As this distribution does not show any significant deviation from the Gaussian, the model is either the truth or the current data cannot make any further distinction between the model and the truth.} 
\label{fig:Residuals}
\end{center}
\end{figure}

In this section, we perform a series of consistency checks in order to assess the validity of our cosmological parameter estimation pipelines and data samples.

\subsubsection{Parameter-dependant theoretical covariance matrix}
We implemented an alternative likelihood based on a theoretical expression for the covariance matrix (Section \ref{Sec:TheoCov}, see Equation \eqref{Eq:TheoVariance}) where the signal angular power spectra, $S_{\ell}$, also depends on the sampled parameters. Most standard cosmological analysis in the literature \citep{2016BOSSCosmology,2017arXiv170801530D,2017MNRAS.465.1454H} assume a covariance matrix which is independent of cosmology and which is estimated for a fiducial simulation. We do not expect to obtain the same cosmological contours from this method as those presented in the sections that follow; however, we do not expect the contours from this method to disagree significantly with the ones obtained with our estimated covariance matrix. Figure \ref{fig:TheoryCovTriangle} shows the results for this test for a $\Lambda$CDM cosmological model. 

\subsubsection{Controlled cosmology pipeline test}\label{Sec:ConsistControl}
For this test, we generated theory auto- and cross- $C_{\ell}$s to mimic the BOSS dataset using a Planck-like cosmology: $h = 0.6725$, $\Omega_b = 0.0492$, $\Omega_{m} = 0.314$, $w_0 = -1.0$, $S_8 = 0.830$, $n_s = 0.96575$. We simulated these fiducial power spectra using the BOSS redshift distribution $n(z)$ from Figure \ref{fig:NZ_BOSS}. We chose the nuisance parameters to match the best fit values obtained in Section \ref{Sec:LCDM} from the combination of the entire cosmological dataset available to us. Using BOSS masks as input, we created \flask mocks like described in Section \ref{Sec:Cov}: generating the mocks at higher resolution, degrading them, and creating galaxy overdensity maps. We applied the PCL estimator on the 13 overdensity maps, calculating the auto- and cross- power spectra as described for the data in Section \ref{Sec:Measurements}.

Finally, we ran a cosmological parameter estimation for a $\Lambda$CDM cosmology, varying also the 28 BOSS nuisance parameters and using the theoretical covariance matrix as in the previous section. The results are shown in Figure \ref{fig:Controlledtest}, where the recovered parameters are within the errors with no indications of biases in the entirety of the pipeline. 

\subsubsection{Internal checks: single redshift bin consistency}
To test the data's internal consistency, we performed a full cosmological analysis in each individual redshift bin from our BOSS samples. The test was performed using a $\Lambda$CDM model, varying the same nuisance parameters as described in Section \ref{Sec:LikelihoodsPriors} for each bin: redshift dispersion, bias, and extra shot noise for the last two CMASS bins. If each individual bin is consistent with all others, this indicates that one can obtain cosmological constraints from the combination of the individual bins. This is shown in Figure \ref{fig:SiingleBinAnaly} for the posterior projections of $\Omega_m$ and $\Omega_b$. In these figures, all contours overlap and, even though some tomographic redshift bins prefer a secondary peak, they are consistent across the redshift bins. This secondary peak is due to a known cosmological parameter degeneracy \citepbox{2001Percival}.

\subsubsection{Distribution of residuals}
For a dataset with uncorrelated errors (diagonal covariance matrix), the vector of normalised residuals is given by:

\begin{ceqn}\begin{equation}
R = \Xi^{-1}(D - T(\vec{\theta}))
\label{Eq:Residuals_1}
\end{equation}\end{ceqn}
where $\Xi$ is a diagonal matrix containing the square roots of the variances, $D$ is the data vector and $T(\vec{\theta})$ is the theory vector for a given parameter vector $\vec{\theta}$. If $T(\vec{\theta})$ represents the true model, and the true errors are known, the residuals are by definition given by a Gaussian with $\mu=0$ and $\sigma=1$ \citepbox{chisq2010}. On the other hand, if the errors are estimated from the data, the residuals are given by a Student's t-distribution. This distribution approaches a Gaussian with increasing number of data points. If this distribution shows a significant deviation from a Gaussian, the model is ruled out. If it follows a Gaussian distribution, either we found the true model, or the current data is not enough to distinguish between the model we found and the true model \citepbox{chisq2010}.

When the covariance matrix is not diagonal (the errors are correlated), Equation \eqref{Eq:Residuals_1} is no longer true and we have to deal with the full covariance matrix. In order to get back to a diagonal matrix, we write the covariance matrix in terms of its eigen-decomposition :

\begin{ceqn}\begin{equation}
C = Q \Lambda Q^{-1}
\end{equation}\end{ceqn}
where $Q$ is the matrix of eigenvectors and $\Lambda$ is the diagonal matrix containing the eigenvalues of $C$. The inverse is then given by $C^{-1} = Q \Lambda^{-1} Q^{-1}$, which transforms the $\chi^2$ into:

\begin{ceqn}\begin{equation}
\chi^2 = \big[ \hat{S}_{\Delta\ell} - S^{th}_{\Delta\ell}(\pmb{\Theta}, \pmb{\nu})\big]^T Q \Lambda^{-1} Q^{-1} \big[ \hat{S}_{\Delta\ell} - S^{th}_{\Delta\ell}(\pmb{\Theta}, \pmb{\nu})\big]\, .
\end{equation}\end{ceqn}
If we treat $\Lambda$ as the new (diagonal) covariance matrix, it follows that the normalised residuals are now given by:

\begin{ceqn}\begin{equation}
R = \Xi^{-1} Q^{-1} \big[ \hat{S}_{\Delta\ell} - S^{th}_{\Delta\ell}(\pmb{\Theta}, \pmb{\nu})\big]
\label{Eq:Residuals_2}
\end{equation}\end{ceqn}
where $\Xi$ now contains the square roots of the eigenvalues.

We use Equation \eqref{Eq:Residuals_2} to calculate the residuals at our best-fit point in a flat $\Lambda$CDM cosmology and plot the results in a histogram (Figure \ref{fig:Residuals}). There are no significant deviations from a Gaussian with $\mu=0$ and $\sigma=1$, which means the model seems to be a very good representation of the data.

\begin{table*}
    \centering 
    \caption{Marginalised cosmological constraints and 68\% credible intervals for the models considered in this work using a variety of datasets and combinations. The contours for these results are shown in Figures \ref{fig:LCDM_Cosmology} for $\Lambda$CDM, \ref{fig:wCDM_Cosmology} for $w$CDM, \ref{fig:nuCDM5pc} for the $\Lambda$CDM + $\sum m_{\nu}$ with $\ell^{5\%}_{max}$ cut, and \ref{fig:nuCDM_Cosmology} for $\Lambda$CDM + $\sum m_{\nu}$ with $\ell_{max}^{10\%}$ cut.}
    \label{tab:LCDM_Constraints}
    
    \begin{tabular}{cccccc}
        \hline
		Model & Parameter & BOSS & BOSS & BOSS + JLA  & Planck \\
       & & & + JLA & + Planck & \\
		\hline
		$\Lambda$CDM & $\Omega_m$ & $0.315^{+0.034}_{-0.033}$ & $0.317^{+0.022}_{-0.021}$ & $ 0.327 \pm 0.008$ &  $0.315\pm 0.011$ \\[0.1cm] 
		& $\Omega_b$ & $0.0404^{+0.010}_{-0.009}$ & $0.0381^{+0.007}_{-0.008}$ & $0.0502 \pm 0.0006$ & $0.0492 \pm 0.0009$ \\[0.1cm]     
		& $S_8$ & $0.715^{+0.072}_{-0.064}$ & $0.745^{+0.059}_{-0.052}$ & $0.862^{+0.015}_{-0.016}$ & $0.850^{+0.023}_{-0.021}$ \\[0.1cm]      
		& $h$ & $0.716^{+0.088}_{-0.069}$ & $0.699\pm 0.039$ & $0.663 \pm 0.005$ & $0.672 \pm 0.008$ \\[0.1cm]      
		& $n_s$ & $0.929^{+0.064}_{-0.045}$ & $0.955^{+0.052}_{-0.048}$ & $0.960 \pm 0.004$ & $0.964 \pm 0.006$ \\[0.1cm]
		
        \hline
        
      $w$CDM & $\Omega_m$ & $0.277^{+0.050}_{-0.042}$ & $0.308^{+0.021}_{-0.018}$  & $0.330 \pm 0.012$ & $0.213^{+0.062}_{-0.039}$\\[0.1cm]
		& $\Omega_b$ & $0.0318^{+0.0117}_{-0.0098}$ & $0.0429 \pm 0.007 $ & $0.0505 \pm 0.002$ & $0.0334^{+0.009}_{-0.006}$\\[0.1cm]      
		& $S_8$ & $0.726^{+0.072}_{-0.061}$ & $0.743^{+0.079}_{-0.068}$  & $0.863\pm 0.016$ & $0.811^{+0.037}_{-0.034}$\\[0.1cm]       
		& $h$ & $0.767^{+0.069}_{-0.091}$ & $0.745^{+0.049}_{-0.052}$ & $0.661\pm 0.012$ & $0.816^{+0.073}_{-0.101}$  \\[0.1cm]      
		& $n_s$ & $0.939^{+0.057}_{-0.049}$ & $0.957^{+0.049}_{-0.050}$ & $0.960 \pm 0.004$ & $ 0.964 \pm 0.006 $ \\[0.1cm]   
		& $w_0$ & $-1.36^{+0.36}_{-0.38}$ & $-1.030^{+0.073}_{-0.076}$ &  $-0.993^{+0.046}_{-0.043}$ & $-1.45^{+0.32}_{-0.23}$ \\[0.1cm]
        
       \hline
       
      $\Lambda$CDM + $\sum m_{\nu}$ & $\Omega_m$ & $0.326^{+0.038}_{-0.035}$ & $0.304^{+0.022}_{-0.021}$ & $0.328 \pm 0.009$ & $0.326^{+0.028}_{-0.021}$\\[0.1cm]        
		[$\ell_{max}^{5\%}$ cut]& $\Omega_b$ & $0.040^{+0.009}_{-0.010} $ & $0.0432 \pm 0.008 $ & $0.05017^{+0.0009}_{-0.0008}$ & $0.0506^{+0.0039}_{-0.0026}$\\[0.1cm]      
		& $S_8$ & $0.723^{+0.069}_{-0.063}$  & $0.700^{+0.065}_{-0.056}$  & $0.862\pm 0.017$  & $0.836^{+0.031}_{-0.035}$\\[0.1cm]      
		& $h$ & $0.730^{+0.075}_{-0.078}$ & $0.814^{+0.054}_{-0.064}$ & $0.663^{+0.006}_{-0.007}$& $0.662^{+0.018}_{-0.026}$ \\[0.1cm]      
		& $n_s$ & $0.933^{+0.066}_{-0.046}$ & $0.941^{+0.055}_{-0.049}$ & $0.960 \pm 0.042$ & $ 0.962^{+0.006}_{-0.007} $ \\[0.1cm]      	 
        & $\sum m_{\nu}$ (95\% CI)[eV] & $ < 0.75 $ & $ < 0.71 $ &  $ < 0.14 $ & $ < 0.76 $ \\[0.1cm]
       
       \hline
      $\Lambda$CDM + $\sum m_{\nu}$ & $\Omega_m$ & $0.345^{+0.033}_{-0.030}$ & $0.324^{+0.034}_{-0.029}$ & $0.333^{+0.014}_{-0.012}$ & $0.326^{+0.050}_{-0.029}$ \\[0.1cm] 
        
		[$\ell_{max}^{10\%}$ cut]& $\Omega_b$ &  $0.045 \pm 0.009$ & $0.040\pm 0.013$ & $ 0.0510^{+0.0016}_{-0.0014}$ & $0.0506^{+0.0069}_{-0.0033}$  \\[0.1cm] 
        
		& $S_8$ & $0.751^{+0.062}_{-0.057}$ & $0.768^{+0.097}_{-0.092}$ & $0.864^{+0.030}_{-0.029}$ & $0.839^{+0.058}_{-0.067}$  \\[0.1cm] 
        
		& $h$ & $0.689^{+0.076}_{-0.066}$ & $0.661^{+0.067}_{-0.063}$ & $0.658^{+0.010}_{-0.011}$ & $0.662^{+0.024}_{-0.044}$ \\[0.1cm] 
        
		& $n_s$ & $0.930^{+0.062}_{-0.044}$ & $1.011^{+0.056}_{-0.086}$ &  $0.958 \pm 0.006$ & $0.962\pm 0.013$ \\[0.1cm]
		
        & $\sum m_{\nu}$ (95\% CI)[eV] & $ < 0.72 $ & $< 0.66 $ &  $ < 0.16 $ & $ < 0.76 $ \\[0.1cm]
		\hline
    \end{tabular}
    
\end{table*}
\subsection{External data}\label{Sec:ExternalData}

We compared and combined our results with results obtained from the Planck satellite CMB experiment \citepbox{PlanckResults2015} and Type Ia Supernovae from the Joint Light curve Analysis (JLA) collaboration \citepbox{JLAdata}. The relevant likelihood codes for these probes were implemented and tested in the \uclci pipeline. We checked that the official cosmological results from the relevant collaborations were recovered in order to use them. 

The CMB data from Planck was added through the Planck likelihood codes \texttt{Commander} and \texttt{Plik} \citepbox{PlanckLikelihood2015}. For low multipoles, in the range $l=2-29$, \texttt{Commander} is used with temperature (TT) and polarization auto- and cross- power spectra (BB, TB, EB). For high multipoles, in the range $l=30-2508$, \texttt{Plik} is used with temperature (TT) and polarization auto- and cross- power spectra (TE, EE). This configuration is commonly referred to as Planck TT,TE,EE+lowP. \texttt{Plik} also introduces 94 additional nuisance parameters. In order to reduce this large parameter space, we use the lite version of the data offered by the Planck Collaboration. This lite version allows us to compute a nuisance marginalized CMB likelihood. The only CMB nuisance parameter left is the Planck absolute calibration parameter ($y_{cal}$). We sample this parameter in all the runs that include Planck data. The Planck likelihood codes were added to \uclci and all the Planck results presented have been obtained using this pipeline. We show in the cosmological contours and in table \ref{tab:LCDM_Constraints} that this pipeline reproduces the cosmological results found by the Planck Collaboration \citepbox{PlanckCosmology2016}.

The SN data from JLA was added to the \uclci pipeline through the likelihood code provided by the JLA Collaboration \citepbox{JLAdata}. This likelihood code needs the luminosity distances to the 740 Supernovae in the sample and 4 nuisance parameters ($\alpha, \beta, M_B, \Delta_M$) described in \cite{JLAdata}. The luminosity distances are calculated by \class \citepbox{Class} using the redshifts of the supernovae within a given cosmology (set by the sampled cosmological parameters). We sample the absolute magnitude at peak brightness ($M_B$) as part of our analysis, and keep the other 3 nuisance parameters fixed to their best-fit values found by \cite{JLAdata} since these have a small impact in the cosmological parameters when combined with the BOSS dataset.

We also implemented a BAO likelihood in order to compare our results with the official \textit{BOSS Consensus BAO} results from \cite{2016BOSSCosmology}. This measurements use 3 redshift bins with $z_{\text{eff}} = 0.38, \, 0.51, \, \text{and} \, 0.61 $, where we used the full shape post-reconstruction measurements from the correlation function and 3D power spectra, which contain additional information from measurements of $f\sigma_8$ (see table 7 from \cite{2016BOSSCosmology}). We have not combined these results with our BOSS results, but we plot them alongside our results alone in the next section to show how our results compare with the BOSS alone results from \cite{2016BOSSCosmology}.


\subsection{Flat $\Lambda$CDM Constraints}\label{Sec:LCDM}
\begin{figure}
\begin{center}
\includegraphics[width=\columnwidth]{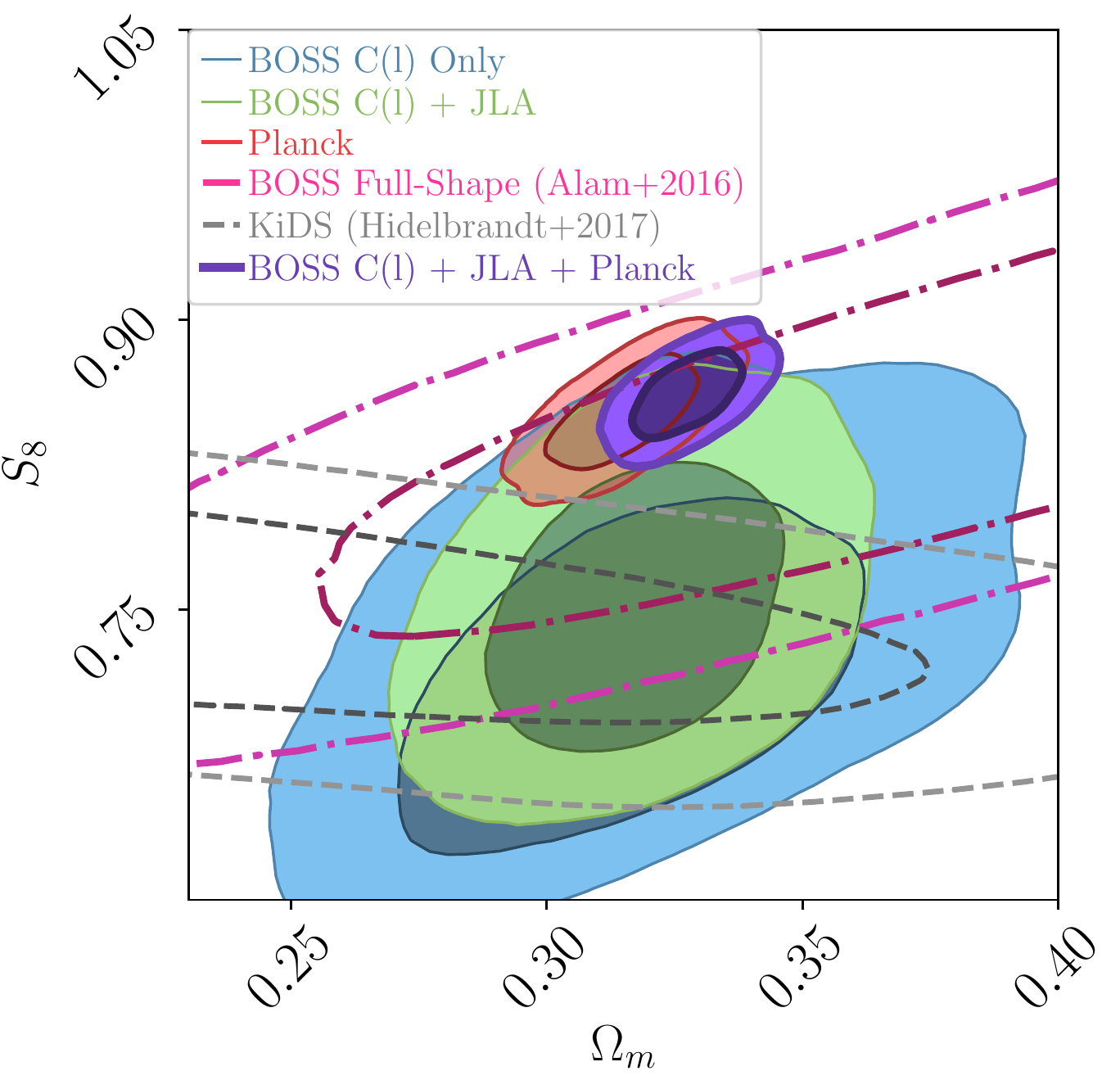}
\caption{2D $\Omega_m \, \times \, S_8$ marginalised credible intervals for a \textbf{$\Lambda$CDM cosmology}. In this figure we show in detail the cosmological results from Section \ref{Sec:LCDM} for BOSS $C_{\ell}$s only \textit{(blue)}; BOSS $C_{\ell}$s plus JLA \textit{(green)}; BOSS $C_{\ell}$s plus JLA and Planck \textit{(purple)}; together with results using the post-reconstruction full shape (incl. $f\sigma_8(z)$) from \protect\cite{2016BOSSCosmology} consensus results \textit{(pink, dot-dashed)}, and Planck alone \textit{(red)}. {In order to compare our results to a weak-lensing probe, we also show the results from \protect\cite{2017MNRAS.465.1454H} \textit{(grey, dashed)}}. For details on the external datasets, see Section \ref{Sec:ExternalData}.}
\label{fig:Om_S8_LCDM}
\end{center}
\end{figure}

\begin{figure*}
\begin{center}
\includegraphics[scale=0.55]{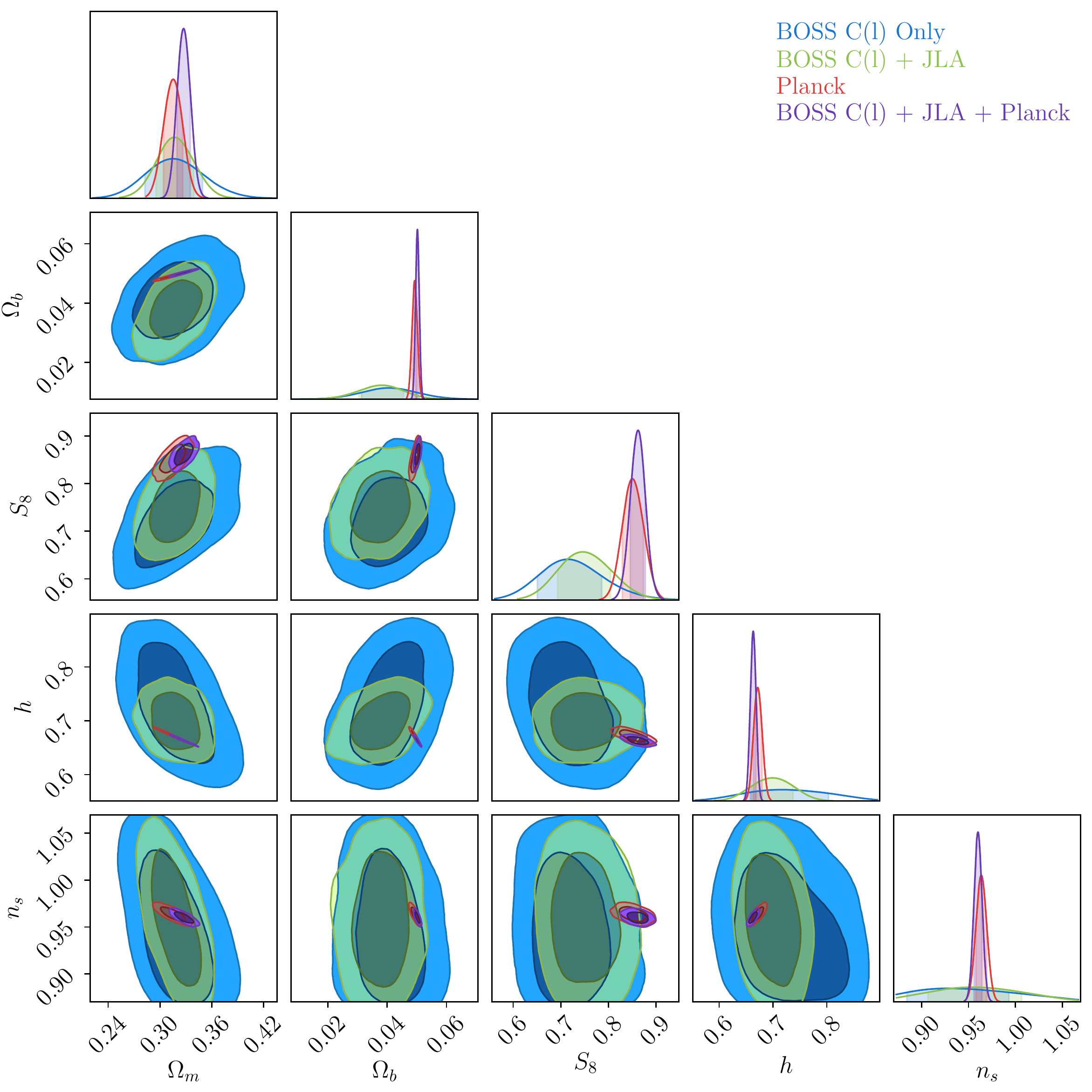}
\caption{Marginalised 1 \& 2D cosmological constraints for the \textbf{$\Lambda$CDM model} varying five cosmological parameters with 1-$\sigma$ (darker) and 2-$\sigma$ (lighter) contour levels. We show here a combination of sampled and relevant derived parameters: $\Omega_m$, $\Omega_b$, $S_8$, $h$, and $n_s$ (marginalising over $\tau_{\text{reio}}$ for the Planck combinations). The \textit{blue contours} where estimated from the BOSS $C_{\ell}$s data alone using the SH16 likelihood; \textit{the green contours} are a combination the BOSS likelihood and JLA data (see Section \ref{Sec:ExternalData}); \textit{the red contours} are the Planck high-$\ell$ TT, TE, EE and low-$\ell$ P likelihood results (see Section \ref{Sec:ExternalData}); finally, the purple contours are a combination of the three probes: BOSS $C_{\ell}$, JLA and Planck (also high-$\ell$ TT, TE, EE and low-$\ell$ P). Note that none of the results here use Planck Lensing data.}
\label{fig:LCDM_Cosmology}
\end{center}
\end{figure*}

\begin{figure*}
\begin{center}
\includegraphics[scale=0.35]{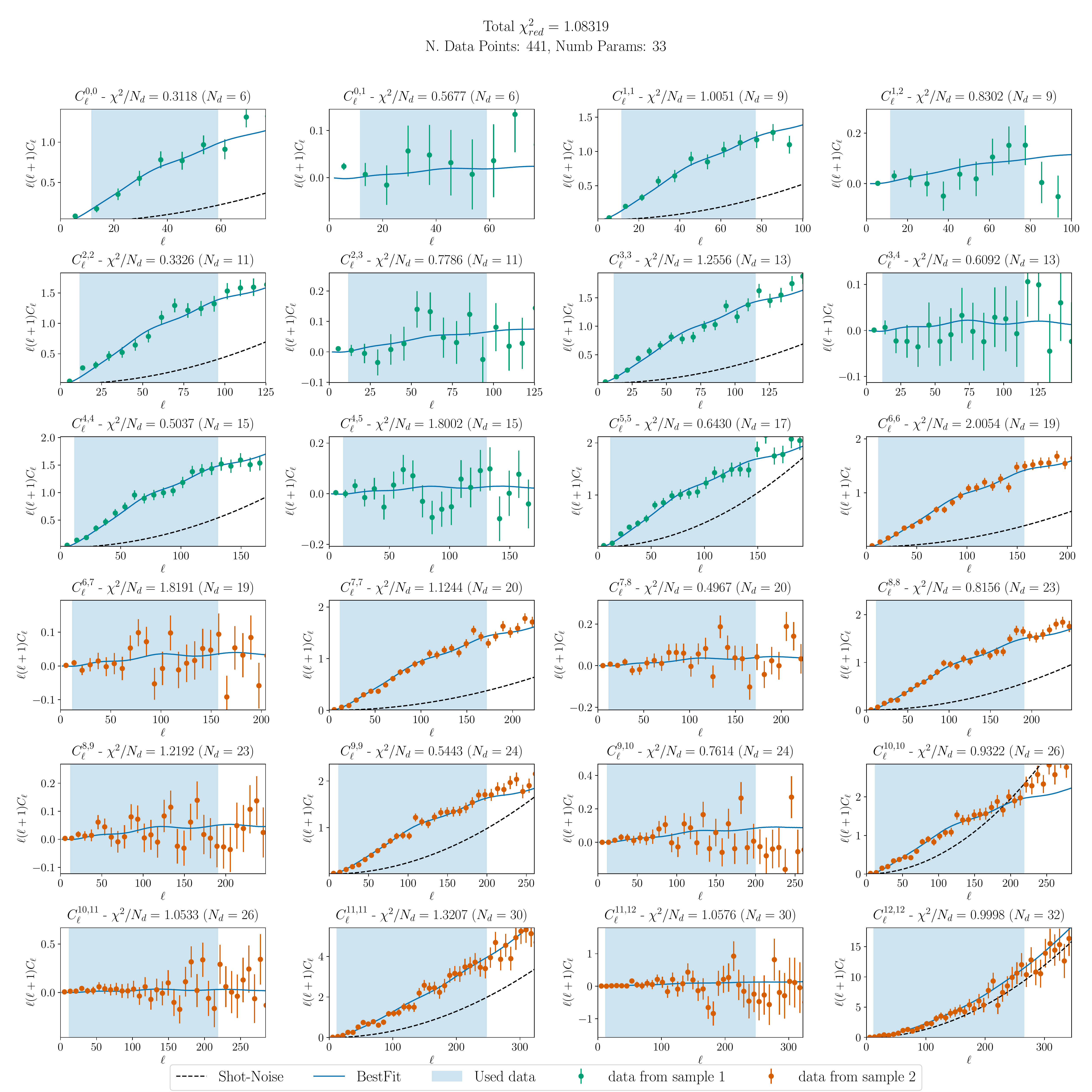}
\caption{Auto- and cross- angular power spectra for the 13 tomographic redshift bins considered for the BOSS DR12 samples: LOWZ (sample 1) and CMASS (sample 2). The shaded blue regions show the scales considered in the cosmological parameter estimation in Section \ref{Sec:CosmoBananas}. The \textit{data points} are the pseudo-$C_{\ell}$ estimates, described in Section \ref{Sec:Measurements}, for LOWZ and CMASS. The \textit{solid blue lines}, generated with \texttt{UCLCL}, reflect the \textit{best fit} auto- and cross-power spectra for the \textbf{$\Lambda$CDM model} estimated in Section \ref{Sec:LCDM}. Finally, the \textit{black dashed lines} show both shot noise and sampled shot noise (for bins 11 and 12). The overall reduced $\chi^2$ for this fit is $\chi^2_{red} \approx 1.08$, where the number of data points is 441 and the total number of sampled parameters is 33 -- 5 cosmological parameters and 28 nuisance parameters. The title on each individual plot reflects the bins \textit{i \& j} for each $C^{ij}_{\ell}$, the $\chi^2$ per data point ($\chi^2/N_d$), and the number of data points for that individual angular power spectrum, $N_d$. The $\ell$-ranges used in this figure correspond to the $\ell_{max}^{5\%}$ in table \ref{Tb:EllCuts}. {Most of the constraining power comes from the auto-power spectra. The cross-power spectra serve to constrain parameters related to the RSD by helping to break the degeneracy between the bias and $A_s$ while also probing the redshift dispersion due to the peculiar motion of galaxies (FoG).}}
\label{fig:Cl_Bestfit}
\end{center}
\end{figure*}
We obtain constraints for the standard cosmological model, a flat $\Lambda$CDM cosmology. We fixed the curvature of the universe to zero, e.g. $\Omega_k = 0$, and varied five cosmological parameters: the baryonic density, $\Omega_b$; the dark matter density, $\Omega_{cdm}$; the amplitude of the primordial power spectra, $A_s$; the spectral index, $n_s$; and the Hubble constant, $h$. As this model considers dark energy as the cosmological constant $\Lambda$, we fixed the $w_0$ parameter to a cosmological constant ($w_0 = -1$); therefore: $\Omega_{\Lambda} = 1 - (\Omega_b + \Omega_{cdm})$. Here, we also fixed the sum of neutrino masses to the minimum found from neutrino oscillation experiments, $\sum m_{\nu} = 0.06 \, eV$ \citepbox{2006NeutrinoReview,2014NeutrinoCosmoPlanck}. From these, we also obtained derived parameters: the total matter density, $\Omega_m \equiv \Omega_b + \Omega_{cdm}$, and the fluctuation of amplitude at 8 $h^{-1}$Mpc, $\sigma_8$ or $S_8 = \sigma_8\sqrt{\Omega_m/0.3}$. Finally, as described in the previous section, we also varied the BOSS, Planck and JLA nuisance parameters. For this analysis, we used the $\ell_{max}^{5\%}$ cuts (see table \ref{Tb:EllCuts}).

We checked the consistency of the datasets by running the same analysis for these probes alone and combined (Planck, JLA, and Planck plus JLA) and calculating respective Bayes factors for these runs. The Bayes factor, Equation \eqref{Eq:BayesFactor},  for combinations of the considered datasets indicates consistency between all three probes:

\begin{ceqn}
\begin{align}
R_{\scriptscriptstyle\text{BOSS+JLA}}^{\Lambda CDM} & \simeq 18  \\
R_{\scriptscriptstyle\text{BOSS+PLANCK}}^{\Lambda CDM} & \simeq 74 \\
R_{\scriptscriptstyle\text{PLANCK+JLA}}^{\Lambda CDM} & \simeq 11 \\
R_{\scriptscriptstyle\text{BOSS+PLANCK+JLA}}^{\Lambda CDM} & \simeq 4 \times 10^4\, ;
\end{align}
\end{ceqn}
these indicate that the datasets are compatible for the considered model, given the chosen priors.

Finally, when considering the combination of BOSS $C_{\ell}$s, Planck and JLA, we find results consistent with \cite{2016BOSSCosmology,2017MNRAS.465.1454H,2017arXiv170801530D}. Figure \ref{fig:Om_S8_LCDM} shows the $\Omega_m \, \times \, S_8 $ 2D plane for this analysis and comparisons with Planck and the BOSS full-shape post-reconstruction from \cite{2016BOSSCosmology}. Despite the Bayes factors showing no significant reason to be concerned about the compatibility of these datasets we see an interesting trend in this figure insofar as a tension and the BOSS dataset in this paper preferring a smaller $S_8$ than the Planck analysis. We argue that this method would prove potentially very useful in resolving any $S_8$ tensions which exist currently between CMB and weak lensing data \citepbox{2015MacCrann,2017Charnock}.

The results for the 1D marginalised cosmological constraints for BOSS $C_{\ell}$ and combinations, together with the 68\% credible intervals, can be found in table \ref{tab:LCDM_Constraints}. The 1-$\sigma$ and 2-$\sigma$ contour levels can be found in Figure \ref{fig:LCDM_Cosmology} where the nuisance parameters have been marginalised over. Figure \ref{fig:Cl_Bestfit} shows the best fit theory $C_{\ell}$ using the parameters estimated from this analysis with a $\chi^2_{red} = 1.08$, which also indicates reliability and robustness of the analysis performed. 

Even though we do not show the results in this work, we performed a cosmological analysis using a $\Lambda$CDM with a fixed zero neutrino mass, $\sum m_{\nu} = 0$ eV. We compared it with the model used in this section, $\Lambda$CDM with $\sum m_{\nu}$ fixed to $0.06$ eV, using the Bayes factor for model selection. Consider $\vec{D}$ representing the combination of data vectors; the Bayes factor is given by

\EQ{}{
R_{A,B} = \frac{P(\vec{D}_{\scriptscriptstyle\text{BOSS+PLANCK+JLA}}|\sum m_{\nu} = 0.06 \text{ eV})}{P(\vec{D}_{\scriptscriptstyle\text{BOSS+PLANCK+JLA}}|\sum m_{\nu} = 0 \text{ eV}) } = 8\times 10^5
}
\noindent This indicates that, for a $\Lambda$CDM model, the data prefers massive neutrinos over no neutrino mass at all.


\subsection{Flat $w$CDM Constraints}\label{Sec:wCDM}

In this section, we allowed the equation of state of dark energy, $w_0$, to vary. This is a trivial extension of the standard model of cosmology with just one extra parameter. If $w_0 = -1$, the solution indicates that the nature of dark energy is actually the cosmological constant, $\Lambda$. The procedure for this analysis followed in similar fashion as the one outlined in Section \ref{Sec:LCDM}, varying six parameters instead of five: $\Omega_b$, $\Omega_{cdm}$, $n_s$, $\ln 10^{10} A_s$, $h$, and the extra $w_0$. Note that, for this case, we are not varying $w_a$, i. e., we do not consider a redshift evolution in the equation of state of dark energy. Again, we fixed the neutrino parameter to $\sum m_{\nu} = 0.06 \, eV$ \citepbox{2006NeutrinoReview,2014NeutrinoCosmoPlanck}. Here, we used the same $\ell_{max}^{5\%}$ cuts as in the last section (see table \ref{Tb:EllCuts}).
\begin{figure}
\begin{center}
\includegraphics[width=\columnwidth]{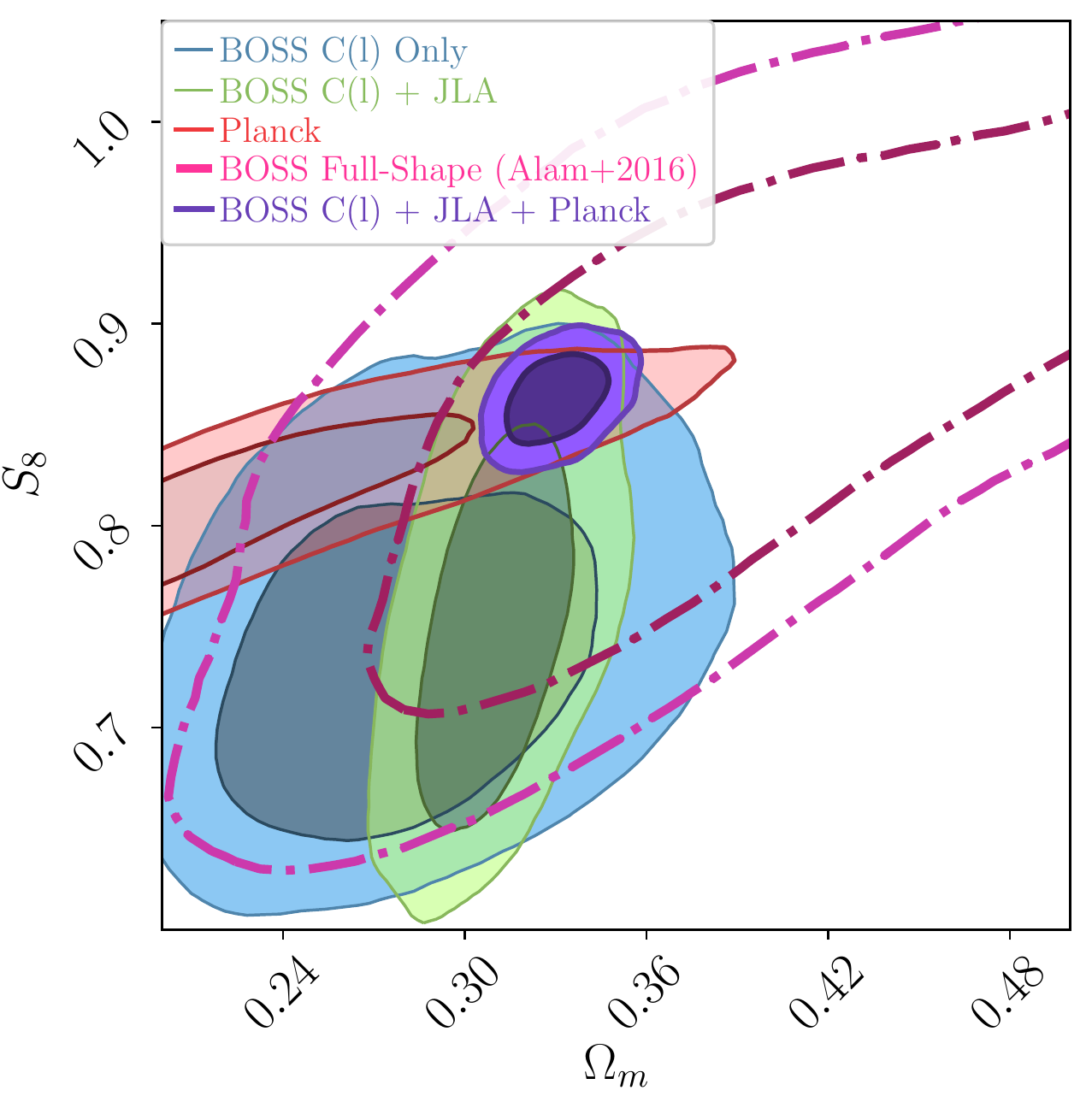}
\caption{2D $\Omega_m \, \times \, S_8$ marginalised credible intervals for a \textbf{$w$CDM Cosmology}. This shows in detail the cosmological results from Section \ref{Sec:wCDM} for BOSS $C_{\ell}$s only \textit{(blue)}; BOSS $C_{\ell}$s plus JLA \textit{(green)}; BOSS $C_{\ell}$s plus JLA and Planck \textit{(purple)}; together with results using just the full shape (pre-reconstruction) from \protect\cite{2016BOSSCosmology} consensus results \textit{(pink)}, and Planck alone \textit{(red)}.}
\label{fig:Om_S8_wCDM}
\end{center}
\end{figure}

\begin{figure}
\begin{center}
\includegraphics[width=\columnwidth]{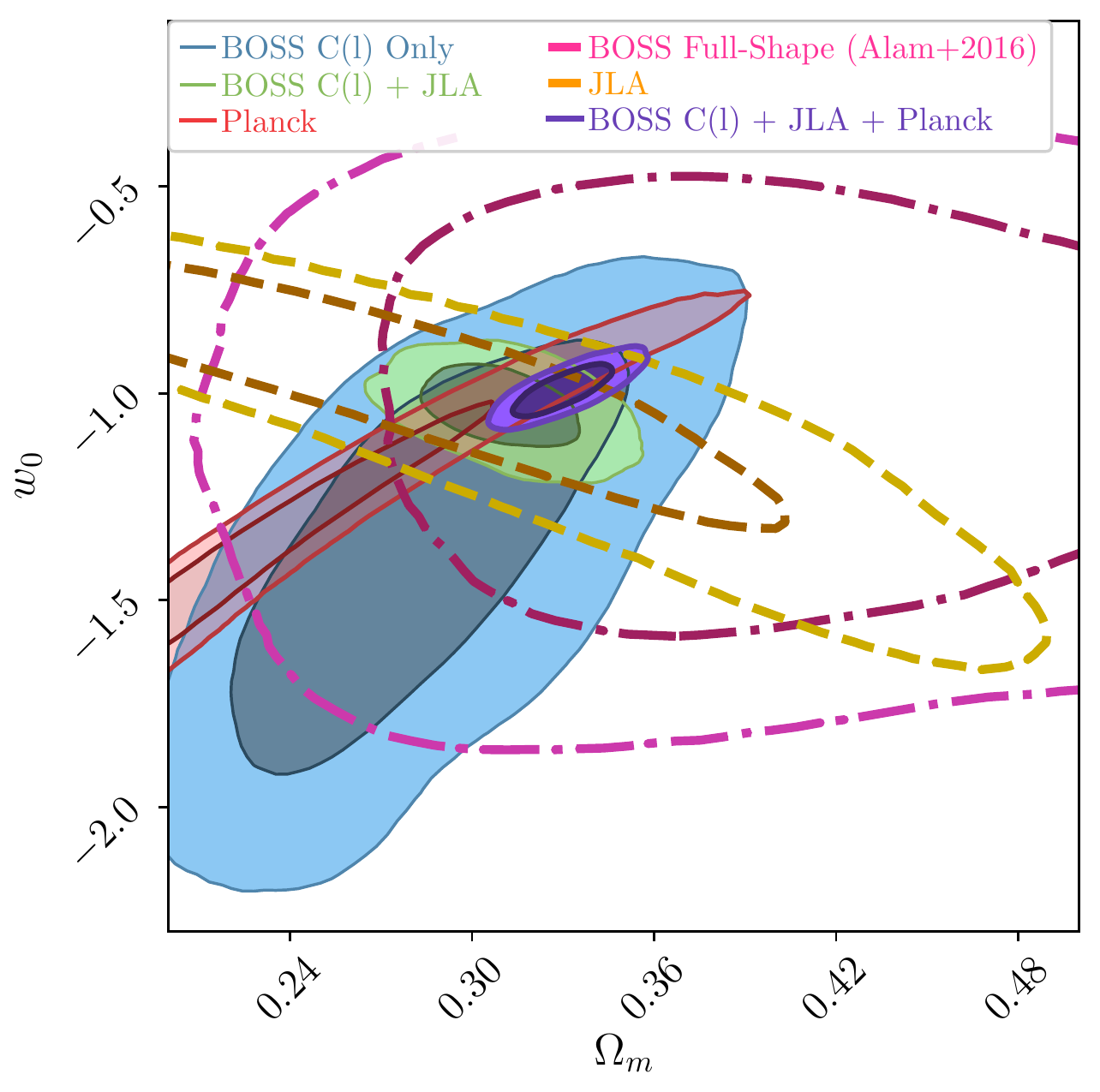}
\caption{2D $w_0 \, \times \, \Omega_m$ marginalised credible intervals for a \textbf{$w$CDM Cosmology}. This shows in detail the cosmological results from Section \ref{Sec:wCDM} for BOSS $C_{\ell}$s only \textit{(blue)}; BOSS $C_{\ell}$s plus JLA \textit{(green)}; BOSS $C_{\ell}$s plus JLA and Planck \textit{(purple)}; together with results using just the full shape (pre-reconstruction) from \protect\cite{2016BOSSCosmology} consensus results \textit{(pink)}, JLA \textit{(yellow)}, and Planck alone \textit{(red)}.}
\label{fig:Om_w0_wCDM}
\end{center}
\end{figure}

\begin{figure*}
\begin{center}
\includegraphics[scale=0.60]{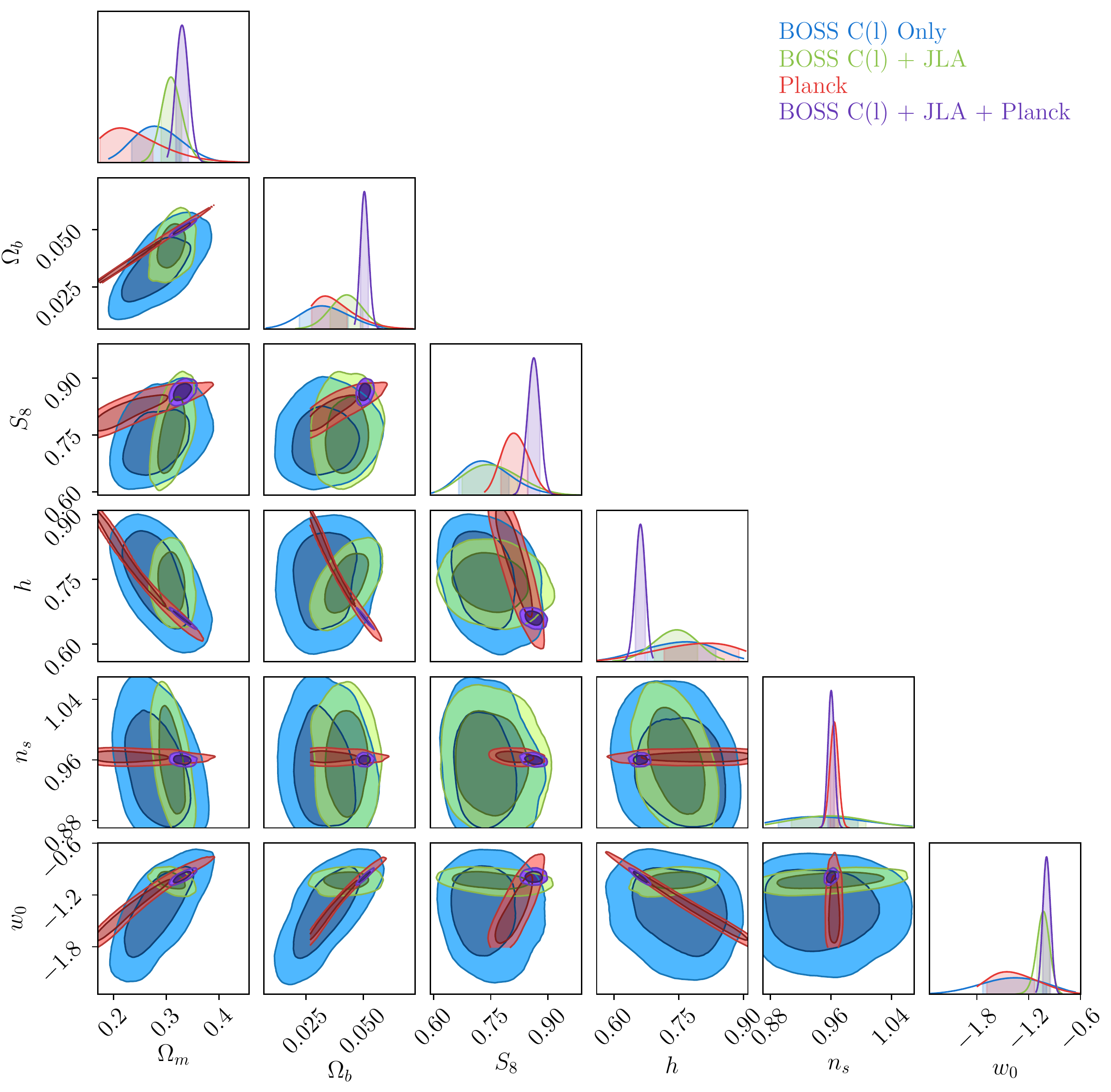}
\caption{Cosmological constraints for the \textbf{$w$CDM model} varying now 6 cosmological parameters. This figure contains a combination of sampled and relevant derived parameters from the chains: $\Omega_m$, $\Omega_b$, $S_8$, $h$, $n_s$, and $w_0$. Note that the Planck chains also varied $\tau_{reio}$. The \textit{blue contours} where estimated from the BOSS $C_{\ell}$s data alone using the SH16 likelihood; \textit{green contours} are a combination the BOSS likelihood and JLA data; \textit{ red contours} are the Planck high-$\ell$ TT, TE, EE and low-$\ell$ P likelihood results; finally, the \textit{purple contours} are a combination of the three probes: BOSS $C_{\ell}$, JLA and Planck (also high-$\ell$ TT, TE, EE and low-$\ell$ P). The apparent cuts in the Planck alone contours are due to the prior in $h$. Note, again, that none of the results here use Planck Lensing data.}
\label{fig:wCDM_Cosmology}
\end{center}
\end{figure*}

Figures \ref{fig:Om_S8_wCDM} and \ref{fig:Om_w0_wCDM} show in detail the contours for $S_8 \, \times \, \Omega_m$ and $w_0 \, \times \, \Omega_m$, respectively, and comparisons with previous measurements in the literature. From the Figure \ref{fig:Om_w0_wCDM} and from the complete set of results in \ref{fig:wCDM_Cosmology} we show that a $\sim 4\%$ error (1-$\sigma$ CI) on the equation of state of dark energy is obtained from this cosmological analysis:

\EQ{w0PlankBOSSJLA}{w_0 = -0.993^{+0.046}_{-0.043}.}
This result is consistent with the $\Lambda$CDM scenario of standard cosmology, i. e., it is consistent with dark energy being a cosmological constant, $\Lambda$. Note from Figure \ref{fig:wCDM_Cosmology} that we find a small value of $h$ (compared to \citealt{PlanckCosmology2016}) when combining BOSS $C_{\ell}$s, Planck, and JLA:

\EQ{}{h^{w\text{CDM}} = 0.661\pm 0.012.} This value is lower than the quoted Planck value alone; this puts further tension in this measurements if compared to the Hubble constant result from Cepheid Variables \citepbox{Riess2016, Riess2018}.

As the model in this section is different from the previous section, we performed an evidence analysis using the Bayes factor, Equation \eqref{Eq:BayesFactor}, in order to be sure that our measurements can be combined with the the external data described in Section \ref{Sec:ExternalData}. The following measurements indicate that such combinations are consistent for a $w$CDM model:

\begin{ceqn}
\begin{align}
R_{\scriptscriptstyle\text{BOSS+JLA}}^{\textit{w}CDM} & \simeq 2 \times 10^2  \\
R_{\scriptscriptstyle\text{BOSS+PLANCK}}^{\textit{w}CDM} & \simeq 4 \times 10^3 \\
R_{\scriptscriptstyle\text{PLANCK+JLA}}^{\textit{w}CDM} & \simeq 2 \\
R_{\scriptscriptstyle\text{BOSS+PLANCK+JLA}}^{\textit{w}CDM} & \simeq 3 \times 10^5\, .
\end{align}
\end{ceqn}

Finally, we used the ratio of the evidences, the Bayes factor, to perform a model selection between $w$CDM and $\Lambda$CDM using the final dataset combination. Assuming $\vec{D}$ to be the combination of data vectors for all the datasets, the Bayes factor between the two models is

\EQ{}{
R_{w,\Lambda} = \frac{P(\vec{D}_{\text{BOSS+Planck+JLA}}|w\text{CDM})}{P(\vec{D}_{\text{BOSS+Planck+JLA}}|\Lambda\text{CDM})} = 0.67
}



\subsection{Flat $\Lambda$CDM + $\sum m_\nu$ Constraints}\label{Sec:nCDM}
\begin{figure}
\begin{center}
\includegraphics[width=\columnwidth]{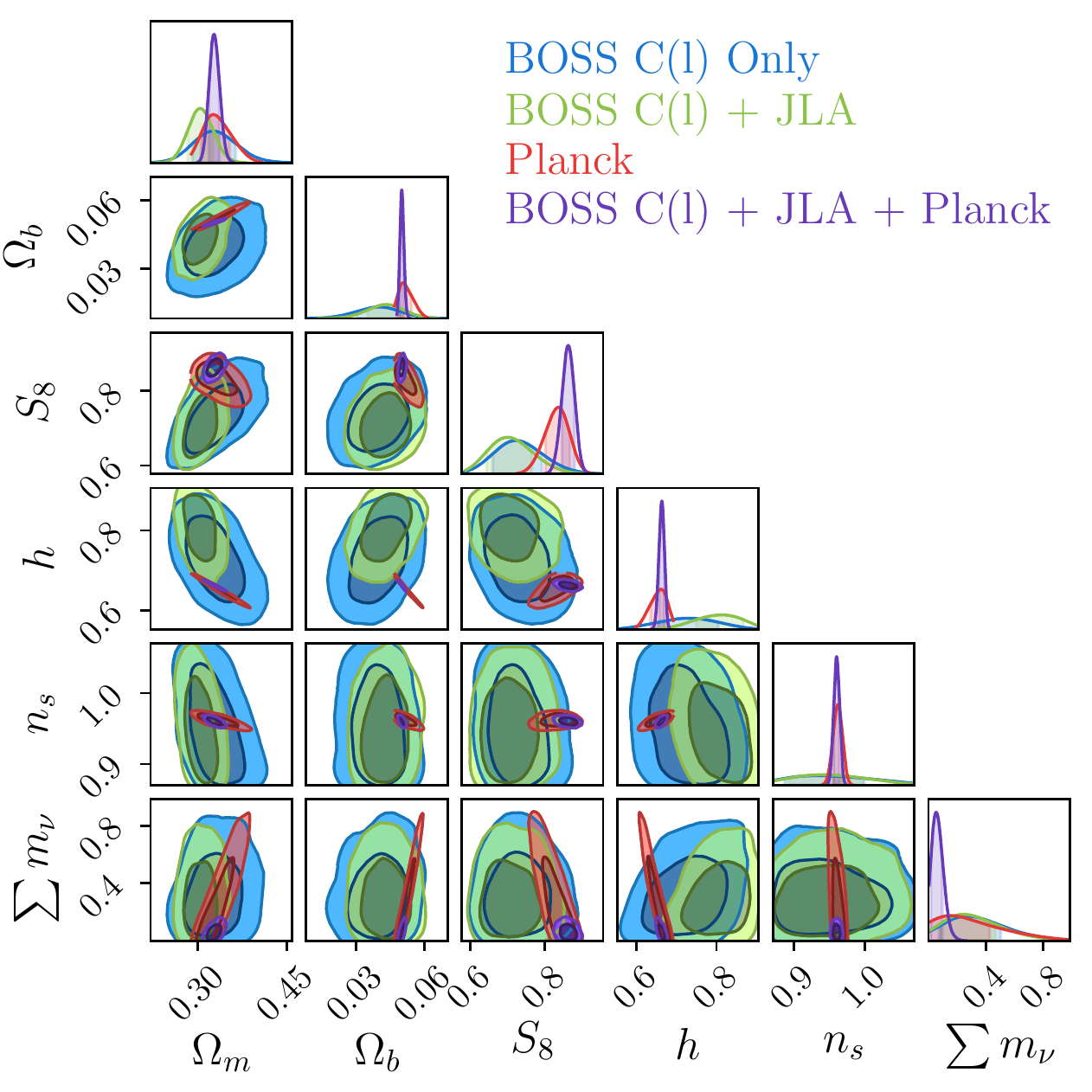}
\caption{1D and 2D marginalised credible intervals for a \textbf{$\Lambda$CDM Cosmology with $\sum m_{\nu}$} when using scales up to $k_{max}\approx 0.07 h$ Mpc$^{-1}$ ($\ell_{max}^{5\%}$ cut). Here we show the $\Omega_m$, $\Omega_b$, $S_8$, $h$, $n_s$, and $\sum m_{\nu}$ contours for BOSS $C_{\ell}$s alone \textit{(blue)}; BOSS $C_{\ell}$s plus JLA \textit{(green)}; Planck high-$\ell$ TT, TE, EE and low-$\ell$ P \textit{(red)}; and BOSS $C_{\ell}$s plus JLA and Planck high-$\ell$ TT, TE, EE and low-$\ell$ P \textit{(purple)}. As most of the scales that contain clean information on the neutrino masses are cut off, the 95\% CI upper bound found is $\sum m_{\nu} < 0.14$ eV.}
\label{fig:nuCDM5pc}
\end{center}
\end{figure}
\begin{figure}
\begin{center}
\includegraphics[width=\columnwidth]{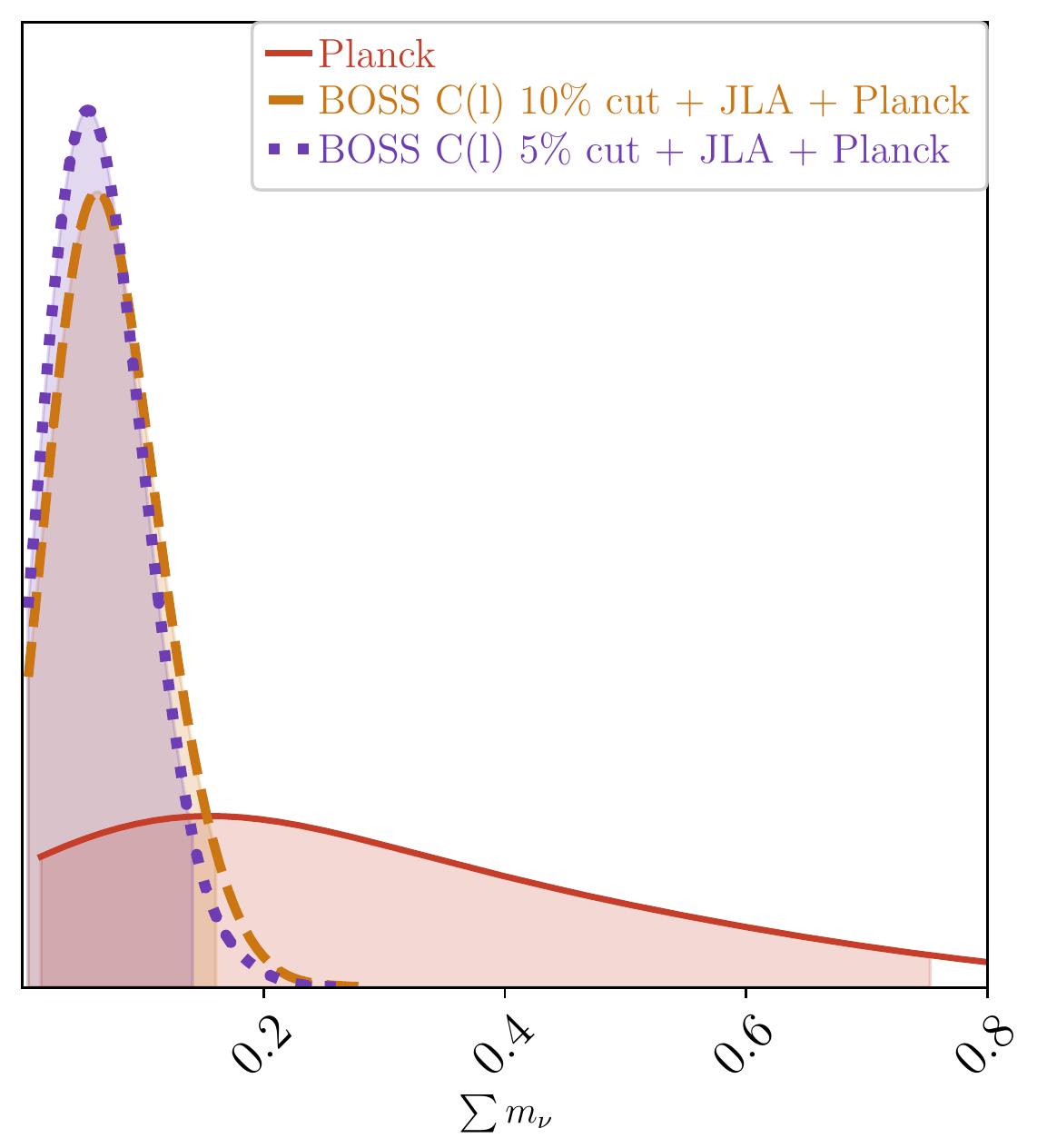}
\caption{1D marginalised with 95\% credible intervals for $\sum m_{\nu}$ in three different cases: \textit{(red solid)} Planck high-$\ell$ TT, TE, EE and low-$\ell$ P, \textit{(yellow dashed)} BOSS $C_{\ell}$s with the $\ell_{max}^{10\%}$ cut plus Planck and JLA, and \textit{(purple dotted)} BOSS $C_{\ell}$s with the $\ell_{max}^{5\%}$ cut plus Planck and JLA. The 95\% upper limit for each case is, respectively: \textit{(red)} 0.76 eV, \textit{(yellow)} 0.16 eV, and \textit{(purple)} 0.14 eV.}
\label{fig:neutrinoCompare1}
\end{center}
\end{figure}

\begin{figure}
\begin{center}
\includegraphics[width=\columnwidth]{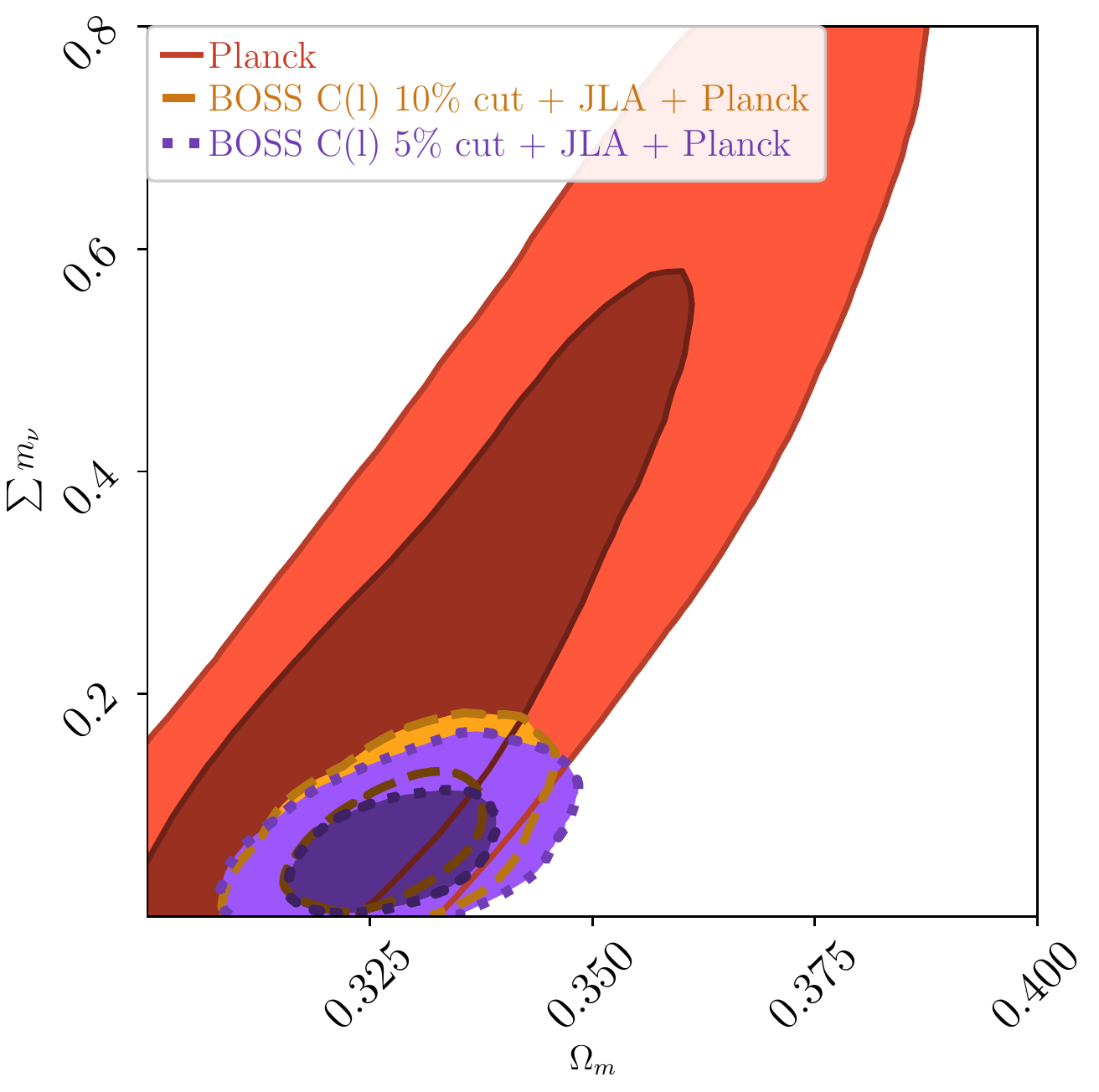}
\caption{2D marginalised 1- and 2-$\sigma$ credible intervals for the $\sum m_{\nu}$-$\Omega_m$ plane for three different cases: \textit{(blue solid)} Planck high-$\ell$ TT, TE, EE and low-$\ell$ P, \textit{(red dashed)} BOSS $C_{\ell}$s with the $\ell_{max}^{10\%}$ cut plus Planck and JLA, and \textit{(orange dotted)} BOSS $C_{\ell}$s with the $\ell_{max}^{5\%}$ cut plus Planck and JLA.}
\label{fig:neutrinoCompare2}
\end{center}
\end{figure}

\begin{figure*}
\begin{center}
\includegraphics[scale=0.65]{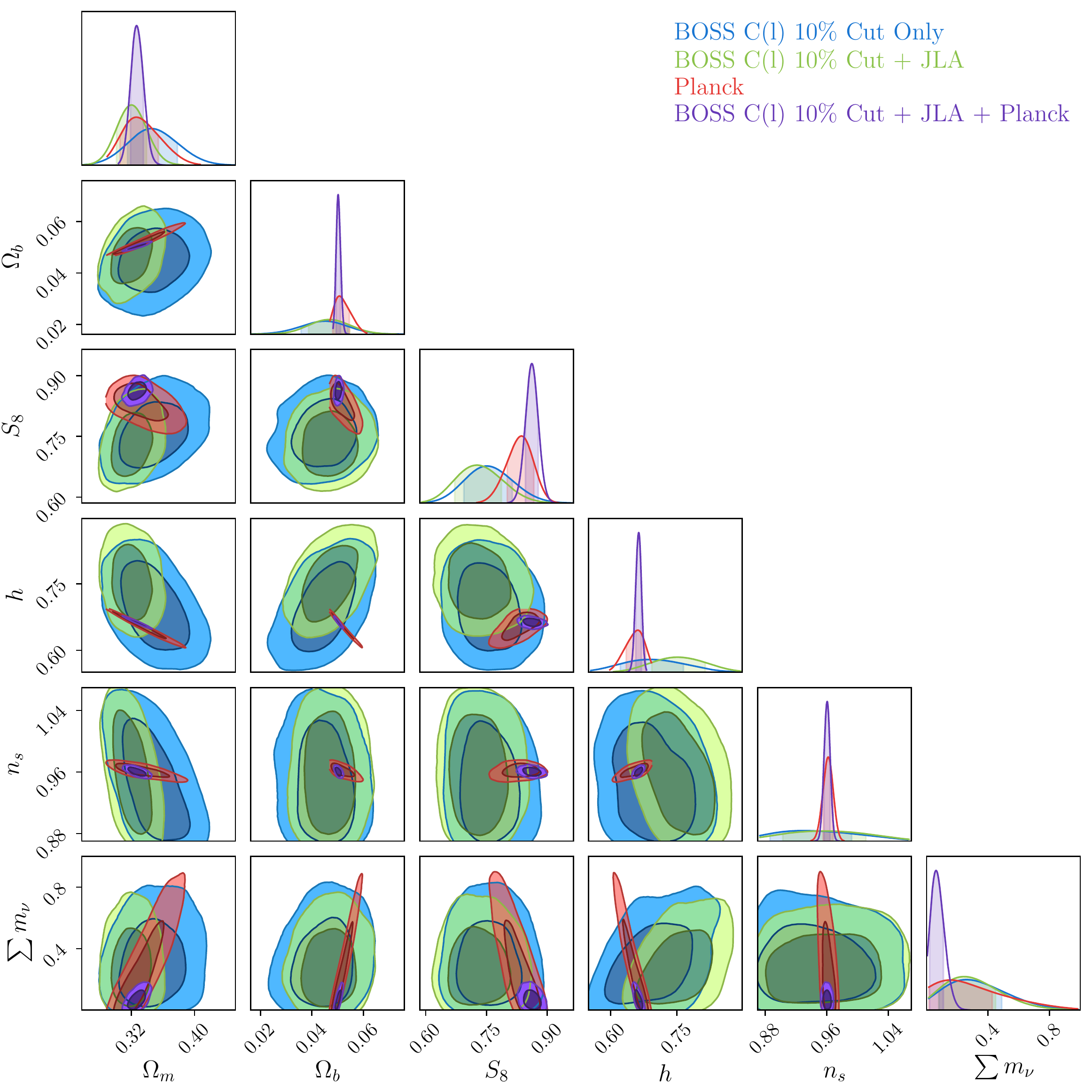}
\caption{Cosmological constraints for the \textbf{$\Lambda$CDM + $\sum m_{\nu}$ model}, using the $\ell_{max}^{10\%}$ cut, varying now 6 cosmological parameters, including the sum of neutrino masses considering only one massive species. This figure contains a combination of sampled and relevant derived parameters from the chains: $\Omega_m$, $\Omega_b$, $S_8$, $h$, $n_s$, and $\sum m_{\nu}$. Note that the Planck chains also varied $\tau_{reio}$. The \textit{blue contours} where estimated from the BOSS $C_{\ell}$s data alone using the SH16 likelihood; \textit{the green contours} are a combination the BOSS likelihood and JLA data; \textit{the red contours} are the Planck high-$\ell$ TT, TE, EE and low-$\ell$ P likelihood results; finally, the purple contours are a combination of the three probes: BOSS $C_{\ell}$, JLA and Planck (also high-$\ell$ TT, TE, EE and low-$\ell$ P). For this scale cut, the combination of datasets yields an upper bound on $\sum m_{\nu} < 0.16$eV.}
\label{fig:nuCDM_Cosmology}
\end{center}
\end{figure*}
For the last model considered in this work, we assume a flat $\Lambda$CDM with variable neutrino masses, varying the sum of the species' masses, $\sum m_{\nu}$. In the previous sections, we fixed the sum of neutrino masses to $\sum m_{\nu} = 0.06 \, eV$ due to results from neutrino oscillation experiments for the lower bound of the normal neutrino mass ordering \citep{2003HannestadNeutrino,2006NeutrinoReview,2016HannestadNeutrino}. 

In this section, we considered one of the two different scenarios regarding different neutrino hierarchies, the \textit{normal hierarchy}. To approximate the normal hierarchy, one can approximate the two lower masses to be zero and vary $\sum m_{\nu}$ for one remaining massive species. We do not investigate details of how the prior on the hierarchy or on the absolute mass changes this result and we leave this to a future analysis. We fix $N_{eff} = 3.046$ by changing the values of massive neutrinos and ultra-relativist particles for the case considered, i. e., $N_{\nu} = 1$ and $N_{ur} = 2.0328$.

We perform an analysis using the same $\ell$-range as in the previous sections, $\ell_{max}^{5\%}$ from table \ref{Tb:EllCuts}. A summary of the marginalised 1D credible intervals from the cosmological estimation made with this cut can be found in the third part of table \ref{tab:LCDM_Constraints} showing the one sigma intervals for the $\Lambda$CDM parameters plus the 95\% upper limit for $\sum m_{\nu}$. The 1D and 2D marginalised credible intervals for this analysis can be found in Figure \ref{fig:nuCDM5pc}. When considering an approximation for the normal hierarchy, for a combination of BOSS $C_{\ell}$s, Planck CMB data, and supernovae data from JLA, the 95\% upper limit for sum of neutrino masses is:

\EQ{}{\sum m_{\nu} < 0.14 \, eV \quad \small\text{(BOSS + Planck + JLA -- $\ell_{max}^{5\%}$ cut)}\, .}

From Figure \ref{fig:neutrinoCompare1} and even more so from Figure \ref{fig:neutrinoCompare2}, one can notice that we are not so far from excluding zero total neutrino mass using cosmological data alone. As the power of such datasets increase we should be able, using the correct analysis and tools, to measure and detect neutrino masses independently from atmospheric experiments.

We then proceed to check the consistency of datasets by using the evidence of each cosmological parameter estimation for these model to calculate the Bayes factor (Equation \ref{Eq:BayesFactor}):

\begin{ceqn}
\begin{align}
R_{\scriptscriptstyle\text{BOSS+JLA}}^{\Lambda CDM + \sum m_{\nu} \, 5\%} & \simeq 1\times 10^2  \\
R_{\scriptscriptstyle\text{BOSS+PLANCK}}^{\Lambda CDM +\sum m_{\nu} \, 5\%} & \simeq 4\times 10^2 \\
R_{\scriptscriptstyle\text{PLANCK+JLA}}^{\Lambda CDM +\sum m_{\nu}} & \simeq 40 \\
R_{\scriptscriptstyle\text{BOSS+PLANCK+JLA}}^{\Lambda CDM +\sum m_{\nu} \, 5\%} & \simeq 3 \times 10^2\, .
\end{align}
\end{ceqn}
These values, again, indicate the consistency of datasets for the considered model. 

For the final analysis in this work, we extended the scales considered for the $\ell_{max}^{10\%}$ cuts (see table \ref{Tb:EllCuts} for details). This allow us to access smaller scales that are still in the beginning of the so-called the weak non-linear regime \citepbox{Thomas2010Neutr,Bird2012}. Note that these scales are still larger than the scales that most collaborations use for power spectra or correlation function cosmological analysis \citep{Ho2012,2016BOSSCosmology,2017MNRAS.465.1454H,2017arXiv170801530D} -- \cite{2017arXiv170801530D}, for example, uses scales up to $0.78\, ,h$ Mpc $^{-1}$. In other words, one can be confident that the $\ell_{max}^{10\%}$ cuts are safe to be used, not using scales outside the weak non-linear regime. 

We then proceed to perform a similar cosmological analysis for a $\Lambda$CDM model with one massive species of neutrino, approximating the normal hierarchy. The 1D and 2D marginalised credible intervals for these final analyses can be found in Figure \ref{fig:nuCDM_Cosmology} and the marginalised 68\% credible intervals for the $\Lambda$CDM parameters and the 95\% credible interval upper limit for $\sum m_{\nu}$ using this cut can be found in table \ref{tab:LCDM_Constraints}. The Bayes factors for this choice are shown below (note that we have 2 further nuisance parameters in the $\ell_{max}^{10\%}$ cut and we checked that failure to add these reduces the Bayes factor significantly).

\begin{ceqn}
\begin{align}
R_{\scriptscriptstyle\text{BOSS+JLA}}^{\Lambda CDM + \sum m_{\nu} \, 10\%} & \simeq 70  \\
R_{\scriptscriptstyle\text{BOSS+PLANCK}}^{\Lambda CDM +\sum m_{\nu} \, 10\%} & \simeq 6 \times 10^2 \\
R_{\scriptscriptstyle\text{BOSS+PLANCK+JLA}}^{\Lambda CDM +\sum m_{\nu} \, 10\%} & \simeq 3 \times 10^5\, .
\end{align}
\end{ceqn}

This extended scale analysis demonstrates the robustness of the results presented in this section as the 95\% CI upper bound for $\sum m_{\nu}$ remains robust to these cuts (see Figures \ref{fig:neutrinoCompare1} and \ref{fig:neutrinoCompare2}):
\EQ{}{\sum m_{\nu} < 0.16 \, eV \quad \small\text{(BOSS + Planck + JLA -- $\ell_{max}^{10\%}$ cut)}.}

\section{Conclusions}
In this work, we have taken a different approach\footnote{Compared to the approaches from the official BOSS Collaboration papers: \cite{2017RossBOSS,2017BeutlerBOSS,2017Beutler2BOSS,2017SatpathyBOSS,2017SanchezBOSS,2017GriebBOSS,2017SalazarBOSSwTheta,2017WangBOSS,2017ZhaoBOSS}.} obtaining galaxy clustering information from the BOSS DR12 large scale structure catalogue \citepbox{BOSSCatalogue2016}. This approach consisted of using a pseudo angular power spectra estimator (PCL) applied to 13 tomographic redshift bins ranging from $0.15 \leq z < 0.8$ with a redshift dependent bias, a redshift dispersion, and extra shot-noise as nuisance parameters to be sampled with the cosmological parameters using \uclci (Cuceu et al., \textit{in prep}). In this approach, we also used splines of the data as input for the simulation used for covariance matrix estimation.

The tomographic approach in redshift space and the covariance matrix estimation method used in this work allowed us to perform a cosmology-free inference from the data. In other words, nowhere in this analysis was a fiducial cosmology assumed. This is, by itself, a great advantage over methods that use $P(k)$ or $\xi(r)$ as these need to assume a fiducial cosmology in order to transform from redshift space to radial distances. The impact of such strong assumption in the cosmological inference is still unknown.

We performed systematic and consistency checks with the data and the method itself with satisfactory results. From the 18 different sources of systematics considered in Section \ref{Sec:Systm}, none demonstrated worrying excess of power in the scales considered in table \ref{Tb:EllCuts}. Consistency checks demonstrated the robustness of: our estimated covariance matrix, since the recovered cosmology was the same under different estimation methods (through simulation and theory); the likelihood, given that it returns the same contours under three different approaches; and of our whole method, since we recovered the right cosmology from a controlled simulation.

Cosmological parameters were obtained for 3 different models: $\Lambda$CDM, $w$CDM, and $\Lambda$CDM with $\sum m_{\nu}$. We highlight the following main points regarding the results obtained in Section \ref{Sec:CosmoBananas}:
\begin{itemize}
\item[\textbf{1.}] The constraints obtained for all three models considered, using a tomographic analysis in harmonic space, are extremely competitive in comparison to the ones obtained by the BOSS Collaboration \citepbox{2016BOSSCosmology} and other recent large collaboration results such as DES \citepbox{2017arXiv170801530D} and KiDS \citepbox{2017MNRAS.465.1454H} with errors as small as those obtained by these collaborations.

\item[\textbf{2.}] Even though information along the line-of-sight is ``washed away" due to projecting the data into tomographic bins, we obtain one of the tightest constraints for the  equation-of-state of dark energy with a $\sim$4\% error when combining BOSS $C_{\ell}$s, Planck CMB, and JLA Supernovae. This has not been achieved before using $C_{\ell}$ with a spectroscopic survey and the constraint is as tight as the one obtained from the state-of-the-art Dark Energy Collaboration analysis, using a combination of DES galaxy clustering \& weak lensing, Planck, JLA, and BAO \citepbox{2017arXiv170801530D}.

\item[\textbf{3.}] For the models and datasets considered, we find very high values for the Bayes factor, $R$, when combining BOSS $C_{\ell}$s, Planck, and JLA. We highlight: $R_{\scriptscriptstyle\text{BOSS+PLANCK+JLA}} \simeq 4 \times 10^4$ for $\Lambda$CDM and $R_{\scriptscriptstyle\text{BOSS-10\%+PLANCK+JLA}} \simeq 3 \times 10^5$ for $\Lambda$CDM varying neutrino masses. 

\item[\textbf{4.}] The Bayes factor can also be used for model selection. Considering the combination of datasets, the Bayes factor between $\Lambda$CDM and $w$CDM is

\EQ{}{R_{w,\Lambda} = \frac{P(\vec{D}_{\text{BOSS+Planck+JLA}}|w\text{CDM})}{P(\vec{D}_{\text{BOSS+Planck+JLA}}|\Lambda\text{CDM})} = 0.67 \nonumber}
where $\vec{D}$ here represents the overall combination of data vectors. This indicates that this combination prefers slightly $\Lambda$CDM to $w$CDM, although no strong conclusion can be made. 

\item[\textbf{5.}] We find a small tension between BOSS $C_{\ell}$s and Planck for $S_8$ in all models considered, with BOSS preferring smaller values. For example, for $\Lambda$CDM: 

\begin{ceqn}
\begin{align*}
S_8 & = 0.715^{+0.072}_{-0.064} \quad(\text{BOSS}) \\ \nonumber
S_8 &  = 0.850^{+0.023}_{-0.021}\quad(\text{Planck}) \nonumber
\end{align*}\end{ceqn}
although the combination of these datasets prefers higher values such as Planck (see table \ref{tab:LCDM_Constraints}) and the Bayes factor suggest the datasets are compatible. We conclude that such tension can be investigated further with this method as LSS data increases in size and depth.

\item[\textbf{6.}] Even though we do not show these results, we performed a cosmological analysis using a $\Lambda$CDM model but fixing $\sum m_{\nu} = 0 eV$ and compared with the $\Lambda$CDM results from Section \ref{Sec:LCDM}, which has a $\sum m_{\nu}$ fixed to $0.06$ eV. Using the Bayes factor for model selection, it is clear that the data prefers massive neutrinos against no neutrino mass at all:

\begin{ceqn}
\begin{align}
R_{0.06 eV, 0} & = \frac{P(\vec{D}_{\tiny\text{BOSS+Planck+JLA}}| \Lambda\text{CDM} + \sum m_{\nu} = 0.06)}{P(\vec{D}_{\tiny\text{BOSS+Planck+JLA}}| \Lambda\text{CDM} + \sum m_{\nu} = 0)} \nonumber \\
& = 8 \times 10^5 \, . \nonumber
\end{align}
\end{ceqn}

\item[\textbf{7.}] The neutrino mass constraints we obtain here can be compared to the tomographic analysis in real space done by \cite{2017SalazarBOSSwTheta}, which obtains an upper bound of $\sum m_{\nu} < 0.474$ eV (95\% CI). The reason we obtain much tighter constraints ($\sum m_{\nu} < 0.14$ eV (95\% CI)), even though we are also performing a tomographic analysis, is due to a series of decisions, including the approach we take to model the redshift dispersion and galaxy ``shell-crossing" (see Section \ref{Sec:SpecNz}), bias, and extra-shot noise. It is possible that the main difference between the results is due to different approach in modelling the neutrino mass hierarchy. \cite{2017SalazarBOSSwTheta} considers a model where the three neutrino species have degenerate mass hierarchy, i. e., the three masses are equal. This is already ruled out by particle physics experiments that measure the mass splitting from neutrino oscillation experiments (see \citealt{2014Gonzalez-GarciaNeutrino} for an update on the neutrino mass splitting fits). The approach we took in this work (see Section \ref{Sec:nCDM}) naturally yields smaller upper bounds in $\sum m_{\nu}$. In \cite{2018LoureiroNeutrino}, we studied the impact of model choice in the sum of neutrino masses and their hierarchy, showing that this upper bound was driven by our model choices.
\end{itemize}

We have shown here that it is possible to recover very powerful and competitive cosmological constraints from spectroscopy alone with a 2+1D analysis -- as good as the constraints from doing a 3D analysis. {There are several reasons why a tomographic analysis in harmonic space performs as well as the standard 3D analysis. The precision of spectroscopic redshifts allows for very fine radial binning, probing the evolution of structure in more detail. The `effective redshift' approximation impacts 3D analyses, complicating the proper treatment of `light-cone' effects. By contrast, this is not a problem with $C_{\ell}$s as we are dealing with data in redshift space using a fine binning. As we are using data in redshift space, no transformation to comoving coordinates is necessary, and hence no fiducial cosmology is required. We argue that it is much simpler in Fourier space than in real space to correct certain types of systematic errors and to perform scale cuts to deal with non-linearities; the arguments here are similar to the ones used to compare $P(k)$ and $\xi(r)$.}

Finally, we point out that the most promising advantage of performing this 2+1D tomographic analysis in harmonic space is the practicality of combining spectroscopic and photometric probes, which include also cosmic shear, as the latter also ``lives" in a 2+1D space. Using the method proposed by \cite{2016McLeod} the spectroscopic sample has the potential of ``fixing" the photometric redshift limitations when probing the photometric clustering redshift distribution together with the cosmological parameters.

With increasingly large future photometric and spectroscopic surveys such as LSST \citepbox{2012arXiv1211.0310L}, DESI \citepbox{2013arXiv1308.0847L}, J-PAS \citepbox{JPAS}, and Euclid \citepbox{2011arXiv1110.3193L}, the future of precision cosmology lies in our ability to combine datasets from across the entire electromagnetic spectrum. The angular power spectrum approach offers a unified framework for coherently combining different datasets in order to obtain maximal information from each \citep{JoachimiBridle2010,Kirk2015,2016McLeod}. We believe that the approach used in this paper, in which cosmological information is extracted from the projected distribution of the galaxies in a spectroscopic survey, is a useful step towards achieving this unified framework. We further claim that this approach leads to a better understanding of the evolution of structure in the Universe as it provides more information on the redshift evolution of galaxy bias.

\section*{Acknowledgements}
We would like to acknowledge the BOSS Collaboration for their great work and for making their data public. AL and BM thank the support of the Brazilian people through the Science without Borders CNPq Brazil fellowships. BM and FBA acknowledge support from the European Community through the DEDALE grant (contract no. 665044) within the H2020 Framework Program of the European Commission. FBA acknowledges the Royal Society for support via a Royal Society URF. AC acknowledges the Royal Astronomical Society for support via a summer bursary. OL acknowledges support from an European Research Council Advanced Grant FP7/291329 and support from the UK Science and Technology Research Council (STFC) Grant No. ST/M001334/1. MM acknowledged the support from the European Union's Horizon 2020 research and innovation programme under Marie Sklodowska-Curie grant agreement No 6655919. We also thank Niall Jeffrey and Andreu Font-Ribera for their very helpful comments on the final draft of this work. All cosmological contour plots were generated using \texttt{ChainConsumer} \citepbox{ChainConsumer}.




\bibliographystyle{mnras}
\interlinepenalty=10000
\bibliography{bibliog} 

\begin{thebibliography}{}
\makeatletter
\relax
\def\mn@urlcharsother{\let\do\@makeother \do\$\do\&\do\#\do\^\do\_\do\%\do\~}
\def\mn@doi{\begingroup\mn@urlcharsother \@ifnextchar [ {\mn@doi@}
  {\mn@doi@[]}}
\def\mn@doi@[#1]#2{\def\@tempa{#1}\ifx\@tempa\@empty \href
  {http://dx.doi.org/#2} {doi:#2}\else \href {http://dx.doi.org/#2} {#1}\fi
  \endgroup}
\def\mn@eprint#1#2{\mn@eprint@#1:#2::\@nil}
\def\mn@eprint@arXiv#1{\href {http://arxiv.org/abs/#1} {{\tt arXiv:#1}}}
\def\mn@eprint@dblp#1{\href {http://dblp.uni-trier.de/rec/bibtex/#1.xml}
  {dblp:#1}}
\def\mn@eprint@#1:#2:#3:#4\@nil{\def\@tempa {#1}\def\@tempb {#2}\def\@tempc
  {#3}\ifx \@tempc \@empty \let \@tempc \@tempb \let \@tempb \@tempa \fi \ifx
  \@tempb \@empty \def\@tempb {arXiv}\fi \@ifundefined
  {mn@eprint@\@tempb}{\@tempb:\@tempc}{\expandafter \expandafter \csname
  mn@eprint@\@tempb\endcsname \expandafter{\@tempc}}}

\bibitem[\protect\citeauthoryear{{Alam} et~al.}{{Alam} et~al.}{2015}]{BOSS2015}
{Alam} S.,  et~al., 2015, \mn@doi [\apjs] {10.1088/0067-0049/219/1/12}, \href
  {http://adsabs.harvard.edu/abs/2015ApJS..219...12A} {219, 12}

\bibitem[\protect\citeauthoryear{{Alam} et~al.,}{{Alam}
  et~al.}{2016}]{2016BOSSCosmology}
{Alam} S.,  et~al., 2016, preprint, \href
  {http://adsabs.harvard.edu/abs/2016arXiv160703155A} {} (\mn@eprint {arXiv}
  {1607.03155})

\bibitem[\protect\citeauthoryear{Anderson}{Anderson}{2003}]{AndersonBook}
Anderson T.~W.,  2003, An Introduction to Multivariate Statistical Analysis,
  3rd edn.
Wiley

\bibitem[\protect\citeauthoryear{{Anderson} et~al.,}{{Anderson}
  et~al.}{2014}]{2014Anderson}
{Anderson} L.,  et~al., 2014, \mn@doi [\mnras] {10.1093/mnras/stu523}, \href
  {http://adsabs.harvard.edu/abs/2014MNRAS.441...24A} {441, 24}

\bibitem[\protect\citeauthoryear{{Andrae}, {Schulze-Hartung}  \&
  {Melchior}}{{Andrae} et~al.}{2010}]{chisq2010}
{Andrae} R.,  {Schulze-Hartung} T.,   {Melchior} P.,  2010, preprint, \href
  {http://adsabs.harvard.edu/abs/2010arXiv1012.3754A} {} (\mn@eprint {arXiv}
  {1012.3754})

\bibitem[\protect\citeauthoryear{{Asorey}, {Crocce}, {Gazta{\~n}aga}  \&
  {Lewis}}{{Asorey} et~al.}{2012}]{Asorey2012}
{Asorey} J.,  {Crocce} M.,  {Gazta{\~n}aga} E.,   {Lewis} A.,  2012, \mn@doi
  [\mnras] {10.1111/j.1365-2966.2012.21972.x}, \href
  {http://adsabs.harvard.edu/abs/2012MNRAS.427.1891A} {427, 1891}

\bibitem[\protect\citeauthoryear{{Balaguera-Antol{\'{\i}}nez}, {Bilicki},
  {Branchini}  \& {Postiglione}}{{Balaguera-Antol{\'{\i}}nez}
  et~al.}{2018}]{2018-FreeCitation}
{Balaguera-Antol{\'{\i}}nez} A.,  {Bilicki} M.,  {Branchini} E.,
  {Postiglione} A.,  2018, \mn@doi [\mnras] {10.1093/mnras/sty262}, \href
  {http://adsabs.harvard.edu/abs/2018MNRAS.476.1050B} {476, 1050}

\bibitem[\protect\citeauthoryear{{Benitez} et~al.}{{Benitez}
  et~al.}{2014}]{JPAS}
{Benitez} N.,  et~al., 2014, preprint, \href
  {http://adsabs.harvard.edu/abs/2014arXiv1403.5237B} {} (\mn@eprint {arXiv}
  {1403.5237})

\bibitem[\protect\citeauthoryear{{Betoule} et~al.,}{{Betoule}
  et~al.}{2014}]{JLAdata}
{Betoule} M.,  et~al., 2014, \mn@doi [\aap] {10.1051/0004-6361/201423413},
  \href {http://adsabs.harvard.edu/abs/2014A%26A...568A..22B} {568, A22}

\bibitem[\protect\citeauthoryear{{Beutler} et~al.,}{{Beutler}
  et~al.}{2017a}]{2017BeutlerBOSS}
{Beutler} F.,  et~al., 2017a, \mn@doi [\mnras] {10.1093/mnras/stw2373}, \href
  {http://adsabs.harvard.edu/abs/2017MNRAS.464.3409B} {464, 3409}

\bibitem[\protect\citeauthoryear{{Beutler} et~al.,}{{Beutler}
  et~al.}{2017b}]{2017Beutler2BOSS}
{Beutler} F.,  et~al., 2017b, \mn@doi [\mnras] {10.1093/mnras/stw3298}, \href
  {http://adsabs.harvard.edu/abs/2017MNRAS.466.2242B} {466, 2242}

\bibitem[\protect\citeauthoryear{{Bird}, {Viel}  \& {Haehnelt}}{{Bird}
  et~al.}{2012}]{Bird2012}
{Bird} S.,  {Viel} M.,   {Haehnelt} M.~G.,  2012, \mn@doi [\mnras]
  {10.1111/j.1365-2966.2011.20222.x}, \href
  {http://adsabs.harvard.edu/abs/2012MNRAS.420.2551B} {420, 2551}

\bibitem[\protect\citeauthoryear{{Blake}, {Ferreira}  \& {Borrill}}{{Blake}
  et~al.}{2004}]{BlakeFerreira2004}
{Blake} C.,  {Ferreira} P.~G.,   {Borrill} J.,  2004, \mn@doi [\mnras]
  {10.1111/j.1365-2966.2004.07831.x}, \href
  {http://adsabs.harvard.edu/abs/2004MNRAS.351..923B} {351, 923}

\bibitem[\protect\citeauthoryear{{Blake}, {Collister}, {Bridle}  \&
  {Lahav}}{{Blake} et~al.}{2007}]{Blake2007}
{Blake} C.,  {Collister} A.,  {Bridle} S.,   {Lahav} O.,  2007, \mn@doi
  [\mnras] {10.1111/j.1365-2966.2006.11263.x}, \href
  {http://adsabs.harvard.edu/abs/2007MNRAS.374.1527B} {374, 1527}

\bibitem[\protect\citeauthoryear{{Blas}, {Lesgourgues}  \& {Tram}}{{Blas}
  et~al.}{2011}]{Class}
{Blas} D.,  {Lesgourgues} J.,   {Tram} T.,  2011, \mn@doi [\jcap]
  {10.1088/1475-7516/2011/07/034}, \href
  {http://adsabs.harvard.edu/abs/2011JCAP...07..034B} {7, 034}

\bibitem[\protect\citeauthoryear{{Brown}, {Castro}  \& {Taylor}}{{Brown}
  et~al.}{2005}]{PolSpice2005}
{Brown} M.~L.,  {Castro} P.~G.,   {Taylor} A.~N.,  2005, \mn@doi [\mnras]
  {10.1111/j.1365-2966.2005.09111.x}, \href
  {http://adsabs.harvard.edu/abs/2005MNRAS.360.1262B} {360, 1262}

\bibitem[\protect\citeauthoryear{{Cannon} et~al.,}{{Cannon}
  et~al.}{2006}]{2006Cannon2Slaq}
{Cannon} R.,  et~al., 2006, \mn@doi [\mnras]
  {10.1111/j.1365-2966.2006.10875.x}, \href
  {http://adsabs.harvard.edu/abs/2006MNRAS.372..425C} {372, 425}

\bibitem[\protect\citeauthoryear{Challinor \& Lewis}{Challinor \&
  Lewis}{2011}]{CambSources}
Challinor A.,  Lewis A.,  2011, \mn@doi [Phys.Rev.]
  {10.1103/PhysRevD.84.043516}, D84, 043516

\bibitem[\protect\citeauthoryear{{Charnock}, {Battye}  \& {Moss}}{{Charnock}
  et~al.}{2017a}]{2017CharnockTension}
{Charnock} T.,  {Battye} R.~A.,   {Moss} A.,  2017a, \mn@doi [\prd]
  {10.1103/PhysRevD.95.123535}, \href
  {http://adsabs.harvard.edu/abs/2017PhRvD..95l3535C} {95, 123535}

\bibitem[\protect\citeauthoryear{{Charnock}, {Battye}  \& {Moss}}{{Charnock}
  et~al.}{2017b}]{2017Charnock}
{Charnock} T.,  {Battye} R.~A.,   {Moss} A.,  2017b, \mn@doi [\prd]
  {10.1103/PhysRevD.95.123535}, \href
  {http://adsabs.harvard.edu/abs/2017PhRvD..95l3535C} {95, 123535}

\bibitem[\protect\citeauthoryear{{Christodoulou} et~al.,}{{Christodoulou}
  et~al.}{2012}]{Christodoulou2012}
{Christodoulou} L.,  et~al., 2012, \mn@doi [\mnras]
  {10.1111/j.1365-2966.2012.21434.x}, \href
  {http://adsabs.harvard.edu/abs/2012MNRAS.425.1527C} {425, 1527}

\bibitem[\protect\citeauthoryear{{DES Collaboration} et~al.,}{{DES
  Collaboration} et~al.}{2017}]{2017arXiv170801530D}
{DES Collaboration} et~al., 2017, preprint, \href
  {http://adsabs.harvard.edu/abs/2017arXiv170801530D} {} (\mn@eprint {arXiv}
  {1708.01530})

\bibitem[\protect\citeauthoryear{{Dahlen} \& {Simons}}{{Dahlen} \&
  {Simons}}{2008}]{2008DahlenSimons}
{Dahlen} F.~A.,  {Simons} F.~J.,  2008, \mn@doi [Geophysical Journal
  International] {10.1111/j.1365-246X.2008.03854.x}, \href
  {http://adsabs.harvard.edu/abs/2008GeoJI.174..774D} {174, 774}

\bibitem[\protect\citeauthoryear{{Dawson} et~al.}{{Dawson} et~al.}{2013}]{BOSS}
{Dawson} K.~S.,  et~al., 2013, \mn@doi [\aj] {10.1088/0004-6256/145/1/10},
  \href {http://adsabs.harvard.edu/abs/2013AJ....145...10D} {145, 10}

\bibitem[\protect\citeauthoryear{{Di Dio}, {Montanari}, {Lesgourgues}  \&
  {Durrer}}{{Di Dio} et~al.}{2013}]{CLASSgal}
{Di Dio} E.,  {Montanari} F.,  {Lesgourgues} J.,   {Durrer} R.,  2013, \mn@doi
  [\jcap] {10.1088/1475-7516/2013/11/044}, \href
  {http://adsabs.harvard.edu/abs/2013JCAP...11..044D} {11, 044}

\bibitem[\protect\citeauthoryear{{Doux}, {Penna-Lima}, {Vitenti},
  {Tr{\'e}guer}, {Aubourg}  \& {Ganga}}{{Doux} et~al.}{2017}]{Doux2017}
{Doux} C.,  {Penna-Lima} M.,  {Vitenti} S.~D.~P.,  {Tr{\'e}guer} J.,  {Aubourg}
  E.,   {Ganga} K.,  2017, preprint, \href
  {http://adsabs.harvard.edu/abs/2017arXiv170604583D} {} (\mn@eprint {arXiv}
  {1706.04583})

\bibitem[\protect\citeauthoryear{{Efstathiou}}{{Efstathiou}}{2004}]{Efstat2004}
{Efstathiou} G.,  2004, \mn@doi [\mnras] {10.1111/j.1365-2966.2004.07530.x},
  \href {http://adsabs.harvard.edu/abs/2004MNRAS.349..603E} {349, 603}

\bibitem[\protect\citeauthoryear{{Efstathiou} \& {Lemos}}{{Efstathiou} \&
  {Lemos}}{2018}]{Efstathiou2018}
{Efstathiou} G.,  {Lemos} P.,  2018, \mn@doi [\mnras] {10.1093/mnras/sty099},
  \href {http://adsabs.harvard.edu/abs/2018MNRAS.476..151E} {476, 151}

\bibitem[\protect\citeauthoryear{{Eisenstein} et~al.,}{{Eisenstein}
  et~al.}{2001}]{2001Eisenstein}
{Eisenstein} D.~J.,  et~al., 2001, \mn@doi [\aj] {10.1086/323717}, \href
  {http://adsabs.harvard.edu/abs/2001AJ....122.2267E} {122, 2267}

\bibitem[\protect\citeauthoryear{{Elsner}, {Leistedt}  \& {Peiris}}{{Elsner}
  et~al.}{2016}]{2016ElsnerSyst1}
{Elsner} F.,  {Leistedt} B.,   {Peiris} H.~V.,  2016, \mn@doi [\mnras]
  {10.1093/mnras/stv2777}, \href
  {http://adsabs.harvard.edu/abs/2016MNRAS.456.2095E} {456, 2095}

\bibitem[\protect\citeauthoryear{{Elsner}, {Leistedt}  \& {Peiris}}{{Elsner}
  et~al.}{2017}]{2017ElsnerPCLModeProj}
{Elsner} F.,  {Leistedt} B.,   {Peiris} H.~V.,  2017, \mn@doi [\mnras]
  {10.1093/mnras/stw2752}, \href
  {http://adsabs.harvard.edu/abs/2017MNRAS.465.1847E} {465, 1847}

\bibitem[\protect\citeauthoryear{{Elvin-Poole} et~al.,}{{Elvin-Poole}
  et~al.}{2017}]{ElvinPoole2017}
{Elvin-Poole} J.,  et~al., 2017, preprint, \href
  {http://adsabs.harvard.edu/abs/2017arXiv170801536E} {} (\mn@eprint {arXiv}
  {1708.01536})

\bibitem[\protect\citeauthoryear{{Eriksen} \& {Gazta{\~n}aga}}{{Eriksen} \&
  {Gazta{\~n}aga}}{2015}]{Eriksen2015}
{Eriksen} M.,  {Gazta{\~n}aga} E.,  2015, \mn@doi [\mnras]
  {10.1093/mnras/stv1075}, \href
  {http://adsabs.harvard.edu/abs/2015MNRAS.452.2168E} {452, 2168}

\bibitem[\protect\citeauthoryear{{Esteban}, {Gonzalez-Garcia}, {Maltoni},
  {Martinez-Soler}  \& {Schwetz}}{{Esteban} et~al.}{2017}]{Esteban2017}
{Esteban} I.,  {Gonzalez-Garcia} M.~C.,  {Maltoni} M.,  {Martinez-Soler} I.,
  {Schwetz} T.,  2017, \mn@doi [Journal of High Energy Physics]
  {10.1007/JHEP01(2017)087}, \href
  {http://adsabs.harvard.edu/abs/2017JHEP...01..087E} {1, 87}

\bibitem[\protect\citeauthoryear{{Feeney}, {Peiris}, {Williamson}, {Nissanke},
  {Mortlock}, {Alsing}  \& {Scolnic}}{{Feeney}
  et~al.}{2018}]{2018FeeneyTension}
{Feeney} S.~M.,  {Peiris} H.~V.,  {Williamson} A.~R.,  {Nissanke} S.~M.,
  {Mortlock} D.~J.,  {Alsing} J.,   {Scolnic} D.,  2018, preprint, \href
  {http://adsabs.harvard.edu/abs/2018arXiv180203404F} {} (\mn@eprint {arXiv}
  {1802.03404})

\bibitem[\protect\citeauthoryear{{Feroz} \& {Hobson}}{{Feroz} \&
  {Hobson}}{2008}]{2008FerozHobson}
{Feroz} F.,  {Hobson} M.~P.,  2008, \mn@doi [\mnras]
  {10.1111/j.1365-2966.2007.12353.x}, \href
  {http://adsabs.harvard.edu/abs/2008MNRAS.384..449F} {384, 449}

\bibitem[\protect\citeauthoryear{{Fisher}, {Scharf}  \& {Lahav}}{{Fisher}
  et~al.}{1994}]{FisherLahav1994}
{Fisher} K.~B.,  {Scharf} C.~A.,   {Lahav} O.,  1994, \mn@doi [\mnras]
  {10.1093/mnras/266.1.219}, \href
  {http://adsabs.harvard.edu/abs/1994MNRAS.266..219F} {266, 219}

\bibitem[\protect\citeauthoryear{{Fixsen}, {Cheng}, {Gales}, {Mather}, {Shafer}
   \& {Wright}}{{Fixsen} et~al.}{1996}]{COBE}
{Fixsen} D.~J.,  {Cheng} E.~S.,  {Gales} J.~M.,  {Mather} J.~C.,  {Shafer}
  R.~A.,   {Wright} E.~L.,  1996, \mn@doi [\apj] {10.1086/178173}, \href
  {http://adsabs.harvard.edu/abs/1996ApJ...473..576F} {473, 576}

\bibitem[\protect\citeauthoryear{{Gazta{\~n}aga}, {Eriksen}, {Crocce},
  {Castander}, {Fosalba}, {Marti}, {Miquel}  \& {Cabr{\'e}}}{{Gazta{\~n}aga}
  et~al.}{2012}]{Gaztanaga2012}
{Gazta{\~n}aga} E.,  {Eriksen} M.,  {Crocce} M.,  {Castander} F.~J.,  {Fosalba}
  P.,  {Marti} P.,  {Miquel} R.,   {Cabr{\'e}} A.,  2012, \mn@doi [\mnras]
  {10.1111/j.1365-2966.2012.20613.x}, \href
  {http://adsabs.harvard.edu/abs/2012MNRAS.422.2904G} {422, 2904}

\bibitem[\protect\citeauthoryear{{Giannantonio} et~al.,}{{Giannantonio}
  et~al.}{2016}]{Giannantonio2016}
{Giannantonio} T.,  et~al., 2016, \mn@doi [\mnras] {10.1093/mnras/stv2678},
  \href {http://adsabs.harvard.edu/abs/2016MNRAS.456.3213G} {456, 3213}

\bibitem[\protect\citeauthoryear{{Gonzalez-Garcia}, {Maltoni}  \&
  {Schwetz}}{{Gonzalez-Garcia} et~al.}{2014}]{2014Gonzalez-GarciaNeutrino}
{Gonzalez-Garcia} M.~C.,  {Maltoni} M.,   {Schwetz} T.,  2014, \mn@doi [Journal
  of High Energy Physics] {10.1007/JHEP11(2014)052}, \href
  {http://adsabs.harvard.edu/abs/2014JHEP...11..052G} {11, 52}

\bibitem[\protect\citeauthoryear{{G{\'o}rski}, {Hivon}, {Banday}, {Wandelt},
  {Hansen}, {Reinecke}  \& {Bartelmann}}{{G{\'o}rski} et~al.}{2005}]{Healpix}
{G{\'o}rski} K.~M.,  {Hivon} E.,  {Banday} A.~J.,  {Wandelt} B.~D.,  {Hansen}
  F.~K.,  {Reinecke} M.,   {Bartelmann} M.,  2005, \mn@doi [\apj]
  {10.1086/427976}, \href {http://adsabs.harvard.edu/abs/2005ApJ...622..759G}
  {622, 759}

\bibitem[\protect\citeauthoryear{{Grieb} et~al.,}{{Grieb}
  et~al.}{2017}]{2017GriebBOSS}
{Grieb} J.~N.,  et~al., 2017, \mn@doi [\mnras] {10.1093/mnras/stw3384}, \href
  {http://adsabs.harvard.edu/abs/2017MNRAS.467.2085G} {467, 2085}

\bibitem[\protect\citeauthoryear{{Hand} et~al.,}{{Hand}
  et~al.}{2015}]{Hand2015}
{Hand} N.,  et~al., 2015, \mn@doi [\prd] {10.1103/PhysRevD.91.062001}, \href
  {http://adsabs.harvard.edu/abs/2015PhRvD..91f2001H} {91, 062001}

\bibitem[\protect\citeauthoryear{{Hannestad}}{{Hannestad}}{2003}]{2003HannestadNeutrino}
{Hannestad} S.,  2003, \mn@doi [\prd] {10.1103/PhysRevD.67.085017}, \href
  {http://adsabs.harvard.edu/abs/2003PhRvD..67h5017H} {67, 085017}

\bibitem[\protect\citeauthoryear{{Hannestad} \& {Schwetz}}{{Hannestad} \&
  {Schwetz}}{2016}]{2016HannestadNeutrino}
{Hannestad} S.,  {Schwetz} T.,  2016, \mn@doi [\jcap]
  {10.1088/1475-7516/2016/11/035}, \href
  {http://adsabs.harvard.edu/abs/2016JCAP...11..035H} {11, 035}

\bibitem[\protect\citeauthoryear{{Hartlap}, {Simon}  \& {Schneider}}{{Hartlap}
  et~al.}{2007}]{Hartlap2007}
{Hartlap} J.,  {Simon} P.,   {Schneider} P.,  2007, \mn@doi [\aap]
  {10.1051/0004-6361:20066170}, \href
  {http://adsabs.harvard.edu/abs/2007A%26A...464..399H} {464, 399}

\bibitem[\protect\citeauthoryear{{Hauser} \& {Peebles}}{{Hauser} \&
  {Peebles}}{1973}]{Peebles1973_2}
{Hauser} M.~G.,  {Peebles} P.~J.~E.,  1973, \mn@doi [\apj] {10.1086/152453},
  \href {http://adsabs.harvard.edu/abs/1973ApJ...185..757H} {185, 757}

\bibitem[\protect\citeauthoryear{{Hildebrandt} et~al.,}{{Hildebrandt}
  et~al.}{2017}]{2017MNRAS.465.1454H}
{Hildebrandt} H.,  et~al., 2017, \mn@doi [MNRAS] {10.1093/mnras/stw2805}, \href
  {http://adsabs.harvard.edu/abs/2017MNRAS.465.1454H} {465, 1454}

\bibitem[\protect\citeauthoryear{{Hinton}}{{Hinton}}{2016}]{ChainConsumer}
{Hinton} S.~R.,  2016, \mn@doi [The Journal of Open Source Software]
  {10.21105/joss.00045}, \href
  {http://adsabs.harvard.edu/abs/2016JOSS....1...45H} {1, 00045}

\bibitem[\protect\citeauthoryear{{Hivon}, {G{\'o}rski}, {Netterfield}, {Crill},
  {Prunet}  \& {Hansen}}{{Hivon} et~al.}{2002}]{PolSpice2001}
{Hivon} E.,  {G{\'o}rski} K.~M.,  {Netterfield} C.~B.,  {Crill} B.~P.,
  {Prunet} S.,   {Hansen} F.,  2002, \mn@doi [\apj] {10.1086/338126}, \href
  {http://adsabs.harvard.edu/abs/2002ApJ...567....2H} {567, 2}

\bibitem[\protect\citeauthoryear{{Ho} et~al.,}{{Ho} et~al.}{2012}]{Ho2012}
{Ho} S.,  et~al., 2012, \mn@doi [\apj] {10.1088/0004-637X/761/1/14}, \href
  {http://adsabs.harvard.edu/abs/2012ApJ...761...14H} {761, 14}

\bibitem[\protect\citeauthoryear{{Huterer}, {Knox}  \& {Nichol}}{{Huterer}
  et~al.}{2001}]{2001Huterer}
{Huterer} D.,  {Knox} L.,   {Nichol} R.~C.,  2001, \mn@doi [\apj]
  {10.1086/323328}, \href {http://adsabs.harvard.edu/abs/2001ApJ...555..547H}
  {555, 547}

\bibitem[\protect\citeauthoryear{{Joachimi} \& {Bridle}}{{Joachimi} \&
  {Bridle}}{2010}]{JoachimiBridle2010}
{Joachimi} B.,  {Bridle} S.~L.,  2010, \mn@doi [\aap]
  {10.1051/0004-6361/200913657}, \href
  {http://adsabs.harvard.edu/abs/2010A%26A...523A...1J} {523, A1}

\bibitem[\protect\citeauthoryear{{Johansson} \& {Forss{\'e}n}}{{Johansson} \&
  {Forss{\'e}n}}{2015}]{Wig3j}
{Johansson} H.~T.,  {Forss{\'e}n} C.,  2015, preprint, \href
  {http://adsabs.harvard.edu/abs/2015arXiv150408329J} {} (\mn@eprint {arXiv}
  {1504.08329})

\bibitem[\protect\citeauthoryear{{Kaiser}}{{Kaiser}}{1987}]{1987Kaiser}
{Kaiser} N.,  1987, \mn@doi [\mnras] {10.1093/mnras/227.1.1}, \href
  {http://adsabs.harvard.edu/abs/1987MNRAS.227....1K} {227, 1}

\bibitem[\protect\citeauthoryear{Kajita}{Kajita}{2016}]{Kajita2016}
Kajita T.,  2016, \mn@doi [Rev. Mod. Phys.] {10.1103/RevModPhys.88.030501}, 88,
  030501

\bibitem[\protect\citeauthoryear{{Kang}, {Jing}, {Mo}  \& {B{\"o}rner}}{{Kang}
  et~al.}{2002}]{Kang2002}
{Kang} X.,  {Jing} Y.~P.,  {Mo} H.~J.,   {B{\"o}rner} G.,  2002, \mn@doi
  [\mnras] {10.1046/j.1365-8711.2002.05828.x}, \href
  {http://adsabs.harvard.edu/abs/2002MNRAS.336..892K} {336, 892}

\bibitem[\protect\citeauthoryear{{Kirk}, {Lahav}, {Bridle}, {Jouvel}, {Abdalla}
   \& {Frieman}}{{Kirk} et~al.}{2015}]{Kirk2015}
{Kirk} D.,  {Lahav} O.,  {Bridle} S.,  {Jouvel} S.,  {Abdalla} F.~B.,
  {Frieman} J.~A.,  2015, \mn@doi [\mnras] {10.1093/mnras/stv1268}, \href
  {http://adsabs.harvard.edu/abs/2015MNRAS.451.4424K} {451, 4424}

\bibitem[\protect\citeauthoryear{{Kitaura} et~al.,}{{Kitaura}
  et~al.}{2016}]{2016BOSSMocks}
{Kitaura} F.-S.,  et~al., 2016, \mn@doi [\mnras] {10.1093/mnras/stv2826}, \href
  {http://adsabs.harvard.edu/abs/2016MNRAS.456.4156K} {456, 4156}

\bibitem[\protect\citeauthoryear{{LSST Dark Energy Science
  Collaboration}}{{LSST Dark Energy Science
  Collaboration}}{2012}]{2012arXiv1211.0310L}
{LSST Dark Energy Science Collaboration} 2012, preprint, \href
  {http://adsabs.harvard.edu/abs/2012arXiv1211.0310L} {} (\mn@eprint {arXiv}
  {1211.0310})

\bibitem[\protect\citeauthoryear{{Laureijs} et~al.,}{{Laureijs}
  et~al.}{2011}]{2011arXiv1110.3193L}
{Laureijs} R.,  et~al., 2011, preprint, \href
  {http://adsabs.harvard.edu/abs/2011arXiv1110.3193L} {} (\mn@eprint {arXiv}
  {1110.3193})

\bibitem[\protect\citeauthoryear{{Leistedt} \& {Peiris}}{{Leistedt} \&
  {Peiris}}{2014}]{2014BorisSyst2}
{Leistedt} B.,  {Peiris} H.~V.,  2014, \mn@doi [\mnras]
  {10.1093/mnras/stu1439}, \href
  {http://adsabs.harvard.edu/abs/2014MNRAS.444....2L} {444, 2}

\bibitem[\protect\citeauthoryear{{Leistedt}, {Peiris}, {Mortlock},
  {Benoit-L{\'e}vy}  \& {Pontzen}}{{Leistedt} et~al.}{2013}]{Boris2013}
{Leistedt} B.,  {Peiris} H.~V.,  {Mortlock} D.~J.,  {Benoit-L{\'e}vy} A.,
  {Pontzen} A.,  2013, \mn@doi [\mnras] {10.1093/mnras/stt1359}, \href
  {http://adsabs.harvard.edu/abs/2013MNRAS.435.1857L} {435, 1857}

\bibitem[\protect\citeauthoryear{{Lesgourgues} \& {Pastor}}{{Lesgourgues} \&
  {Pastor}}{2006}]{2006NeutrinoReview}
{Lesgourgues} J.,  {Pastor} S.,  2006, \mn@doi [\physrep]
  {10.1016/j.physrep.2006.04.001}, \href
  {http://adsabs.harvard.edu/abs/2006PhR...429..307L} {429, 307}

\bibitem[\protect\citeauthoryear{{Lesgourgues} \& {Pastor}}{{Lesgourgues} \&
  {Pastor}}{2014}]{2014NeutrinoCosmoPlanck}
{Lesgourgues} J.,  {Pastor} S.,  2014, \mn@doi [New Journal of Physics]
  {10.1088/1367-2630/16/6/065002}, \href
  {http://adsabs.harvard.edu/abs/2014NJPh...16f5002L} {16, 065002}

\bibitem[\protect\citeauthoryear{Lesgourgues, Mangano, Miele  \&
  Pastor}{Lesgourgues et~al.}{2013}]{Lesgourgues2013}
Lesgourgues J.,  Mangano G.,  Miele G.,   Pastor S.,  2013, Neutrino Cosmology.
Cambridge University Press, \mn@doi{10.1017/CBO9781139012874}

\bibitem[\protect\citeauthoryear{{Levi} et~al.,}{{Levi}
  et~al.}{2013}]{2013arXiv1308.0847L}
{Levi} M.,  et~al., 2013, preprint, \href
  {http://adsabs.harvard.edu/abs/2013arXiv1308.0847L} {} (\mn@eprint {arXiv}
  {1308.0847})

\bibitem[\protect\citeauthoryear{Lewis, Challinor  \& Lasenby}{Lewis
  et~al.}{2000}]{CAMB}
Lewis A.,  Challinor A.,   Lasenby A.,  2000, \mn@doi [Astrophys. J.]
  {10.1086/309179}, 538, 473

\bibitem[\protect\citeauthoryear{{Loureiro} et~al.,}{{Loureiro}
  et~al.}{2018}]{2018LoureiroNeutrino}
{Loureiro} A.,  et~al., 2018, preprint, \href
  {http://adsabs.harvard.edu/abs/2018arXiv181102578L} {} (\mn@eprint {arXiv}
  {1811.02578})

\bibitem[\protect\citeauthoryear{{MacCrann}, {Zuntz}, {Bridle}, {Jain}  \&
  {Becker}}{{MacCrann} et~al.}{2015}]{2015MacCrann}
{MacCrann} N.,  {Zuntz} J.,  {Bridle} S.,  {Jain} B.,   {Becker} M.~R.,  2015,
  \mn@doi [\mnras] {10.1093/mnras/stv1154}, \href
  {http://adsabs.harvard.edu/abs/2015MNRAS.451.2877M} {451, 2877}

\bibitem[\protect\citeauthoryear{{Manera} et~al.}{{Manera}
  et~al.}{2013}]{Manera2013}
{Manera} M.,  et~al., 2013, \mn@doi [\mnras] {10.1093/mnras/sts084}, \href
  {http://adsabs.harvard.edu/abs/2013MNRAS.428.1036M} {428, 1036}

\bibitem[\protect\citeauthoryear{McDonald}{McDonald}{2016}]{McDonald2016}
McDonald A.~B.,  2016, \mn@doi [Rev. Mod. Phys.]
  {10.1103/RevModPhys.88.030502}, 88, 030502

\bibitem[\protect\citeauthoryear{{McLeod}, {Balan}  \& {Abdalla}}{{McLeod}
  et~al.}{2017}]{2016McLeod}
{McLeod} M.,  {Balan} S.~T.,   {Abdalla} F.~B.,  2017, \mn@doi [\mnras]
  {10.1093/mnras/stw2989}, \href
  {http://adsabs.harvard.edu/abs/2017MNRAS.466.3558M} {466, 3558}

\bibitem[\protect\citeauthoryear{{McQuinn} \& {White}}{{McQuinn} \&
  {White}}{2013}]{McQuinnWhite2013}
{McQuinn} M.,  {White} M.,  2013, \mn@doi [\mnras] {10.1093/mnras/stt914},
  \href {http://adsabs.harvard.edu/abs/2013MNRAS.433.2857M} {433, 2857}

\bibitem[\protect\citeauthoryear{{Nicola}, {Refregier}  \& {Amara}}{{Nicola}
  et~al.}{2016a}]{2016NicolaA}
{Nicola} A.,  {Refregier} A.,   {Amara} A.,  2016a, \mn@doi [\prd]
  {10.1103/PhysRevD.94.083517}, \href
  {http://adsabs.harvard.edu/abs/2016PhRvD..94h3517N} {94, 083517}

\bibitem[\protect\citeauthoryear{{Nicola}, {Refregier}  \& {Amara}}{{Nicola}
  et~al.}{2016b}]{2016Nicola}
{Nicola} A.,  {Refregier} A.,   {Amara} A.,  2016b, \mn@doi [\prd]
  {10.1103/PhysRevD.94.083517}, \href
  {http://adsabs.harvard.edu/abs/2016PhRvD..94h3517N} {94, 083517}

\bibitem[\protect\citeauthoryear{{Nicola}, {Refregier}  \& {Amara}}{{Nicola}
  et~al.}{2017}]{2017Nicola}
{Nicola} A.,  {Refregier} A.,   {Amara} A.,  2017, \mn@doi [\prd]
  {10.1103/PhysRevD.95.083523}, \href
  {http://adsabs.harvard.edu/abs/2017PhRvD..95h3523N} {95, 083523}

\bibitem[\protect\citeauthoryear{{Padmanabhan} et~al.}{{Padmanabhan}
  et~al.}{2007}]{Padm2007}
{Padmanabhan} N.,  et~al., 2007, \mn@doi [\mnras]
  {10.1111/j.1365-2966.2007.11593.x}, \href
  {http://adsabs.harvard.edu/abs/2007MNRAS.378..852P} {378, 852}

\bibitem[\protect\citeauthoryear{{Peebles}}{{Peebles}}{1973}]{Peebles1973}
{Peebles} P.~J.~E.,  1973, \mn@doi [\apj] {10.1086/152431}, \href
  {http://adsabs.harvard.edu/abs/1973ApJ...185..413P} {185, 413}

\bibitem[\protect\citeauthoryear{{Percival} et~al.,}{{Percival}
  et~al.}{2001}]{2001Percival}
{Percival} W.~J.,  et~al., 2001, \mn@doi [\mnras]
  {10.1046/j.1365-8711.2001.04827.x}, \href
  {http://adsabs.harvard.edu/abs/2001MNRAS.327.1297P} {327, 1297}

\bibitem[\protect\citeauthoryear{{Percival}, {Samushia}, {Ross}, {Shapiro}  \&
  {Raccanelli}}{{Percival} et~al.}{2011}]{Percival-FoG2011}
{Percival} W.~J.,  {Samushia} L.,  {Ross} A.~J.,  {Shapiro} C.,   {Raccanelli}
  A.,  2011, \mn@doi [Philosophical Transactions of the Royal Society of London
  Series A] {10.1098/rsta.2011.0370}, \href
  {http://adsabs.harvard.edu/abs/2011RSPTA.369.5058P} {369, 5058}

\bibitem[\protect\citeauthoryear{{Perlmutter} et~al.}{{Perlmutter}
  et~al.}{1999}]{1999Perlmutter}
{Perlmutter} S.,  et~al., 1999, Astrophysical Journal, 517, 565

\bibitem[\protect\citeauthoryear{{Planck Collaboration} et~al.,}{{Planck
  Collaboration} et~al.}{2016a}]{PlanckResults2015}
{Planck Collaboration} et~al., 2016a, \mn@doi [\aap]
  {10.1051/0004-6361/201527101}, \href
  {https://ui.adsabs.harvard.edu/#abs/2016A&A...594A...1P} {594, A1}

\bibitem[\protect\citeauthoryear{{Planck Collaboration} et~al.,}{{Planck
  Collaboration} et~al.}{2016b}]{PlanckLikelihood2015}
{Planck Collaboration} et~al., 2016b, \mn@doi [\aap]
  {10.1051/0004-6361/201526926}, \href
  {https://ui.adsabs.harvard.edu/#abs/2016A&A...594A..11P} {594, A11}

\bibitem[\protect\citeauthoryear{{Planck Collaboration} et~al.,}{{Planck
  Collaboration} et~al.}{2016c}]{PlanckCosmology2016}
{Planck Collaboration} et~al., 2016c, \mn@doi [\aap]
  {10.1051/0004-6361/201525830}, \href
  {http://adsabs.harvard.edu/abs/2016A%26A...594A..13P} {594, A13}

\bibitem[\protect\citeauthoryear{{Planck Collaboration} et~al.,}{{Planck
  Collaboration} et~al.}{2018}]{PlanckCosmology2018}
{Planck Collaboration} et~al., 2018, preprint, \href
  {http://adsabs.harvard.edu/abs/2018arXiv180706209P} {} (\mn@eprint {arXiv}
  {1807.06209})

\bibitem[\protect\citeauthoryear{{Rassat}, {Land}, {Lahav}  \&
  {Abdalla}}{{Rassat} et~al.}{2007}]{Rassat2007}
{Rassat} A.,  {Land} K.,  {Lahav} O.,   {Abdalla} F.~B.,  2007, \mn@doi
  [\mnras] {10.1111/j.1365-2966.2007.11538.x}, \href
  {http://adsabs.harvard.edu/abs/2007MNRAS.377.1085R} {377, 1085}

\bibitem[\protect\citeauthoryear{{Raveri} \& {Hu}}{{Raveri} \&
  {Hu}}{2018}]{2018HuTension}
{Raveri} M.,  {Hu} W.,  2018, preprint, \href
  {http://adsabs.harvard.edu/abs/2018arXiv180604649R} {} (\mn@eprint {arXiv}
  {1806.04649})

\bibitem[\protect\citeauthoryear{{Reid} et~al.,}{{Reid}
  et~al.}{2016}]{BOSSCatalogue2016}
{Reid} B.,  et~al., 2016, \mn@doi [\mnras] {10.1093/mnras/stv2382}, \href
  {http://adsabs.harvard.edu/abs/2016MNRAS.455.1553R} {455, 1553}

\bibitem[\protect\citeauthoryear{{Riess} et~al.}{{Riess}
  et~al.}{1998}]{1998Riess}
{Riess} A.~G.,  et~al., 1998, \mn@doi [The Astronomical Journal]
  {10.1086/300499}, \href {http://adsabs.harvard.edu/abs/1998AJ....116.1009R}
  {116, 1009}

\bibitem[\protect\citeauthoryear{{Riess} et~al.,}{{Riess}
  et~al.}{2016}]{Riess2016}
{Riess} A.~G.,  et~al., 2016, \mn@doi [\apj] {10.3847/0004-637X/826/1/56},
  \href {http://adsabs.harvard.edu/abs/2016ApJ...826...56R} {826, 56}

\bibitem[\protect\citeauthoryear{{Riess} et~al.,}{{Riess}
  et~al.}{2018}]{Riess2018}
{Riess} A.~G.,  et~al., 2018, \mn@doi [\apj] {10.3847/1538-4357/aac82e}, \href
  {http://adsabs.harvard.edu/abs/2018ApJ...861..126R} {861, 126}

\bibitem[\protect\citeauthoryear{Rollins}{Rollins}{2015}]{PlinyRichardThesis}
Rollins R.~P.,  2015, PhD thesis, University College London, London, UK

\bibitem[\protect\citeauthoryear{{Ross} et~al.,}{{Ross}
  et~al.}{2011}]{2011MNRAS.417.1350R}
{Ross} A.~J.,  et~al., 2011, \mn@doi [\mnras]
  {10.1111/j.1365-2966.2011.19351.x}, \href
  {http://adsabs.harvard.edu/abs/2011MNRAS.417.1350R} {417, 1350}

\bibitem[\protect\citeauthoryear{{Ross} et~al.,}{{Ross}
  et~al.}{2013}]{2013ROSS}
{Ross} A.~J.,  et~al., 2013, \mn@doi [\mnras] {10.1093/mnras/sts094}, \href
  {http://adsabs.harvard.edu/abs/2013MNRAS.428.1116R} {428, 1116}

\bibitem[\protect\citeauthoryear{{Ross} et~al.,}{{Ross}
  et~al.}{2017}]{2017RossBOSS}
{Ross} A.~J.,  et~al., 2017, \mn@doi [\mnras] {10.1093/mnras/stw2372}, \href
  {http://adsabs.harvard.edu/abs/2017MNRAS.464.1168R} {464, 1168}

\bibitem[\protect\citeauthoryear{{Salazar-Albornoz} et~al.,}{{Salazar-Albornoz}
  et~al.}{2017}]{2017SalazarBOSSwTheta}
{Salazar-Albornoz} S.,  et~al., 2017, \mn@doi [\mnras] {10.1093/mnras/stx633},
  \href {http://adsabs.harvard.edu/abs/2017MNRAS.468.2938S} {468, 2938}

\bibitem[\protect\citeauthoryear{{S{\'a}nchez} et~al.,}{{S{\'a}nchez}
  et~al.}{2017}]{2017SanchezBOSS}
{S{\'a}nchez} A.~G.,  et~al., 2017, \mn@doi [\mnras] {10.1093/mnras/stw2443},
  \href {http://adsabs.harvard.edu/abs/2017MNRAS.464.1640S} {464, 1640}

\bibitem[\protect\citeauthoryear{{Satpathy} et~al.,}{{Satpathy}
  et~al.}{2017}]{2017SatpathyBOSS}
{Satpathy} S.,  et~al., 2017, \mn@doi [\mnras] {10.1093/mnras/stx883}, \href
  {http://adsabs.harvard.edu/abs/2017MNRAS.469.1369S} {469, 1369}

\bibitem[\protect\citeauthoryear{{Scharf}, {Hoffman}, {Lahav}  \&
  {Lynden-Bell}}{{Scharf} et~al.}{1992}]{ScharfLahav1992}
{Scharf} C.,  {Hoffman} Y.,  {Lahav} O.,   {Lynden-Bell} D.,  1992, \mn@doi
  [\mnras] {10.1093/mnras/256.2.229}, \href
  {http://adsabs.harvard.edu/abs/1992MNRAS.256..229S} {256, 229}

\bibitem[\protect\citeauthoryear{{Schlegel}, {Finkbeiner}  \&
  {Davis}}{{Schlegel} et~al.}{1998}]{Schlegel1998}
{Schlegel} D.~J.,  {Finkbeiner} D.~P.,   {Davis} M.,  1998, \mn@doi [\apj]
  {10.1086/305772}, \href {http://adsabs.harvard.edu/abs/1998ApJ...500..525S}
  {500, 525}

\bibitem[\protect\citeauthoryear{{Scolnic} et~al.,}{{Scolnic}
  et~al.}{2018}]{Scolnic2018}
{Scolnic} D.~M.,  et~al., 2018, \mn@doi [\apj] {10.3847/1538-4357/aab9bb},
  \href {http://adsabs.harvard.edu/abs/2018ApJ...859..101S} {859, 101}

\bibitem[\protect\citeauthoryear{{Sellentin} \& {Heavens}}{{Sellentin} \&
  {Heavens}}{2016}]{2016SellentinHeavens}
{Sellentin} E.,  {Heavens} A.~F.,  2016, \mn@doi [\mnras]
  {10.1093/mnrasl/slv190}, \href
  {http://adsabs.harvard.edu/abs/2016MNRAS.456L.132S} {456, L132}

\bibitem[\protect\citeauthoryear{{Swanson}, {Tegmark}, {Hamilton}  \&
  {Hill}}{{Swanson} et~al.}{2008}]{2008Mangle}
{Swanson} M.~E.~C.,  {Tegmark} M.,  {Hamilton} A.~J.~S.,   {Hill} J.~C.,  2008,
  \mn@doi [\mnras] {10.1111/j.1365-2966.2008.13296.x}, \href
  {http://adsabs.harvard.edu/abs/2008MNRAS.387.1391S} {387, 1391}

\bibitem[\protect\citeauthoryear{{Szapudi}, {Prunet}, {Pogosyan}, {Szalay}  \&
  {Bond}}{{Szapudi} et~al.}{2001}]{PolSpiceSzapudi2001}
{Szapudi} I.,  {Prunet} S.,  {Pogosyan} D.,  {Szalay} A.~S.,   {Bond} J.~R.,
  2001, \mn@doi [\apjl] {10.1086/319105}, \href
  {http://adsabs.harvard.edu/abs/2001ApJ...548L.115S} {548, L115}

\bibitem[\protect\citeauthoryear{{Takahashi}, {Sato}, {Nishimichi}, {Taruya}
  \& {Oguri}}{{Takahashi} et~al.}{2012}]{Takahashi2012}
{Takahashi} R.,  {Sato} M.,  {Nishimichi} T.,  {Taruya} A.,   {Oguri} M.,
  2012, \mn@doi [\apj] {10.1088/0004-637X/761/2/152}, \href
  {http://adsabs.harvard.edu/abs/2012ApJ...761..152T} {761, 152}

\bibitem[\protect\citeauthoryear{{Thomas}, {Abdalla}  \& {Lahav}}{{Thomas}
  et~al.}{2010}]{Thomas2010Neutr}
{Thomas} S.~A.,  {Abdalla} F.~B.,   {Lahav} O.,  2010, \mn@doi [Physical Review
  Letters] {10.1103/PhysRevLett.105.031301}, \href
  {http://adsabs.harvard.edu/abs/2010PhRvL.105c1301T} {105, 031301}

\bibitem[\protect\citeauthoryear{{Thomas}, {Abdalla}  \& {Lahav}}{{Thomas}
  et~al.}{2011a}]{Thomas2011b}
{Thomas} S.~A.,  {Abdalla} F.~B.,   {Lahav} O.,  2011a, \mn@doi [Physical
  Review Letters] {10.1103/PhysRevLett.106.241301}, \href
  {http://adsabs.harvard.edu/abs/2011PhRvL.106x1301T} {106, 241301}

\bibitem[\protect\citeauthoryear{{Thomas}, {Abdalla}  \& {Lahav}}{{Thomas}
  et~al.}{2011b}]{Thomas2011}
{Thomas} S.~A.,  {Abdalla} F.~B.,   {Lahav} O.,  2011b, \mn@doi [\mnras]
  {10.1111/j.1365-2966.2010.18004.x}, \href
  {http://adsabs.harvard.edu/abs/2011MNRAS.412.1669T} {412, 1669}

\bibitem[\protect\citeauthoryear{{Wandelt}, {Hivon}  \& {G{\'o}rski}}{{Wandelt}
  et~al.}{2001}]{Polspice0}
{Wandelt} B.~D.,  {Hivon} E.,   {G{\'o}rski} K.~M.,  2001, \mn@doi [\prd]
  {10.1103/PhysRevD.64.083003}, \href
  {http://adsabs.harvard.edu/abs/2001PhRvD..64h3003W} {64, 083003}

\bibitem[\protect\citeauthoryear{{Wang}}{{Wang}}{2008}]{2008PhRvD..77l3525W}
{Wang} Y.,  2008, \mn@doi [\prd] {10.1103/PhysRevD.77.123525}, \href
  {http://adsabs.harvard.edu/abs/2008PhRvD..77l3525W} {77, 123525}

\bibitem[\protect\citeauthoryear{{Wang} et~al.,}{{Wang}
  et~al.}{2017}]{2017WangBOSS}
{Wang} Y.,  et~al., 2017, \mn@doi [\mnras] {10.1093/mnras/stx1090}, \href
  {http://adsabs.harvard.edu/abs/2017MNRAS.469.3762W} {469, 3762}

\bibitem[\protect\citeauthoryear{{Wright}, {Smoot}, {Bennett}  \&
  {Lubin}}{{Wright} et~al.}{1994}]{Wright1994}
{Wright} E.~L.,  {Smoot} G.~F.,  {Bennett} C.~L.,   {Lubin} P.~M.,  1994,
  \mn@doi [\apj] {10.1086/174919}, \href
  {http://adsabs.harvard.edu/abs/1994ApJ...436..443W} {436, 443}

\bibitem[\protect\citeauthoryear{{Xavier}, {Abdalla}  \& {Joachimi}}{{Xavier}
  et~al.}{2016}]{Flask2016}
{Xavier} H.~S.,  {Abdalla} F.~B.,   {Joachimi} B.,  2016, \mn@doi [\mnras]
  {10.1093/mnras/stw874}, \href
  {http://adsabs.harvard.edu/abs/2016MNRAS.459.3693X} {459, 3693}

\bibitem[\protect\citeauthoryear{{Zhao} et~al.,}{{Zhao}
  et~al.}{2017}]{2017ZhaoBOSS}
{Zhao} G.-B.,  et~al., 2017, \mn@doi [\mnras] {10.1093/mnras/stw3199}, \href
  {http://adsabs.harvard.edu/abs/2017MNRAS.466..762Z} {466, 762}

\bibitem[\protect\citeauthoryear{{de Haan} et~al.,}{{de Haan}
  et~al.}{2016}]{deHaan2016}
{de Haan} T.,  et~al., 2016, \mn@doi [\apj] {10.3847/0004-637X/832/1/95}, \href
  {http://adsabs.harvard.edu/abs/2016ApJ...832...95D} {832, 95}

\makeatother
\end{thebibliography}



\appendix
\section{The Overdensity Pseudo-$C_{\ell}$ Estimator}\label{Apx:PCL1}

Our aim is to measure the angular power spectrum of the galaxy overdensity field, $\delta^g$. Let $\bar{\rho}^g$ be the average of $\rho^g$ over the sky and define the galaxy overdensity field to be  

\EQ{}{
\delta^g = \frac{\rho^g - \bar\rho^g }{\bar\rho^g } = \frac{\rho^g }{\bar\rho^g} -1.}

\noindent This field may be represented using spherical harmonic expansion:

\EQ{}{
\delta^g(\theta, \phi) = \sum_{\ell=0}^{\ell_{max}} \sum_{m=-\ell}^{\ell} d_{\ell m} Y_{\ell m}(\theta, \phi),}

\noindent where the spherical harmonic coefficients $d_{\ell m}$ are defined by

\EQ{}{
d_{\ell m} = \int \delta^g(\theta,\phi) Y_{\ell m}^*(\theta,\phi) d\Omega.}

\noindent Here and in what follows we have fixed a coordinate system $(\theta, \phi)$ for the celestial sphere; the spherical harmonic functions are defined with respect to this coordinate system. Our estimator of the angular power spectrum of the data is then

\EQ{}{
\hat{D}_{\ell} = \frac{1}{2\ell +1} \sum_{m = -\ell}^{\ell} d_{\ell m}^{} d_{\ell m}^{\ast}.}

\noindent The averaging over $m$ is motivated by the assumed isotropy of the probability distribution governing the location of galaxies.

To handle the partial-sky case, let $\Omega_{tot}$ be the survey region and define

\EQ{Jlm}{
J_{\ell m} = \int_{\Omega_{tot}} \left|Y_{\ell m} \right|^2d\Omega \ .}

\noindent This is a normalization factor due to the average of modes in the partial sky coverage; note that $J_{\ell m}=1$ for a full-sky survey. There will also be a term correcting for bias introduced by the partial sky measurement. However this term is proportional to the average field value; in our case this average vanishes, so the bias correction need not be made. See Appendix \ref{Apx:PCL2} for details. 

We can repeat this analysis for galaxy density fields $\rho^{g, i}$ and $\rho^{g, j}$ defined in tomographic bins $i$ and $j$. Combining the partial sky effect and tomographic binning results in an estimator $\hat{D}_{\ell}^{i j}$ for the cross- $(i \neq j)$ or auto- $(i = j)$ power spectrum of the data

\EQ{D_hat}{
\hat{D}_{\ell}^{i j} = \frac{1}{2\ell +1} \sum_{m = -\ell}^{\ell}D^{ij}_{\ell m}}

\noindent where

\EQ{D_lm_ij} {
D^{ij}_{\ell m} = \frac{\Re(d_{\ell m}^i d_{\ell m}^{j \ast})}{J_{\ell m}}.
}

\noindent Here we take the real part $\Re()$ of a quantity whose expectation value will have no imaginary part.


In reality we work with a pixelised celestial sphere and we measure not $\rho^g$ but rather a galaxy count $n^g_p$ per pixel $p$. From this we derive the per-pixel galaxy overdensity

\EQ{define_delta_p}{
\delta^g_p = \frac{n^g_p}{\Delta \Omega_p} \frac{\Delta \Omega_{tot}}{n^g_{tot}} - 1,}

\noindent where $n^g_{tot}$ is the total galaxy count, $\Delta \Omega_p$ the solid angle subtended by pixel $p$, and $\Delta \Omega_{tot}$ the total solid angle of the survey region $\Omega_{tot}$.

On the pixelated sphere, the spherical harmonic coefficients are estimated by

\EQ{AlmPix}{
d_{\ell m} = \sum_p \delta^g_p Y_{\ell m}^{\ast}(\theta_p,\phi_p) \Delta\Omega_p ,}

\noindent where $(\theta_p,\phi_p)$ are the coordinates at the centre of pixel $p$, $\Delta\Omega_p$ is the area of $p$, and the sum is over all pixels in the survey region.

Pixelisation is a smoothing operator, and hence suppresses power at small scales. We summarise here the standard treatment of this effect; see \cite{Healpix,Boris2013} and the \healpix documentation for details.\footnote{For details on the pixel window function: \url{https://healpix.jpl.nasa.gov/html/intronode14.htm}} Consider the contribution of a given pixel $p$ to $d_{\ell m}$ both for the (measured) pixelised field and for the (desired) ideal continuous field; the ratio of these quantities is 

\EQ{}{
w_{\ell m}^p = \frac{\int_p Y_{\ell m}(\theta,\phi) d\Omega}{Y_{\ell m}(\theta_p,\phi_p) \Delta \Omega_p}.}

\noindent This quantity depends sensitively on $\ell$: for small $\ell$, $Y_{\ell m}$ is slowly varying and hence $w_{\ell m}^p$ will be close to unity while for large $\ell$ the rapidly varying $Y_{\ell m}$ will have vanishing integral over $p$. However the dependence on $m$ and $p$ will be small and can be averaged out (in quadrature), yielding:

\EQ{pixwin}{
w_{\ell}^2 = \frac{1}{N_{pix}(2\ell+1)} \sum_{p,m} \left| w_{\ell m}^p \right|^2.}

\noindent The ratio of the power spectra of the (measured) pixelised overdensity field to that of the (desired) continuous field will then be $w_{\ell}^2$. This study uses a \healpix resolution of $N_{side} = 512$; this means that at $\ell_{max} = 510$ this ratio of powers ($C_{\ell}^{pix}/C_{\ell}^{unpix}$) due to the pixel window function  is then $0.911$.

\section{Correspondence between overdensity and number counts Pseudo-$C_{\ell}$ estimators}\label{Apx:PCL2}
In Section \ref{Sec:Measurements}, we showed the Pseudo-$C_{\ell}$ estimator for galaxy overdensity maps. To link this with what is most commonly done in the literature, one can show that this galaxy overdensity measure is closely related to the more familiar galaxy number counts estimator as seen in \cite{Peebles1973,ScharfLahav1992,FisherLahav1994,Blake2007,Thomas2011}. For the purpose of this section, we define the galaxy overdensity quantities with an upper $\delta$ index and the number counts quantities with a $n$ upper index. Example: the galaxy overdensity angular power spectra is represented by $C^{\delta}_{\ell}$.

We start the derivation by multiplying the overdensity spherical harmonics coefficients from Equation \eqref{Eq:AlmPix} by $\bar n^g_i  = n^g_{tot,i}/\Delta \Omega_{tot}$. Equation \eqref{Eq:D_lm_ij} then becomes:

\begin{ceqn}
\begin{align}
\label{doesnt_look_helpful_but_is}
{\bar n^g_i \bar n^g_j} D^{\delta, ij}_{\ell m} \approx & \left[ \sum_p^{N_{pix}} \delta_{p,i}^g \Delta\Omega_p \bar n^g_i Y_{\ell m}^{\ast}(\theta_p,\phi_p) \right] \\ \nonumber
&\times \left[ \sum_p^{N_{pix}} \delta_{p,j}^g \Delta\Omega_p \bar n^g_j Y_{\ell m}^{\ast}(\theta_p,\phi_p) \right]
\end{align}
\end{ceqn} 
where we bear in mind that different subsamples $ i $ and $j$ can have different total numbers of galaxies and different galaxies in each pixel, but use the same pixels. 

Using Equation \eqref{Eq:define_delta_p} we can write

\EQ{}{
\delta_p^g \Delta\Omega_p \bar n^g = n_p^g - \Delta\Omega_p \bar n^g.
}

\noindent Now, we can use the above expression to rewrite Equation \ref{doesnt_look_helpful_but_is} as:

\begin{ceqn}
\begin{align}
{\bar n^g_i \bar n^g_j} D^{\delta, ij}_{\ell m} & \approx  \left[ \sum_p^{N_{pix}} \left( n_{p,i}^g - \bar n^g_i \Delta\Omega_p \right) Y_{\ell m}^{\ast}(\theta_p,\phi_p) \right] \\ \nonumber
&\times  \left[ \sum_p^{N_{pix}}  \left( n_{p,j}^g - \bar n^g_j \Delta\Omega_p \right) Y_{\ell m}^{\ast}(\theta_p,\phi_p) \right] .
\end{align}
\end{ceqn} 

Let us take one of the square brackets and analyse the sub-terms individually:

\EQ{}{
\sum_p^{N_{pix}} Y_{\ell m}(\theta_p,\phi_p)\Delta\Omega_p \approx \int Y_{\ell m}^{\ast}(\theta,\phi) d\Omega \equiv I_{\ell m}\, .
}
We can therefore see that this term is approximately equivalent to the shot noise correction term $I_{\ell m}$ from \cite{Blake2007,Thomas2011}. The second term can also be re-expressed as:

\begin{ceqn}
\begin{align}
\sum_p^{N_{pix}} Y_{\ell m}(\theta_p,\phi_p) n_p^g &\approx \sum_{g^\prime} Y_{\ell m}(\theta_{g^\prime},\phi_{g^\prime}) \\ \nonumber
								&= \sum_{g^\prime} \int \delta_D(x_{g^\prime}-x) Y_{\ell m}(\theta,\phi) d\Omega
\end{align}
\end{ceqn}

\noindent where the index $g^\prime$ runs over galaxies in the sample that have not been excluded by the mask and $\delta_D(x)$ is the Dirac delta function. We can reverse the order of summation and integration, and express the number count function as:

\EQ{}{
\sigma_1 = \sum_{g^\prime} \delta(x_{g^\prime}-x),
} 

\noindent i.e., the galaxy distribution is a sum of delta functions at the locations of the galaxies, and hence the integral over this function is the total number of galaxies in that area. The function $\sigma_1$ is the filtered galaxy distribution, which has been masked. It is related to the full galaxy distribution $\sigma_0$ by:

\EQ{}{
\sigma_1(\theta,\phi) = \sigma_0(\theta,\phi) W(\theta,\phi),
}

\noindent where $W: S^2 \rightarrow \mathbb{B}$ is a binary filter, and:

\EQ{}{
\sigma_0 = \sum_g \delta(x_g - x)
}

\noindent runs over the full underlying set of galaxies.

We can therefore write:

\begin{ceqn}
\begin{align}
\sum_p^{N_{pix}} Y_{\ell m}(\theta_p,\phi_p)n_p^g  & \approx \int \sigma_0(\theta,\phi)W(\theta,\phi)Y_{\ell m}(\theta,\phi)d\Omega \\ & = a_{\ell m}
\end{align}
\end{ceqn}

\noindent where $a_{\ell m}$ are the spherical harmonic coefficients of the filtered galaxy number count field. Finally, we end up with 

\begin{ceqn}
\begin{align}
\bar n^g_i \bar n^g_j D^{\delta, ij}_{\ell m} & \approx  \frac{\left[ a_{\ell m}^i - \bar n_i^g I_{\ell m} \right] \left[ a_{\ell m}^j - \bar n_j^g I_{\ell m} \right]}{J_{\ell m}} \\ & = D^{n,ij}_{\ell m}\, ;
\end{align}
\end{ceqn}

\noindent in other words, the overdensity and number count power spectra differ only by a factor of the number density of galaxies in each tomographic bin involved.

\section{Code comparison}\label{Apx:Code_Comparison}
We produce our results using \class \citepbox{Class} (background evolution and perturbations) and the {\textit{Unified Cosmological Library for $C_{\ell}$s}} code, \uclcl (projected statistics). Here, we show a comparison for $C_{\ell}$s calculated with both \class (integrated functionality from the former \texttt{CLASSGAL} code \citepbox{CLASSgal}) and \texttt{CAMBSources} \citepbox{CambSources}, matching cosmologies as closely as possible. We also show the derivatives calculated with respect to key cosmological parameters. In this comparison we used Gaussian redshift bins, since this is the functionality provided in \class and \texttt{CAMBSources}. Two redshift bins are chosen with $\bar{z} = \{0.5,0.6\}$ and $\sigma_z = 0.05$ to be of comparable size to the redshift bins used in the body of the paper; auto and cross-correlations are calculated. Codes are run with their default accuracy parameters. 

\subsection{Auto- and cross-correlation precision}
\begin{figure}
\begin{center}
\includegraphics[width=\columnwidth]{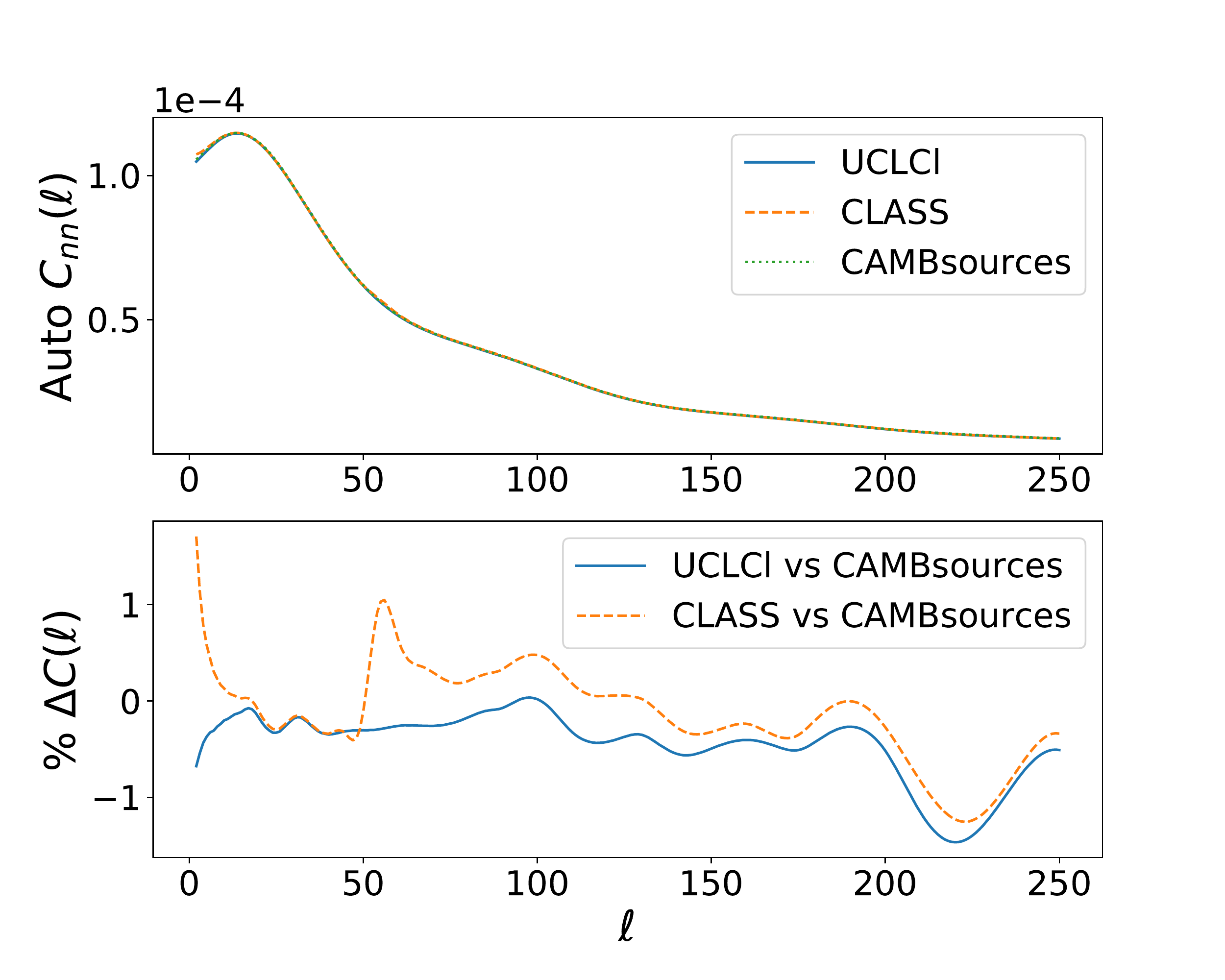}
\caption{Auto-correlation $C_{\ell}$ ($\bar z = 0.5, \sigma_z = 0.05$) comparison the three codes \uclcl, \class, and \texttt{CAMBSources}. The upper panel shows the three $C(\ell)$s over-plotted, whilst the lower panel shows the percentage difference between \uclcl / \class compared to \texttt{CAMBSources}: $\frac{C_\camb-C_{\uclcl / \class}}{C_{\camb}}\times 100$.}
\label{fig:Auto_Precision}
\end{center}
\end{figure}

The auto-power spectrum for a bin with $\bar{z} = 0.5$ and $\sigma_z = 0.05$ is shown in Figure \ref{fig:Auto_Precision}, calculated in each of the three codes for a flat $\Lambda$CDM cosmology with $\Omega_b = 0.05$, $\Omega_{cdm} = 0.25$, $h = 0.67$, $\log(A_s \times 10^{-10}) = 3.2$, $n_s = 0.95$. Codes are in sub-percent level agreement up to $\ell \approx 200$ (although the \class low $\ell$ RSDs disagree to a slightly larger extent), after which there is a small discrepancy between \class and \texttt{CAMBSources} non-linear density perturbations. As one might expect, the differences between \textsc{uclcl} and \texttt{CAMBSources} trace the differences between \class and \texttt{CAMBSources} as the former two share the same perturbations, i.e. $P(k)$.

\begin{figure}
\begin{center}
\includegraphics[width=\columnwidth]{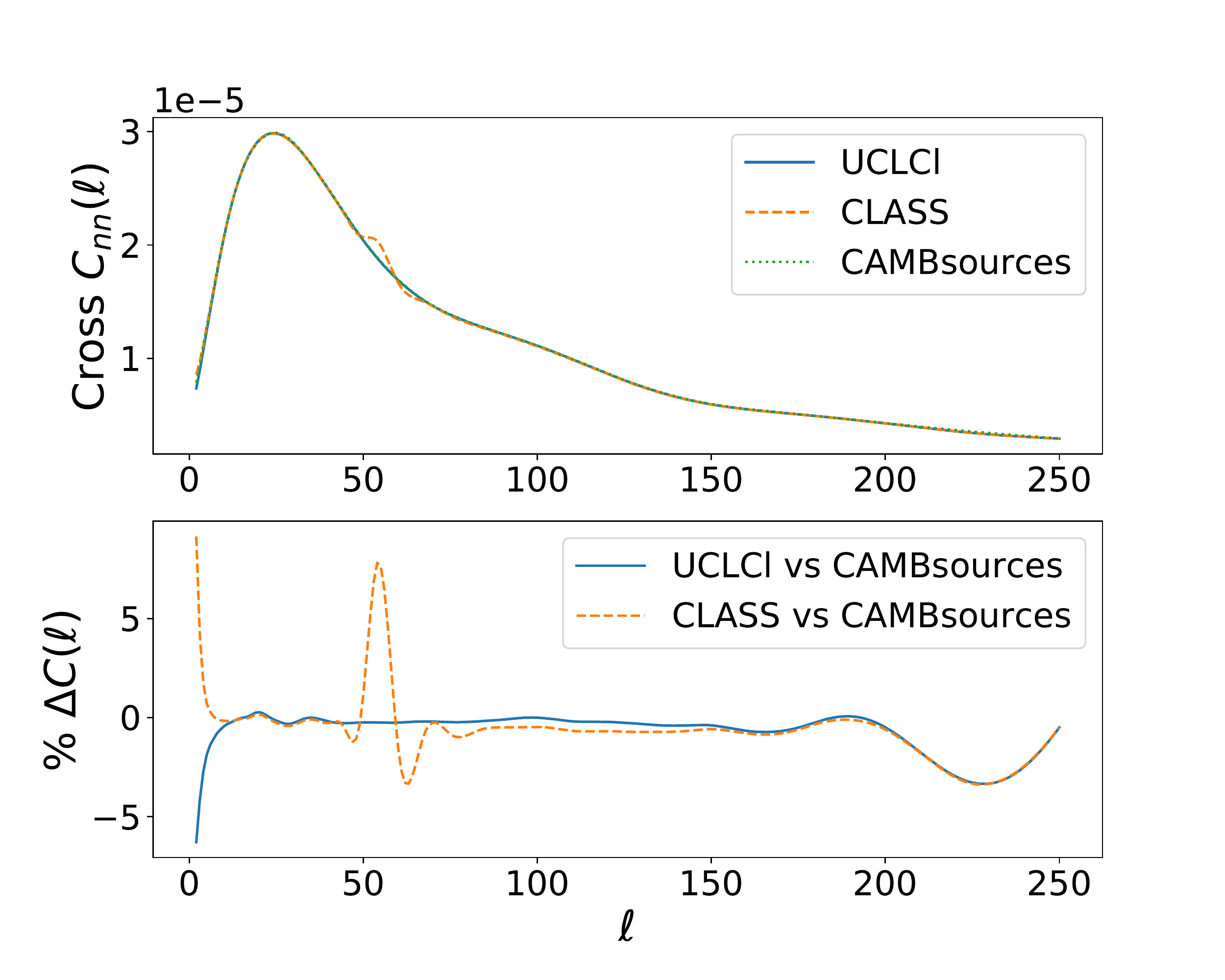}
\caption{Cross-correlation $C_{\ell}$ ($\bar z^i = 0.5, \bar z^j = 0.6, \sigma_z = 0.05$) comparison the three codes \uclcl, \class, and \texttt{CAMBSources}. The upper panel shows the three $C_{\ell}$s over-plotted, whilst the lower panel shows the percentage difference between \uclcl / \class compared to \texttt{CAMBSources}: ${[C_{\camb}-C_{(\uclcl / \class)}]}/{C_{\camb}}\times 100$. Again \uclcl follows \class closely, except in the RSDs and in a distinctive wobble around $l \approx 50$ where \class is transitioning away from the Limber approximation scheme.}
\label{fig:Cross_Precision}
\end{center}
\end{figure}

The same trend is observed in the cross-correlations in Fig \ref{fig:Cross_Precision}, with a notable wobble in the \class cross-correlation presumably when transitioning between approximation schemes (and thus possibly remedied by adjusting accuracy parameters away from the default). 

\subsection{Sensitivity to cosmological parameters} \label{App_accuracy}

In order to check that the accuracy of the codes is not strongly cosmology dependent, the comparison are also made for variations on $h$ and $w_0$ over sensible ranges of the parameters. It is crucial that the sensitivity to the cosmological parameters not be overwhelmed by the (approximately percent level) uncertainty in the $C_{\ell}$ calculation itself. It is also important to check that the derivatives w.r.t. the cosmological parameters are consistent between the codes, as this will ensure the $C_{\ell}$s change consistently as one moves away from the fiducial cosmology.

\begin{figure}
\begin{center}
\includegraphics[width=\columnwidth]{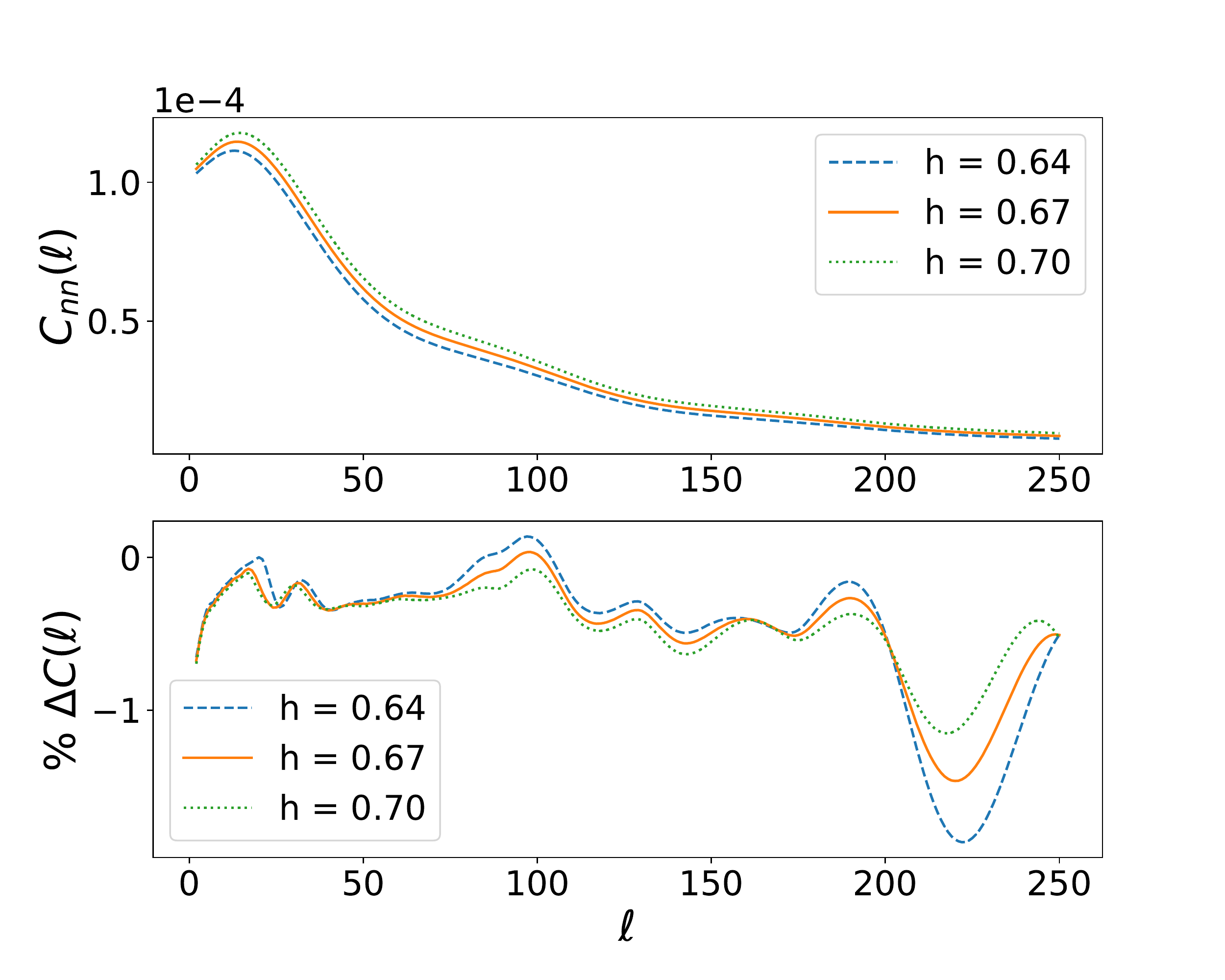}
\caption{The top panel shows the auto-correlations ($\bar z = 0.5$) for three values of $h$ calculated in \uclcl. The lower panel shows the percentage difference of each of these $C_{\ell}$s with the corresponding $C_{\ell}$s from \texttt{CAMBSources} (matching values of $h$).}
\label{fig:h_Precision}
\end{center}
\end{figure}

In Figure \ref{fig:h_Precision} one can see that the $C_{\ell}$s for $h = 0.64, 0.67, 0.70$ are clearly delineated and their differences significantly larger than the differences between the $C_{\ell}$s from different codes. With $w$ over the range $-1.1$ to $-0.9$, shown in Figure \ref{fig:w_Precision}, one can see that at low $\ell$ the $C_{\ell}$s are well distinguished from each other, but at high $\ell$ $w_0$ has little effect, and thus is unlikely to be distinguished from the uncertainties inherent in the non-linear regime. In Figure \ref{fig:Diff_w} one can also see that the variation at high $\ell$ is significantly different for \uclcl and \texttt{CAMBSources}, likely originating from the difference in the perturbations between \class and \texttt{CAMBSources}. Nevertheless, the shape of the derivatives w.r.t. to $w_0$ up to $\ell \approx 200$, and w.r.t. $h$ throughout the $\ell$ range, look consistent with \texttt{CAMBSources}. This shows that the $C_{\ell}$s are changing in the correct way around this fiducial cosmology, and will yield the correct shape of posterior contours. 

\begin{figure}
\begin{center}
\includegraphics[width=\columnwidth]{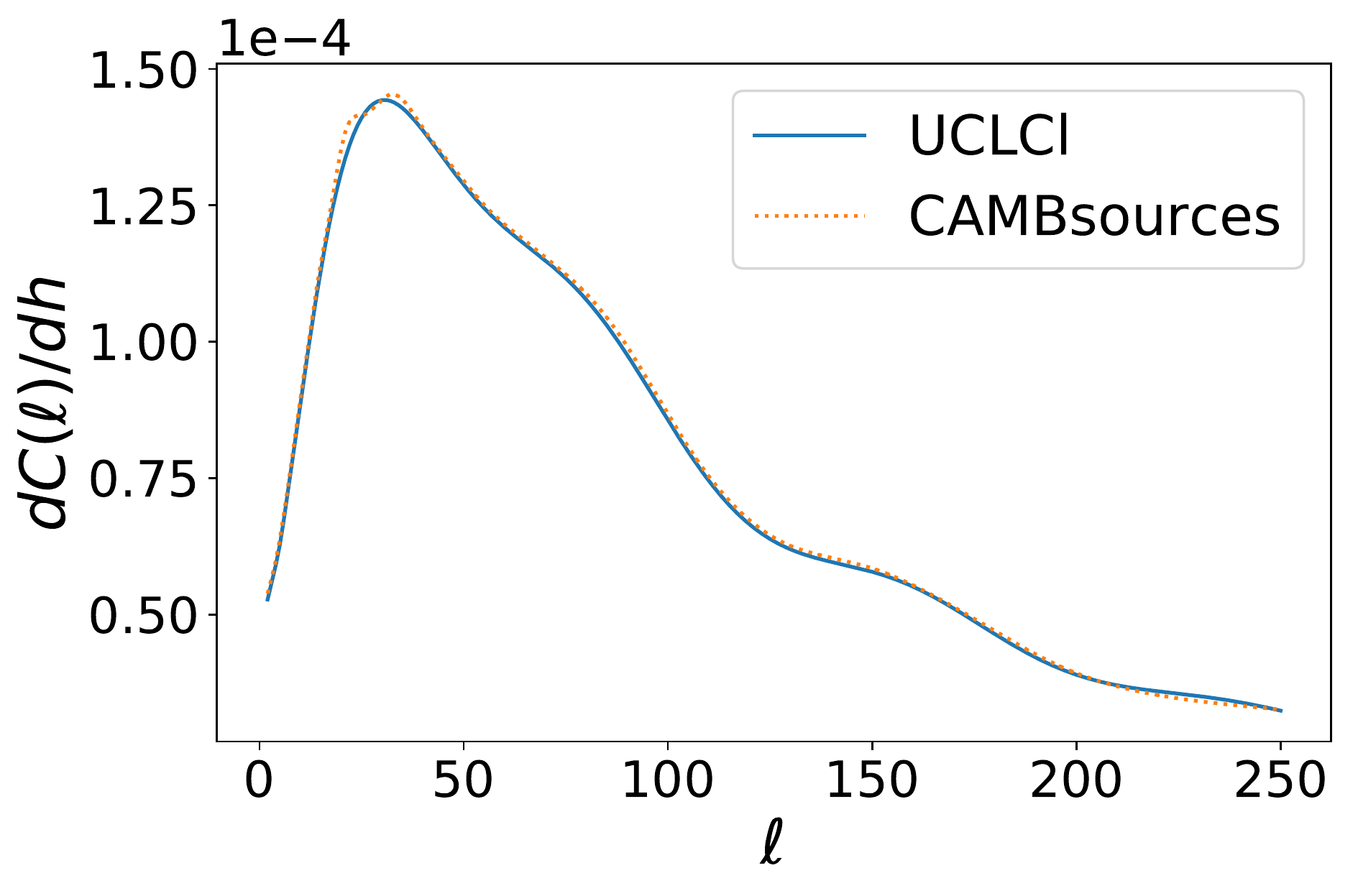}
\caption{Comparison of $C_{\ell}$ derivatives $\frac{d C_{\ell}}{dh}$ between \uclcl and \texttt{CAMBSources}.}
\label{fig:Diff_h}
\end{center}
\end{figure}

\begin{figure}
\begin{center}
\includegraphics[width=\columnwidth]{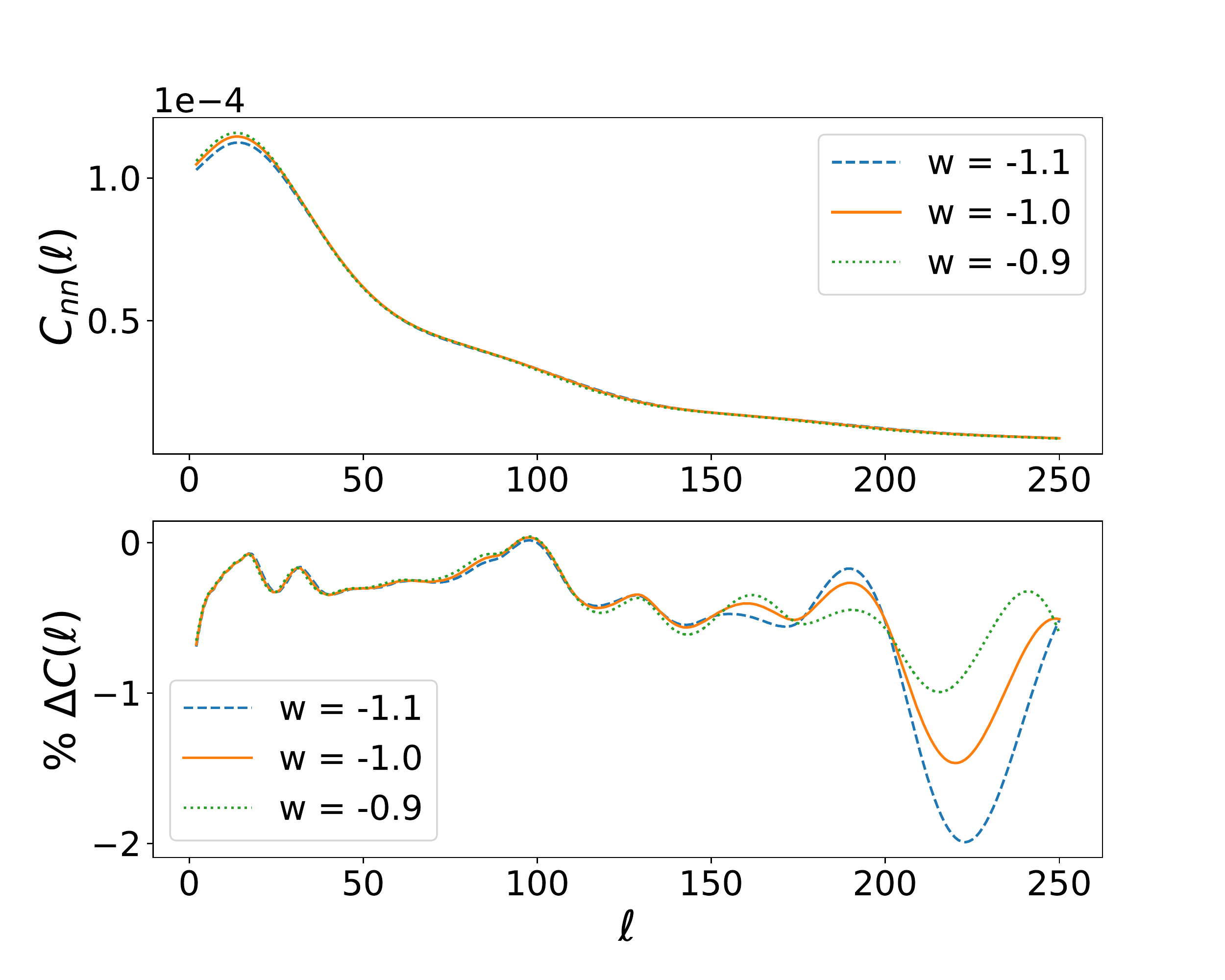}
\caption{The top panel shows the auto-correlations ($\bar z = 0.5$) for three values of $w$ calculated in \uclcl. The lower panel shows the percentage difference of each of these $C_{\ell}$s with the corresponding $C_{\ell}$s from \texttt{CAMBSources} (matching values of $w$).}
\label{fig:w_Precision}
\end{center}
\end{figure}

\begin{figure}
\begin{center}
\includegraphics[width=\columnwidth]{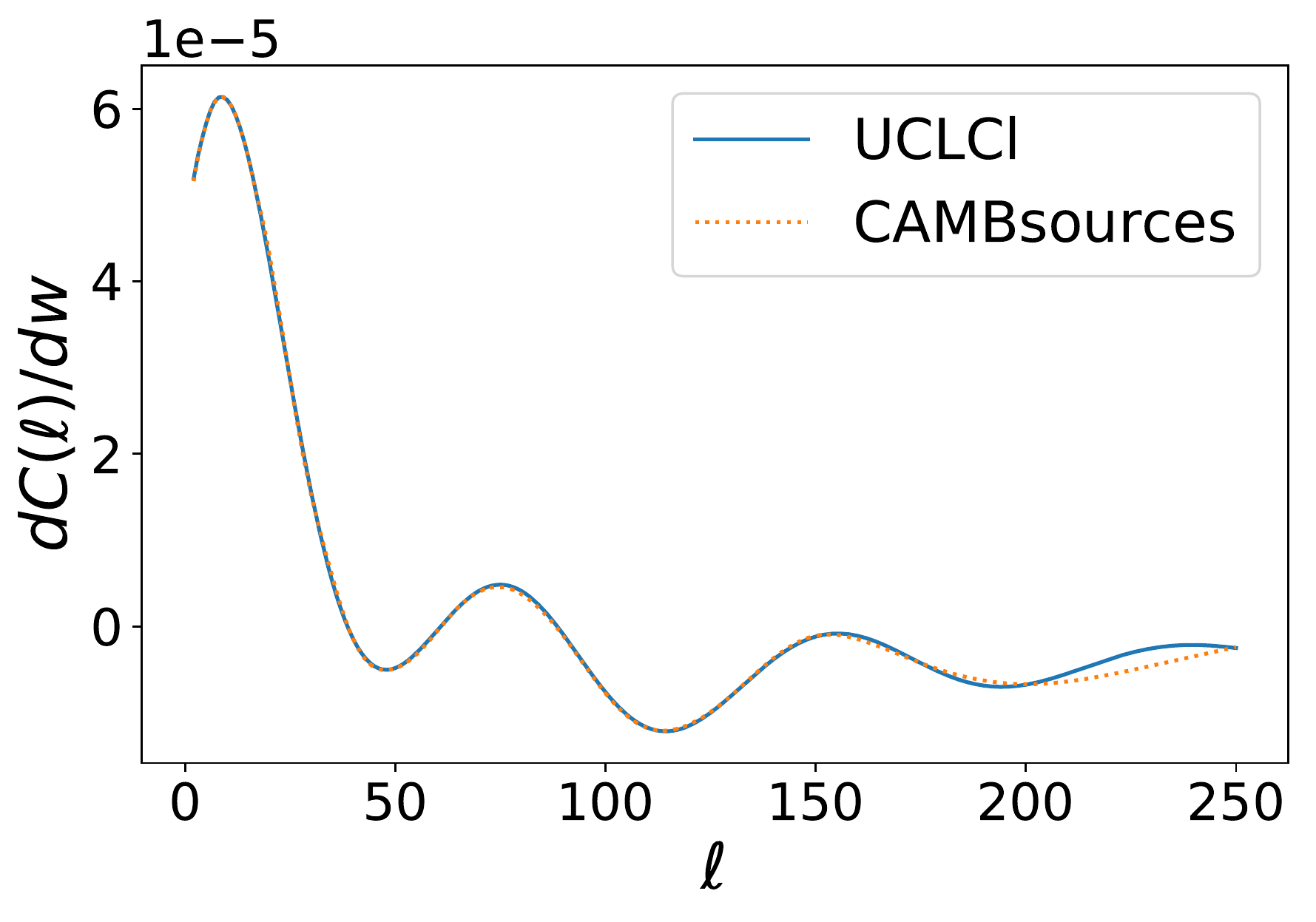}
\caption{Comparison of $C_{\ell}$ derivatives $\frac{d C_{\ell}}{dw}$ between \uclcl and \texttt{CAMBSources}. Here we see a more significant difference at high $\ell$, which can also been seen in Figure \ref{fig:w_Precision}. This characteristic bump appears to come from a difference in the \class and \texttt{CAMBSources} non-linear perturbations.}
\label{fig:Diff_w}
\end{center}
\end{figure}

\bsp	
\label{lastpage}
\end{document}